\documentclass{article}
\usepackage[utf8]{inputenc}
\usepackage{graphicx}
\usepackage{cite}
\usepackage{comment}
\usepackage{appendix}
\usepackage{color}
\usepackage{amsfonts,amsmath,amssymb,amsthm}
\theoremstyle{Plain}
\usepackage{float}
\newtheorem{theorem}{Theorem}

\usepackage{lipsum}
\usepackage{caption}
\usepackage{subcaption}
\usepackage{multicol}
\usepackage[left=1.5 cm,right=1.5 cm,top=1.5cm,bottom=1.5cm]{geometry}
\numberwithin{equation}{section}
\title{Stage structured prey-predator model incorporating mortal peril consequential to inefficiency and habitat complexity in juvenile hunting}
\author{Debasish Bhattacharjee$^1$, Tapasvini Roy$^2$, Santanu Acharjee$^3$,Tarini Kumar Dutta$^4$ \\$^{1,2,3}$Department of Mathematics, Gauhati University, Assam, India\\
$^4$Department of Mathematics, Assam Don Bosco University, Assam, India\\
e-mails:$^1$debabh2@gmail.com,$^2$tapasviniroy@gmail.com,\\$^3$sacharjee326@gmail.com,$^4$tkdutta2001@yahoo.com}
\date{}
\begin{document}
\maketitle

\begin{abstract}
    Dynamic exploration for a predator-prey bio-system of two species with ratio-dependent functional response is carried out, where the capability to predate in both the stages of the predator, the juvenile and the matured, is taken into account. But, only the matured predators are inferred to be efficient in killing the prey without any negative repercussions. The mortality risks for the juvenile predators are attributable to the inefficiency rate of juveniles coupled with habitat complexity which is either in the form of anti-predator behavior of the prey taken with the aid of their habitat or in the form of a territorial generalist mesopredator. So as to avoid extinction of either of the species and to preserve the food chain of the ecological system, the results pertaining to the existence and stability of all the equilibrium points of the bio-system along with permanence, transcritical and Hopf bifurcation has been thoroughly studied. Corroboration of the results along with the dependence of the biosystem on some crucial parameters is done through numerical simulation. It is found that juvenile predators' inefficiency relative to the resistance confronted, plays a crucial role to control each species density of the ecosystem, as an intriguing limit cycle between the trivial and axial equilibriums of the proposed system along with the co-existing periodic point, because of some ineffeciency parametric value of the juvenile predator has been witnessed.
\end{abstract}

\textbf{2020 AMS classifications:} 34C23,92D25,92D40.\\

\begin{multicols}{2}
\section{Introduction}
Predator-prey dynamics are one of the most studied synergy in population ecology. The interactions between the prey and the predator can be studied with the help of mathematical models; the first ecological mathematical model was proposed by Lotka and Volterra in the first quarter of the $20^{th}$ century\cite{apr2,lot,vol}. For over a century, many scientists and researchers are engaged in coming up with various alterations to the foundational Lotka-Volterra model, to comply with more realistic physical scenarios. One such modifiable reality in the model is the risk involvement in predation, which is dependent on types of prey species\cite{muk}.\\
\par  In nature, the line dividing the hunter and the hunted are often known to be blurred. The scenario witnessed between lizards and scorpions, lobsters and whelks is that of role reversal between prey and its predator\cite{lob,liz}. The exploration done in this very field of role-reversal leaves much to be desired. The prey, when faced with their impending demise, more often than not turn towards counter-attacking its predator. Pumas are intermittently killed as a consequence of hunting porcupines or large ungulates, by being pierced or crushed by antlers or horns or being bashed into trees or punctured by tree limbs\cite{elb}. However, it is to be noted that prey are seldom successful in killing the adult predator by reason of them being much more equipped to kill juvenile or sub-adult predators. It has been observed that the juvenile prey that were able to elude predation, eradicated juvenile predators expeditely after maturity\cite{srep}.  Hence, shielding both themselves and their offsprings from predation.  Few Mathematical models have been proposed taking into account the role reversal or where predators are at risks due to the anti-predator behaviour of the prey\cite{ratio,rol,rol1,rol2}.
\vspace{6pt}\par Contrary to popular assumption, it has evidenced that the juveniles of the predators can provide for themselves(i.e hunt). Snakes, juvenile walleye, snails, fishes, Clinus superciliosus are among some of such predators that hunt during their juvenile period of life\cite{snake, snail, wall, csup} . During predation,  juvenile predators are at mortal peril when they engage with dangerous prey due to them being inexperienced and having smaller body sizes than the adults, which is more or less same to that of the adult prey.
The role reversal scenerio between adult prey S.celarius and juvenile Phytoseiid predator can be taken as examplar. It is recognized that immature phytoseiid predator may feed on adult spider mites; also the other way around it is divulged that the immature stages of the phytoseiid predator are sometimes killed by the prey spider mites, S.celarius \cite{phy}. It is interesting to see that adults of the predator were not harmed, and could kill and feed on spider mite females and males. Moreover, researchers \cite{sam1,sam2,sam3} studied this particular aspect of swapping of role between juvenile predator and the prey,( i.e. during predation by juveniles either one of the  prey or the juvenile may be killed) while the adult predator can successfully kill the prey without any negative ramification; with holling type1 functional response with and without delay.\\
\par Another scenerio where prey can be termed as "dangerous prey" is diseased prey. Infectious disease is known to play an important role in nature. A lot of research work has been done in the theory and application of epidemiology modelling to the predator-prey population after the work of Anderson and May(1986), Hadeler and Freedman(1989), Chattopadhyay and Arino(1999)\cite{inf1,inf2,inf3}. Infectious disease crossing the species barrier and becoming fatal for the other species has been extensively studied. Several studies revealed that the infected prey can be slayed more effortlessly compared to that of non-infected prey and that the predator too would come to be infected after the consumption of these prey. Thus, the infected prey can be termed as ‘dangerous’ for the predators. The less-experienced predator would plausibly pursue the infected prey. But, we shall not foray into this particular road for our paper. \\
\par We shall concentrate on assembling a bio-system where both juvenile and matured  predators have the ability to hunt but only matured predators are competent enough to be immune towards prey’s efforts to thwart them. Negative repercussions of juvenile hunting has been probed into; the prey taking the aid of their habitat for anti-predator behavior has not yet been reckoned with. In our paper, we shall study implicitly the scenerio where the juveniles are at mortal peril owing to their inefficiency and prey's anti-predator deportment which is their habitat, bereft of which the prey are not by any manner of means baneful; with the help of ratio dependent functional response, making it quite disparate from \cite{sam2, sam3} .\\ 
\par Most of the researches advocate the concept that structurally complex habitat often impedes with the persuance of predator’s forage, thereby enhancing survivorship of the prey. Alteration in foraging competence in the midst of development of anatomy in different habitats within the population of same species has been documented in perch, stickleback, and bluegill sunfish \cite{refu,refu1}. Therefore, it cannot be denied that the extent of fruition of prey surviving through complex habitat (can also be termed as refuge) is dependent upon the predator.\\
\par Prey females of S.nanjingensis has been documented to oftentimes use their nests made of silk web with the intention of locking out the immature predators (T.bambusae), where they subsequently died of starvation. Nonetheless, the adult predator could not be detained from invading their nests\cite{phy2}.
Bats, such as golden tipped bats, myotis emarginatus and the likes, are known as spider-specialists as more than 75 percent of their diets consists of web building spiders. There are instances of bats getting ensnared in the web of spiders and dying of exhaustion, starvation, dehydration, and/ or hyperthermia (the spider may or may not directly kill/eat them); it is espied that some of the captured bats were juveniles and sub-adults, the large(adult) bats being capable of flying right through or avoid the web . Large orb weaving spider, such as Nephila spp  are sometimes known to feed on bats entangled on their webs  and on contrary golden tipped bats feed primarily on the orb weavers.\\ Also, bats that feed on insects ensnared in webs, while hovering in front of them may sometimes get entangled after inadvertently bumping into the web \cite{bat}.\\
\par Another picture that can be painted with our proposed model is that between top-predator, meso-predator, and the prey. Both the predators are territorial and the top-predator is a specialist predator while meso-predator is a generalist one. During ontogenesis, juveniles of the top-predator are not invulnerable to meso-predators, hence when they step into meso-predator’s territory to forage for its prey, they may get killed, while the adults has no such constraints as they can overpower the meso-predator. For example, wolves and coyotes can be taken as the top and  intermediate predators respectively. Wolves feed specifically on white-tailed deer;  coyotes,  as a generalist omnivore, has a much more variable diet.  Coyotes are known to attack and harass wolves in neutral or their own territories (both of their territories are sometimes known to overlap) \cite{coyo1,coyo2}. For our model, we can take wolves as our predator and white tailed deer as our prey\\
\par Eliciting from all these theoretical results, we shall discuss the mathematical model with stage structure in the predator, where owing to inefficiency and habitat complexity juvenile predators are racked  from mortal risks while hunting its prey. Construction of model is discussed in section 2. In section 3 positivity and boundedness of the solution has been investigated. In section 4, the equilibrium points are discussed along with the local and global stabilities of the system at these points. In section 5, Uniform persistance is analysed. In section 6, transcritical and hopf-bifurcation is studied. Finally, Numerical simulation are presented followed by conclusion.

\section{Mathematical Construction of the Bio-system}
For the construction of the biosystem concerning two interacting species, the prey and the predator, the prey is taken to have a logistic growth rate in the absence of the predator, owing to limited resources in nature.
Let X(T) and Y(T) be biomass densities of the prey and the predator respectively, r is the intrinsic growth rate and K the carrying capacity of environment. Thus the logistic equation is given below
$$\dot{X}= X r(1-\frac{X}{K}). \hspace{10pt} X(0)>0$$
Functional Response is a key element in the mathematical model governing the predator-prey  interaction. All the functional responses can be grouped into three different types of functional responses : 
\begin{enumerate}
  \item Prey Dependent
 \item Predator Dependent
 \item  Multi-species Dependent
\end{enumerate}
In prey dependent, the functional response is affected only by prey. Holling I,II, III,IV, Ivlev type and Rosenzweig type are such types of functional responses. In the simplest form, a prey-dependent predator-prey model is given as:
\begin{align*}
    \Dot{X}&= X f(X/K) -Y g(X)\\
    \Dot{Y}&= \mu_1 Y g(X) - DY 
\end{align*}
with $X(0)>0,  Y(0)>0$\\
Here, D is the natural death rate of the predators, $\mu_1$ is the conversion rate of killed prey into juvenile predator. \\ 
Beddington–DeAngelis type, Crowley–Martin type, Hassell–Varley type are the examples of predator-dependent functional responses. Such type of functional responses are affected by both predator and prey population. Generally, they are of the form
\begin{align*}
    \Dot{X}&= X f(X/K) -Y h(X, Y)\\
    \Dot{Y}&= \mu_1 Y h(X, Y) - DY
\end{align*}
with $X(0)>0,  Y(0)>0$\\
In multi-species dependent functional response, not just the focal prey and predator species, but other species and subsidiary elements that may be of some consequence to the functional response are incorporated.It is generally used when three or more species are involved. Modeling multispecies systems is a complex work, and in these models, the functional response has a particularly important part to play\cite{mults}. \\
\par  In our model, we employ ratio-dependent functional response. This is a particular type of predator dependent functional response which relies on the ratio of prey population size to the predator population size, rather than the definite number of either of the species. This functional response is a finer fit for our model as search and conquer rate is supposed to be at a substaintial position to affect the interaction between the predator and the prey. Many biologists accredits ratio-dependent functional response to be more propitious in depicting real life scenerios\cite{ardi-1,ardi-2,ar-3}. The general form of ratio-dependent functional response is as follows:
\begin{align*}
    \Dot{X}&= X f(X/K) - p(Y/X)\\
    \Dot{Y}&= \mu_1 p(Y/X) - DY
\end{align*}
with $X(0)>0,  Y(0)>0.$\\
\par  Partitioning the predator species into juvenile and matured stage, $Y_j$, $Y_m$ are their biomass densities respectively. Both juvenile and matured stages possesses the ability of predation and follows the ratio dependent functional response, $\phi_1(\frac{Y_j}{X},\frac{Y_m}{X})$, $\phi_2(\frac{Y_j}{X},\frac{Y_m}{X})$ respectively. During ontogenesis, the juveniles reproductive system is not fully developed, and hence only the adults reproduce. Predation by adults corresponds to continuation of the species, i.e. reproduction, thereby enhancing the biomass density of their youngs. While successful predation by juveniles conforms to their own survival and has no impact on the changes in the biomass density. 
\begin{align*}
\dot{X}&=X r(1-X/K)-Y_j\phi_1(\frac{Y_j}{X},\frac{Y_m}{X})-Y_m\phi_2(\frac{Y_j}{X},\frac{Y_m}{X})\\
\dot{Y_j}&=\mu_1\phi_2(\frac{Y_j}{X},\frac{Y_m}{X})Y_m-CY_j-D_1Y_J\\
\dot{Y_m}&=CY_j-D_2Y_m
\end{align*}
$\text{with }X(0)>0,Y_j(0)>0,Y_m(0)>0.$\\
Here, C is the maturation rate of juvenile predator, $D_1,D_2$ are natural death rate of juvenile and matured predator respectively.\\
The success of juvenile predators in killing the prey is substantially interconnected to the prey being outside of their habitat or the juvenile predator staying out of the meso-predator’s territory (for simplification we shall go by habitat complexity). Assimilating habitat complexity leaves \textit{X(1-n)} of the prey that are available for juvenile predator to hunt. The functional response of juveniles is transformed to $\phi_1(\frac{Y_j}{X(1-n)},\frac{Y_m}{X(1-n)})$, where $n\in[0,1]$ is the habitat complexity.\\
In addition to the lack of robustness, juvenile predators has scant knowledge and experience in predation, making them inefficient. The mortal peril due to this inefficiency has been factored into our biosystem as $\phi_3(\frac{Y_j}{X},\frac{Y_m}{X})$.
The negative repercussions of predation by juveniles when coupled with the habitat complexity is remoulded to $n\phi_3(\frac{Y_j}{X},\frac{Y_m}{X})$.  Habitat complexity is omitted in the functional response, since it doesn’t hinder the juvenile predator from getting killed, but rather aids, and hence the multiplication. The root cause that warrants the deaths of juvenile predators is not the prey themselves but their habitat complexity, in the absense of which ($n=0$), the juveniles would be at complete liberty to hunt each and every prey individuals without experiencing any ramification. Also, in the course of events where juveniles are fully efficient($B=0$), the presence of habitat complexity would not result in their deaths, a lot like the scenerio with the matured predators.
$\phi_1=A_1/(m(\frac{Y_j}{X(1-n)}+\frac{Y_m}{X(1-n)})+1),$\\
$\phi_2=A_2/(m(\frac{Y_j}{X}+\frac{Y_m}{X})+1),$\\ $\phi_3=B/(m(\frac{Y_j}{X}+\frac{Y_m}{X})+1)$\\
where \textit{m} is the average search and conquer rate of juvenile and matured predator, $A_1,A_2$ being the predation rates of juvenile and matured predator respectively, and \textit{B} is the inefficiency rate of juvenile predators.\\ \\
Our  biosystem is:\\
\begin{equation}
  \begin{aligned}  
  \frac{dX}{dT} &=rX(1 -\frac{X}{K}) -\frac{ A_1(1 - n)XY_j}{(m(Y_j+Y_m) +X(1 - n))}\\
  &-\frac{A_2 XY_m}{(m(Y_j+Y_m)+X)}\\
  \frac{dY_j}{dT} &=\frac{\mu_1A_2XY_m}{(m(Y_j+Y_m)+X)}-\frac{BnXY_j}{(m(Y_j+Y_m)+X)}\\
  &-CY_j-D_1Y_j \\
  \frac{dY_m}{dT} &= CY_j-D_2Y_m \\
  \end{aligned}
\label{eq1}
\end{equation}
with initial condition $$X(0)>0, Y_j(0)>0, Y_m(0)>0.$$ 
Now, simplifying the biosystem by reducing the number of parameters by taking  
$t=rT,\hspace{5pt} x =X/K,\hspace{5pt} y =Y_j/K,\hspace{5pt} z =Y_m/K,\hspace{5pt} A_3=A_2\mu_1$,
the equations are now transformed to:
\begin{equation}
 \begin{aligned}
\frac{dx}{dt} &= x(1-x)-\frac{a_1(1-n)xy}{(m(y+z)+(1-n)x)}\\
&-\frac{a_2xz}{(m(y+z)+x)}\\
\frac{dy}{dt} &= \frac{a_3xz}{(m(y+z)+x)}-\frac{bnxy}{(m(y+z)+x)}\\
&-cy-d_1y\\
\frac{dz}{dt} &= cy-d_2z
\end{aligned}
\label{eq2}
\end{equation}
with initial condition:
\begin{equation} \label{eq2.1}
    x(0)=x_0>0,  y(0)=y_0>0, z(0)=z_0>0
\end{equation}
where, $a_1=A_1/r,\hspace{5pt} a_2=A_2/r,\hspace{5pt} a_3=A_3/r,\hspace{5pt} b=B/r,\hspace{5pt} c=C_1/r,\hspace{5pt} d_1=D_1/r,\hspace{5pt} d_2=D_2/r.$
\end{multicols}
\begin{figure}[H]
    \centering
    \begin{subfigure}[b]{0.45\textwidth}
         \centering
    \includegraphics[width=\textwidth]{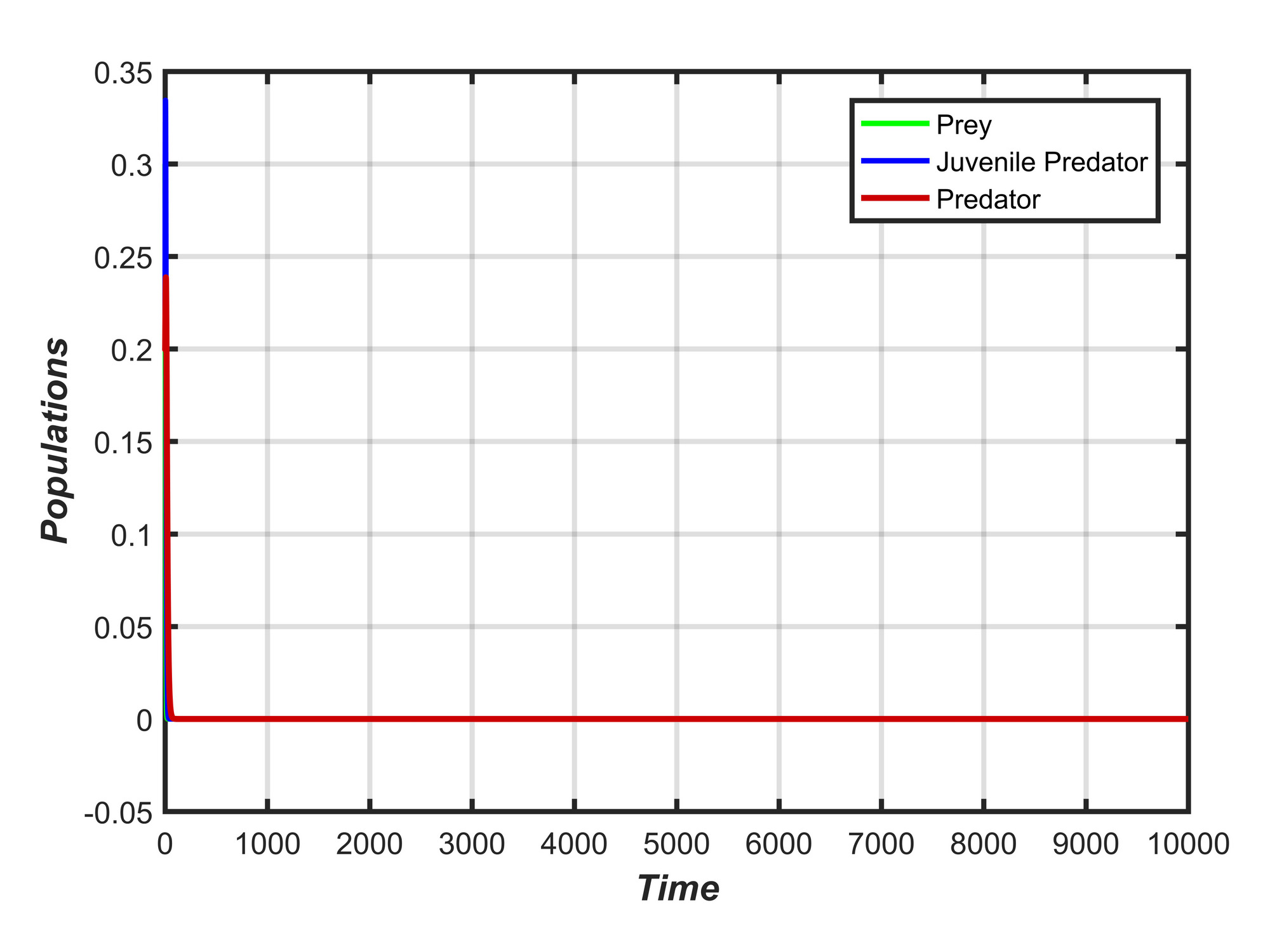}
    \caption{at n=0}
    \label{a}
    \end{subfigure}\hfill
     \begin{subfigure}[b]{0.45\textwidth}
         \centering
    \includegraphics[width=\textwidth]{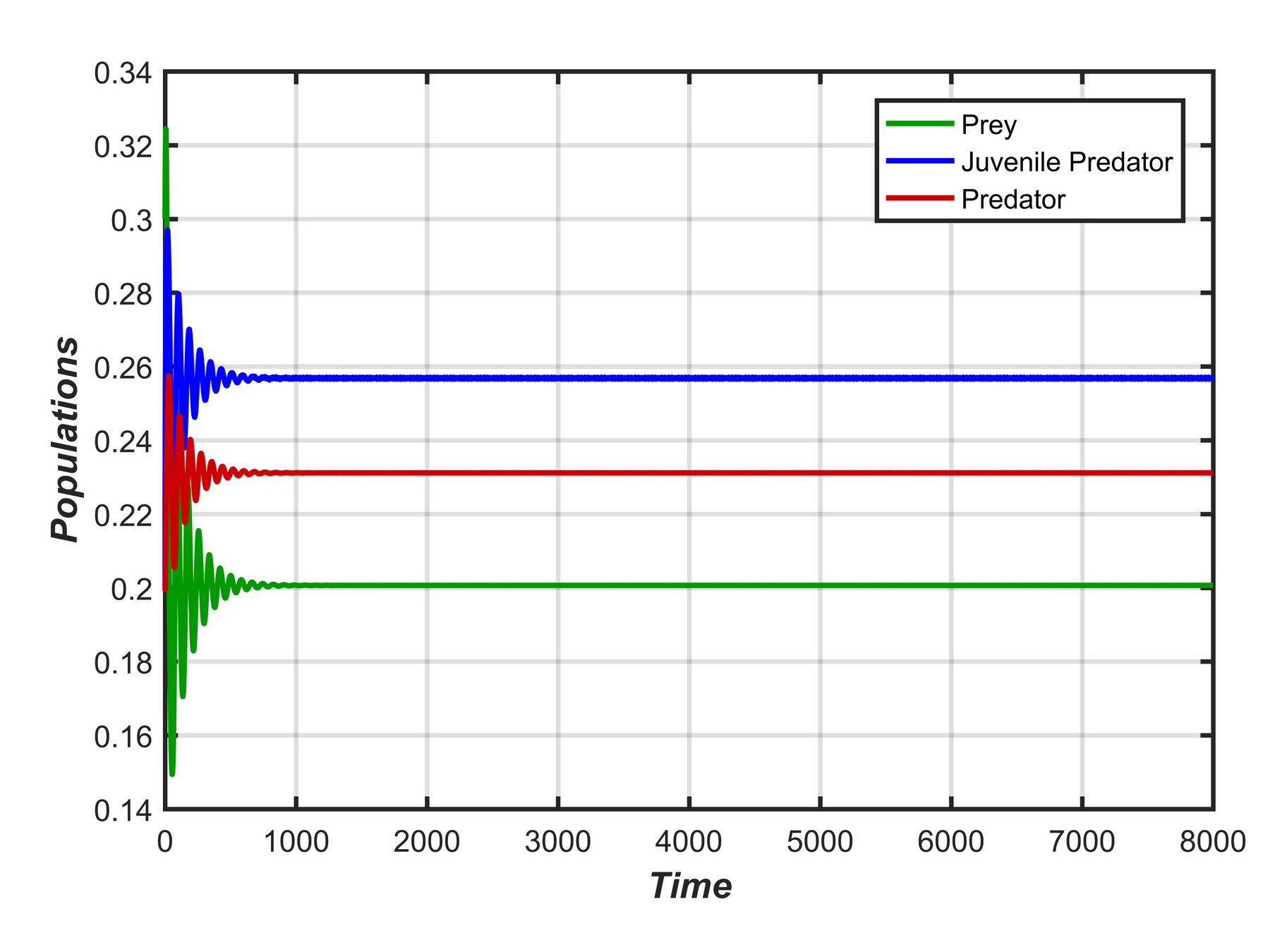}
    \caption{at n=0.44}
    \label{b}
    \end{subfigure}\hfill
     \begin{subfigure}[b]{0.45\textwidth}
         \centering
    \includegraphics[width=\textwidth]{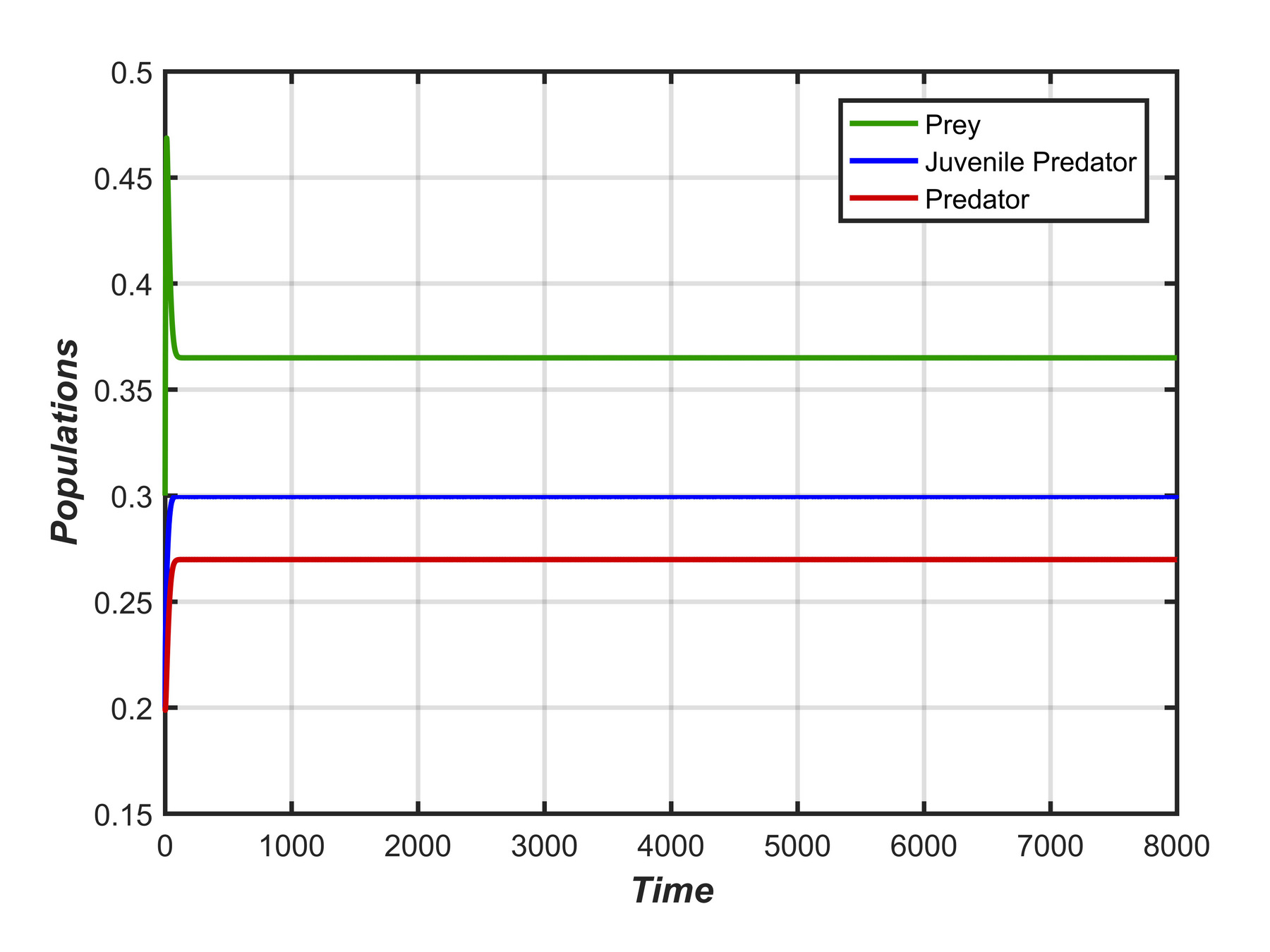}
    \caption{at n=0.65}
    \label{c}
     \end{subfigure}\hfill
      \begin{subfigure}[b]{0.45\textwidth}
         \centering
    \includegraphics[width=\textwidth]{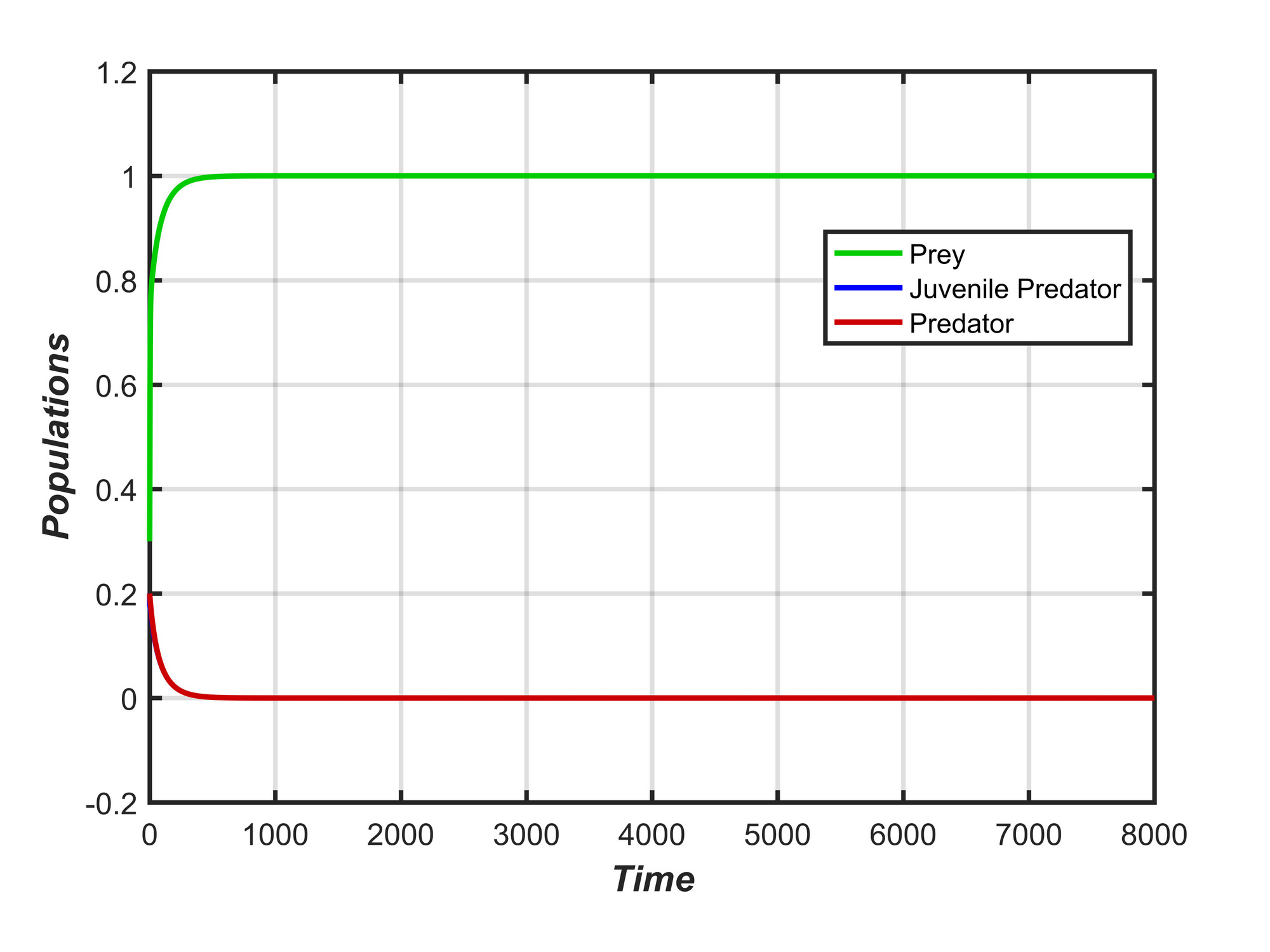}
    \caption{at n=1}
    \label{d}
     \end{subfigure}\hfill
    \caption{Alteration in populations dynamics with varying value of \textit{n} (anti-predator behaviour of prey due to habitat complexity),
parameter values: $a_1=1.25,b=1.5,a_2=1.4,a_3=1.8,m=0.95,c=0.09,d_1=0.2,d_2=0.1$}
    \label{n-time}
\end{figure}
\begin{multicols}{2}
\section{Positivity and Boundedness}
From biological point of view, the prey having logical growth rate, cannot grow exponentially and due to the reliance of the predator population on the prey population, they too are confronted with an identical scenario. The following theorems attests to this very fact of the biosystem \ref{eq2} being restrained within a particular region in the positive octant, numerically.
\begin{theorem}
 Every single one of the solutions of the biosystem \ref{eq2} with initial conditions \ref{eq2.1} are non-negative.
 \end{theorem}
\begin{theorem} \label{bound}
 The solutions of the bio-system \ref{eq2} along with initial condition \ref{eq2.1} are uniformly bounded in $\{(x,y,z)\in R^3_+:p=\frac{a_3(1+d_3)^2}{4d_3}-\tau, \text{ for any } \tau>0\}$, the values of p, $d_3$ are $a_3x+a_2(y+z)$ and $min\{d_1,d_2\}$ respectively
 \end{theorem}
 Refer to appendix (\ref{proof-posi}) and (\ref{proof-bound}) for the proofs.
\section{The Equilibrium points and their Stability}
The ecological system experiences three types of equilibrium which are bionomically realizable. One of them is the obvious vanishing equilibrium $E_1(0,0,0)$. When the preys are not disrupted by the predators, they cherish in the ecological system to the full potential of the system as depicted by the model. Therefore, the long term state of the ecological system would be the equilibrium point $E_2(1,0,0)$. As both the stages of predators are specialist predators, the existence of other types of axial equilibrium are inconceivable. One of the best scenarios of an ecological system is the existence and stability of the co-extant equilibrium $E_3(x^*,y^*,z^*)$ 
In this section, the requirements for all the equilibrium points to be locally and globally stable have been investigated.
\begin{enumerate}
    \item \textbf{The Vanishing Equilibrium point $E_1(0, 0, 0)$}\\
The trivial solution (0, 0, 0) is a saddle point. The stable and unstable manifolds have been investigated along the directions of different axes and planes. The trajectories advancing through either y-axis, z-axis or yz-plane end up at (0,0,0).\\
Refer to appendix(\ref{proof-vanish}) for the proof.
\item \textbf{The Axial Equilibrium or the predator free equilibrium point $E_2(1, 0, 0)$}
\begin{theorem} \label{loca-sta1} The system around the axial point (1, 0, 0) is locally asymptotically stable if
$$0 < a_3 < \frac{(c d_2 + d_1 d_2 + b d_2 n)}{c}$$
\end{theorem}
\begin{theorem} \label{glo-axial}
The sufficient condition for the axial equilibrium point of the bio-system \ref{eq2} alongwith the initial values to be globally asymptotically stable is\\
$a_3z-bnxy<0$ and $a_1a_3(1-n)(1-x)<a_2m(d_1y+d_2z).$
\end{theorem}
Refer to appendix (\ref{proof-stab-axi}) and (\ref{proof-glo-axi}) respectively for the proofs.
\item \textbf{The Co-extant equilibrium point $E_3(x^*, y^*, z^*)$}
The equilibrium point $(x^*,y^*,z^*)$ is given by\\
\begin{align*}
    x^*=&\frac{u}{(c+d_2)m(a_3c-b d_2n)v},\\
    y^*=&\frac{c (a_3 c - d_2 (c + d_1 + b n)) u}{m^2 v (c+d_1) (c+d_2)^2 (a_3 c-b d_2 n)},\\
    z^*=&\frac{c u (a_3 c-d_2 (b n+c+d_1))}{d_2 m^2 v (c+d_1) (c+d_2)^2 (a_3 c-bd_2 n)}
\end{align*}
where,\\ $u=a_2 c [a_3 c - (b + c + d_1) d_2 n] [a_3 c - d_2 (c + d_1 + b n)] - (a_3 c - 
b d_2 n) [(c + d_2) m [a_3 c - (b + c + d_1) d-2 n] - 
a_1 d_2 (-1 + n) (-a_3 c + d-2 (c + d_1 + b n))] $\\
and $v=-a_3 c + (b + c + d_1) d_2 n.$
\begin{theorem} \label{exis}
The necessary and sufficient conditions for the co-extant equilibrium point to exists are
\begin{enumerate}
    \item $(a_3 c) >d_2( c + d_1 + b n)$
    \item $a_2 c [a_3 c-d_2(b n+c+d_1)]<m (c+d_2) (a_3 c-b d_2 n)$
    \item$ 0<a_1<[\{a_3 c-d_2 n (b+c+d_1)\} \{a_2 c (a_3 c-d_2 (b n+c+d_1))-m (c+d_2) (a_3 c-b d_2 n)\}]/[d_2(n-1) (b d_2 n-a_3 c)
    (d_2 (b n+c+d_1)-a_3 c)].$
    \end{enumerate}
\end{theorem}
The proof of this is obvious and hence we omitted it.\\
Existence of co-extant equilibrium point necessitates unstability of predator free equilibrium point, as the first condition given above is in contradiction to theorem\ref{loca-sta1}.
\begin{theorem} \label{loca-sta2} The Equilibrium point $E_3(x^*, y^*, z^*)$ is locally asymptotically stable if  $\chi_1>0, \chi_3>0, \chi_4>0$, where the values of $\chi_1, \chi_3, \chi_4$ are given within the proof.
\end{theorem}
One can see appendix \ref{proof-sta-co}
\begin{theorem} 
\label{glo-coex} The system of equations \ref{eq2} will be globally asymptotically stable around co-extant equilibrium point $E_3 (x^*, y^*, z^*)$, if $\vartheta<0$ where the value of $\vartheta$ is given within the proof.
\end{theorem}
The global stability of the co-extant equilibrium point has been discussed (in appendix \ref{proof-glo-co}) with the help of geometric approach given by  Li and Muldowney\cite{glo-equ} with a parallel technique as given in\cite{ratio}.
\end{enumerate}
\section{Perseverance}
A system of equations can be asserted to persevere for a aeons if there is an existence of a compact region $E \subset {R^3}_+$ with the property that all the solutions of the system with its initial condition eventually enters and resides inside E.\\
For the perseverance of the bio-system \ref{eq2} with initial condition \ref{eq2.1}, we need positive constants $ 0 <\alpha \leq \beta$  such that
\begin{align}
    max \{\limsup_{t \to \infty}x(t), \limsup_{t \to \infty}y(t), \limsup_{t \to \infty}z(t) \} \leq \beta \label{p-eq1}\\
    min \{\liminf_{t \to \infty} x(t),\liminf_{t \to \infty} y(t),\liminf_{t \to \infty} z(t)\} \geq \alpha \label{p-eq}
\end{align}
Equation (\ref{p-eq1}) is already proven through boundedness of the biosystem in the theorem\ref{bound}\\
So, only equation (\ref{p-eq}) is proved in appendix \ref{proof-perse}.
\begin{theorem} \label{persev} The biosystem \ref{eq2} perseveres if $ a_3 c > d_2(bn+c+d_1)$ and $ m>a_1(1-n)+a_2$. \end{theorem}
\section{Bifurcation}{\label{hopf}}
\subsection{Hopf bifurcation}
\textbf{Analysis of Hopf Bifurcation}\\
We orient towards constituting the criteria for Hopf-Bifurcation around the co-extant equilibrium point $E_3(x^*, y^*, z^*) $ with respect to a parameter. Suppose \textit{$n_h$} is the point of bifurcation for the parameter \textit{n},(habitat complexity). The necessary and sufficient  condition for \textit{$n_h$} to be Hopf-bifurcation point is:\\
\begin{enumerate}
    \item$\chi_i (n_h) >0$ for i=1, 2, 3,
    \item$\chi_1 (n_h) \chi_2 (n_h) = \chi_3 (n_h)$,
    \item$[\frac{d}{dn} (\chi_1 \chi_2 -\chi_3)]_{n=n_h} \neq 0.$ 
\end{enumerate}
\textbf{Direction and stability of bifurcating periodic solution} \begin{theorem} \label{ho-thm}
The sign of $\mu_2$ clarifies the trajectory of Hopf bifurcation. The biosystem \ref{eq2} encounters supercritical bifurcation for a positive $\mu_2$  and subcritical bifurcation when $\mu_2<0$.  $\beta_2$ indicates the stability of the bifurcating periodic solution; it is stable for negative $\beta_2$ and unstable for a positive $\beta_2$. The period of bifurcating periodic solution escalates when $T_2>0$ and diminishes when $T_2<0$.
\end{theorem}
\subsection{Trancritical Bifurcation}
\begin{theorem} The biosystem \ref{eq2} is subjected to transcritical bifurcation around the predator free equilibrium point $E_1$ concerning  $b=b_t=(a_3c-cd_2-d_1d_2)/nd_2.$ \end{theorem}
It is to be noted that we can also take other parameters as bifurcating parameter.
\section{Numerical Simulation}
In this section, we perform rigorous numerical simulations taking the assistance of MATLAB software using odes45, MATCONT\cite{matlab} and MATHEMATICA software. The intent for the numerical simulation is to both authenticate the theoretical results developed in the previous sections as well as explicate the rich dynamical scenario under the ascendency of different parameters. Some hypothetical biotically feasible values of parameter have been considered as shown in the following table:
\end{multicols}
\begin{table}[H]
    \centering
    \begin{tabular}{|c|c|c|}
    \hline
   Physical meaning& Parameter&value\\
   \hline
       Predation rate by Juvenile-predator& $a_1$ &1.25 \\
        \hline
        Predation rate by matured predator& $a_2$&1.4\\
         \hline
         Conversion rate of juveniles &$a_3$& 1.8\\
         \hline
         Inefficiency rate of predation by juveniles &$b$& 1.015\\
         \hline
         Search and Conquer rate&$m$&0.95\\
         \hline
        mortality risk of juvenile due to habitat complexity& $n$&0.5\\
         \hline
         transformation rate to adult& $c$&0.09\\
         \hline
         natural death rate of juvenile &$d_1$&0.2\\
         \hline
         natural death rate of juvenile &$d_2$&0.1\\
         \hline
    \end{tabular}
    \caption{Parameter values }
    \label{table}
\end{table}
The parametric values would change when given otherwise.\\
\begin{figure}[H]
    \centering
    \includegraphics[width=10cm]{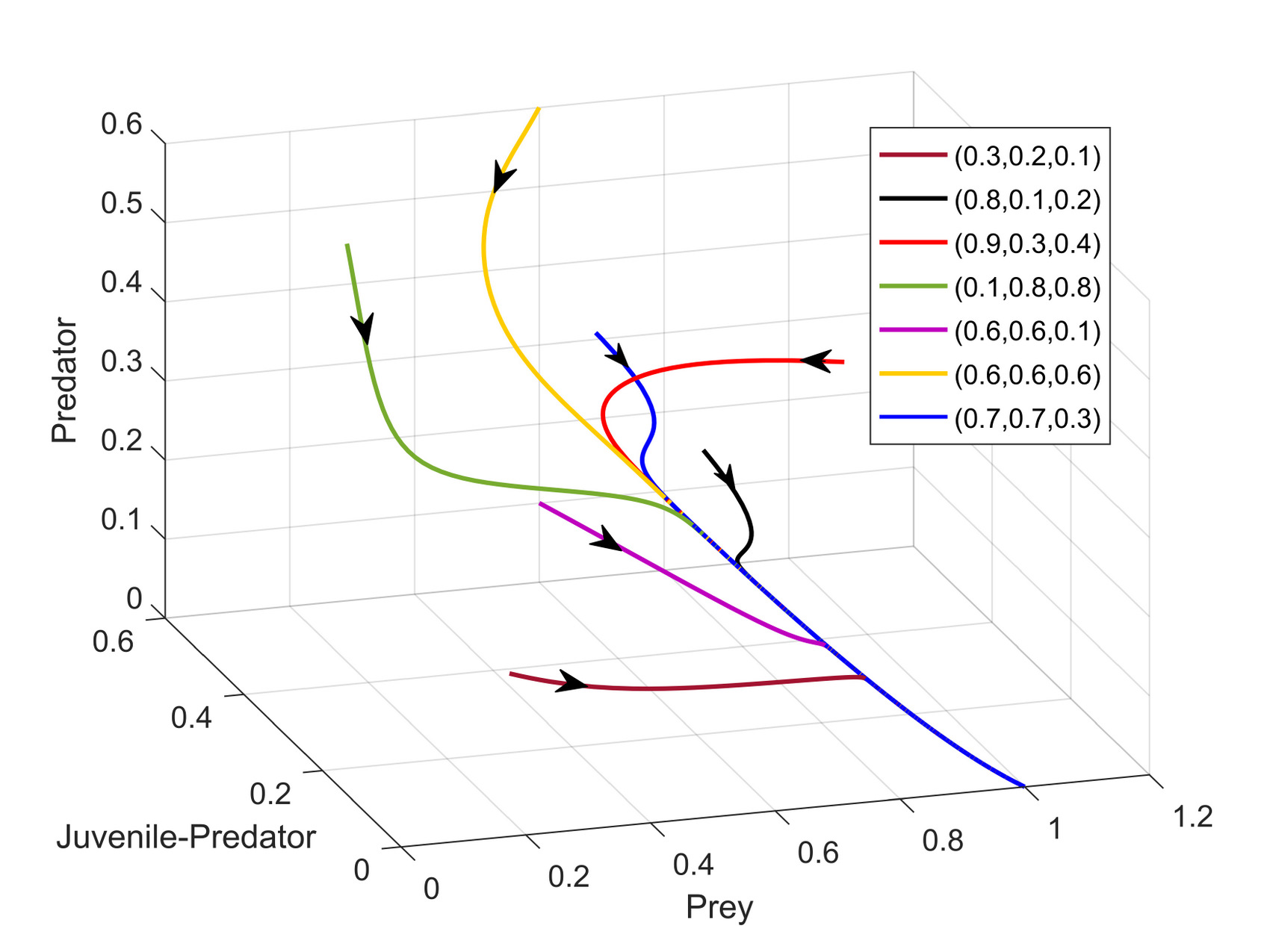}
    \caption{Axial Equilibrium point $E_2(1,0,0)$ being globally asymptotically stable}
    \label{lo-axial}
\end{figure}
\begin{figure}[H]
    \centering
    \includegraphics[width=8cm]{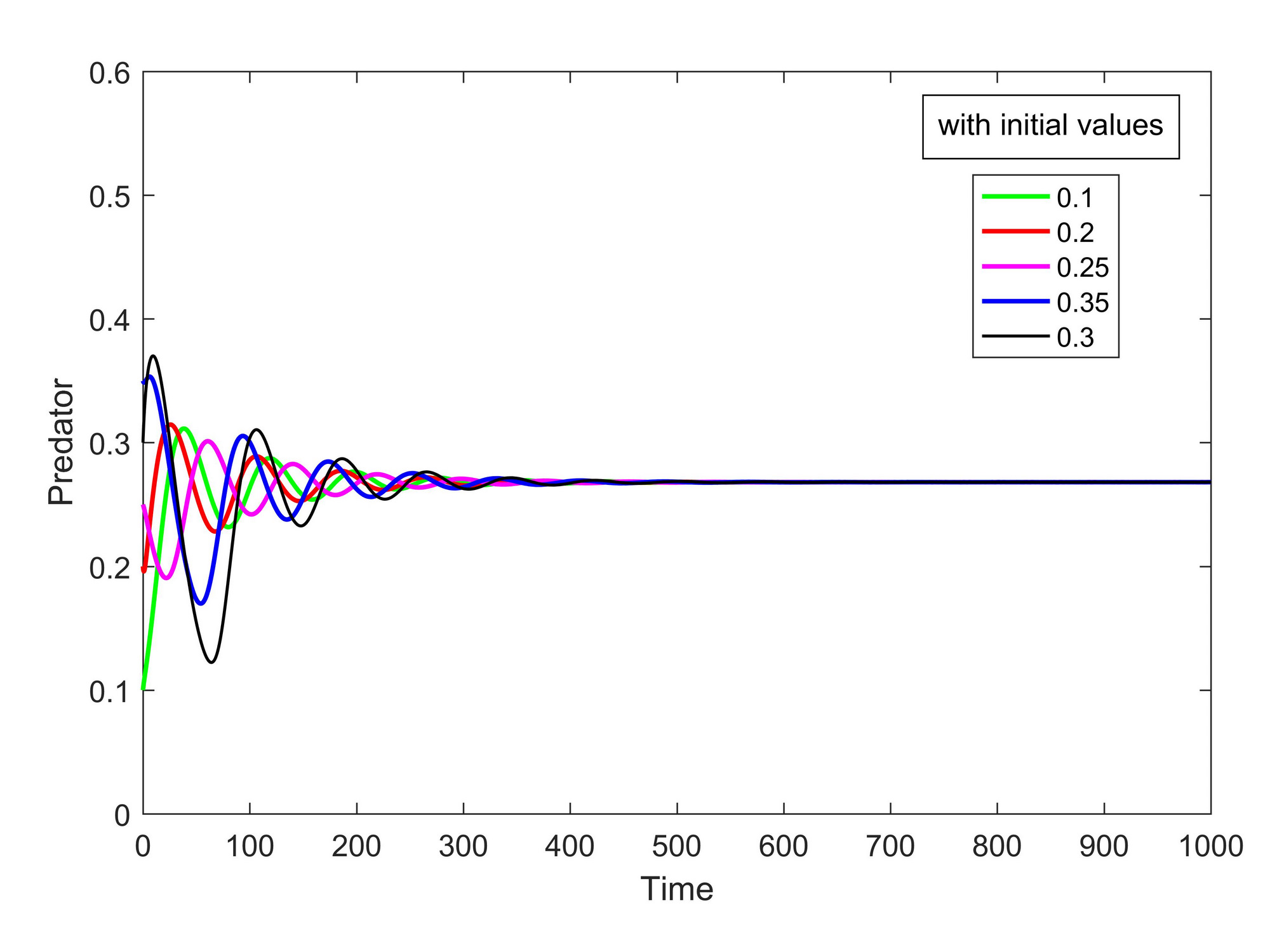}
    \includegraphics[width=8cm]{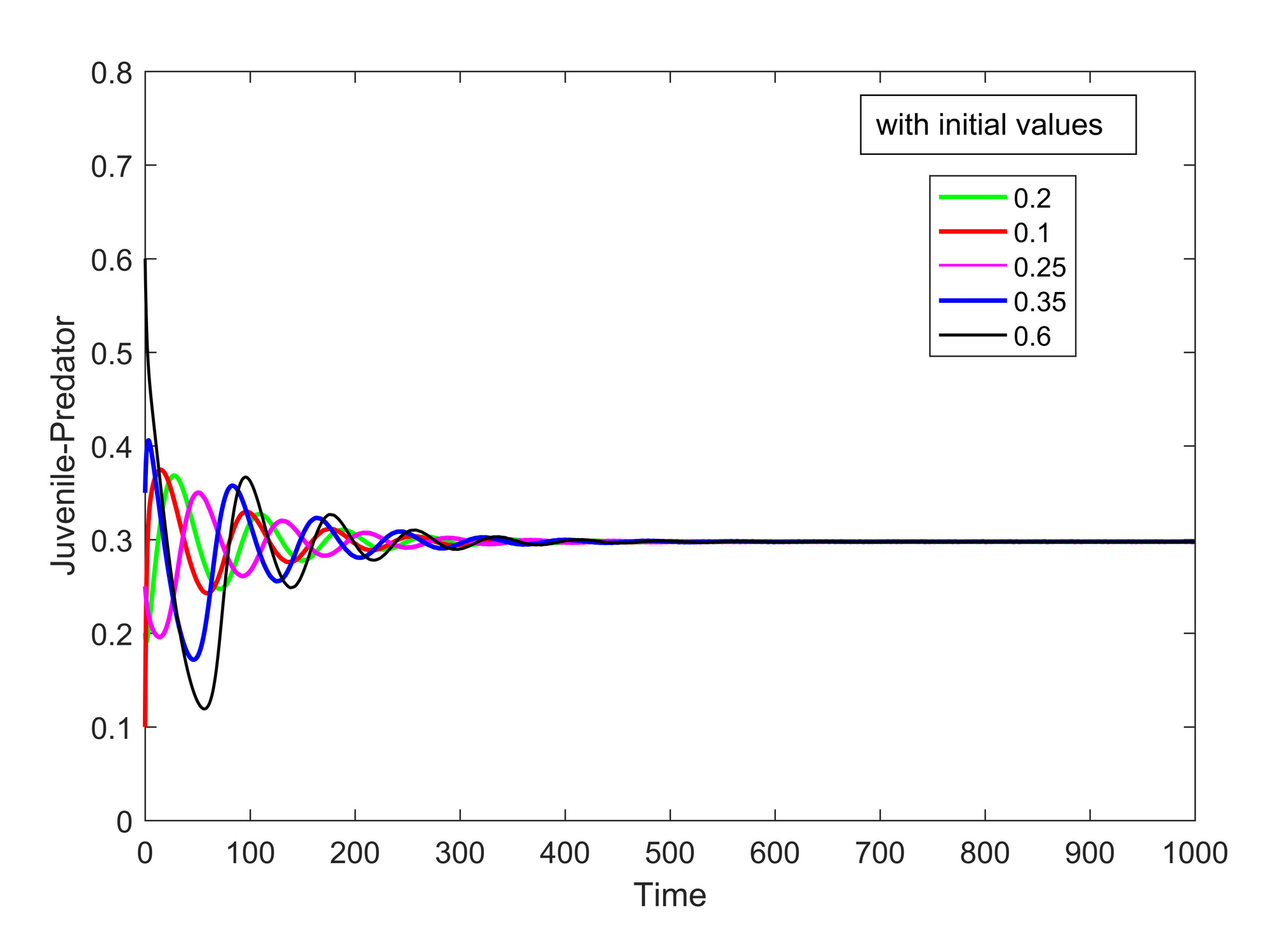}\\
    \includegraphics[width=8cm]{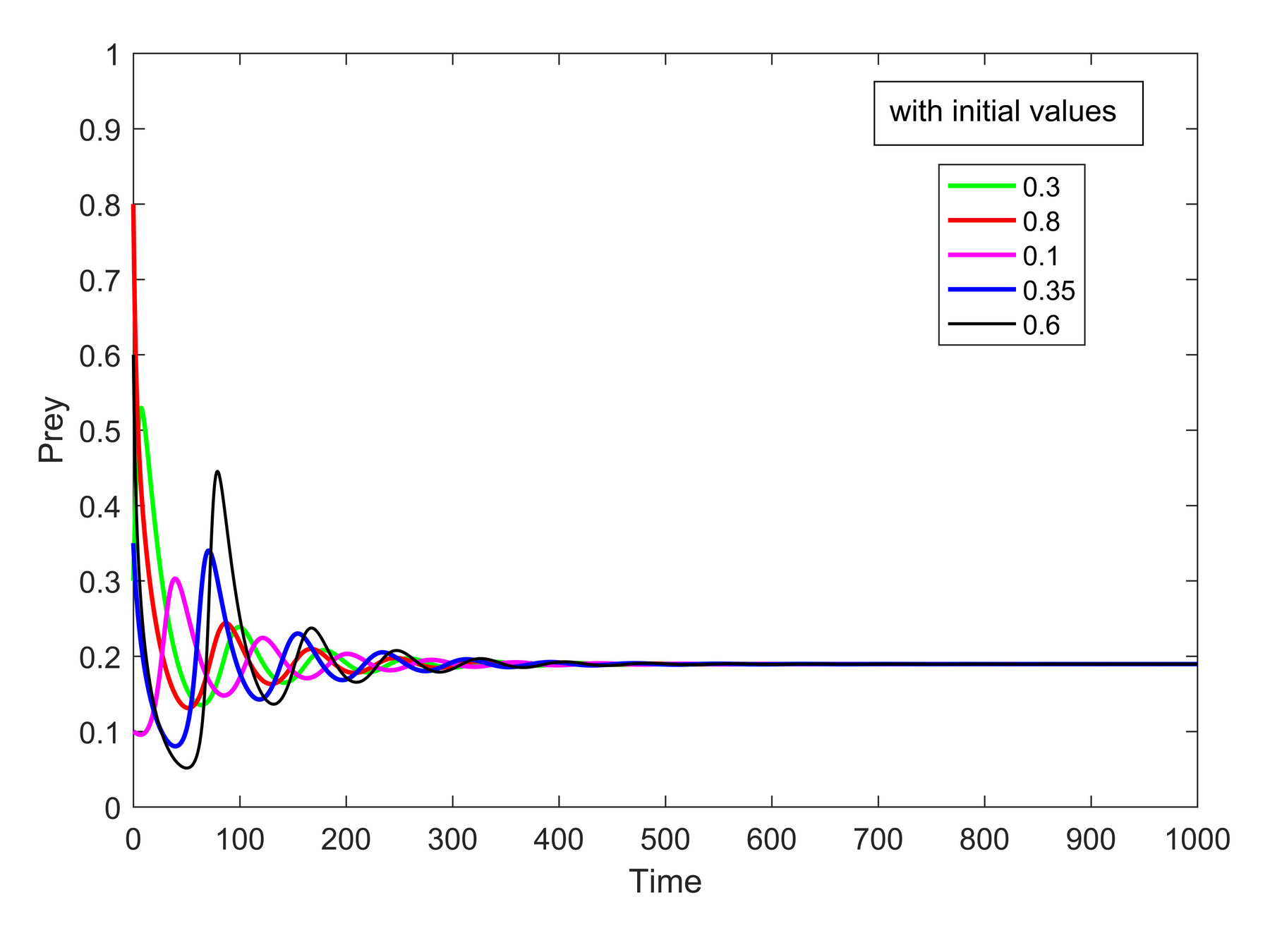}
    \includegraphics[width=8cm]{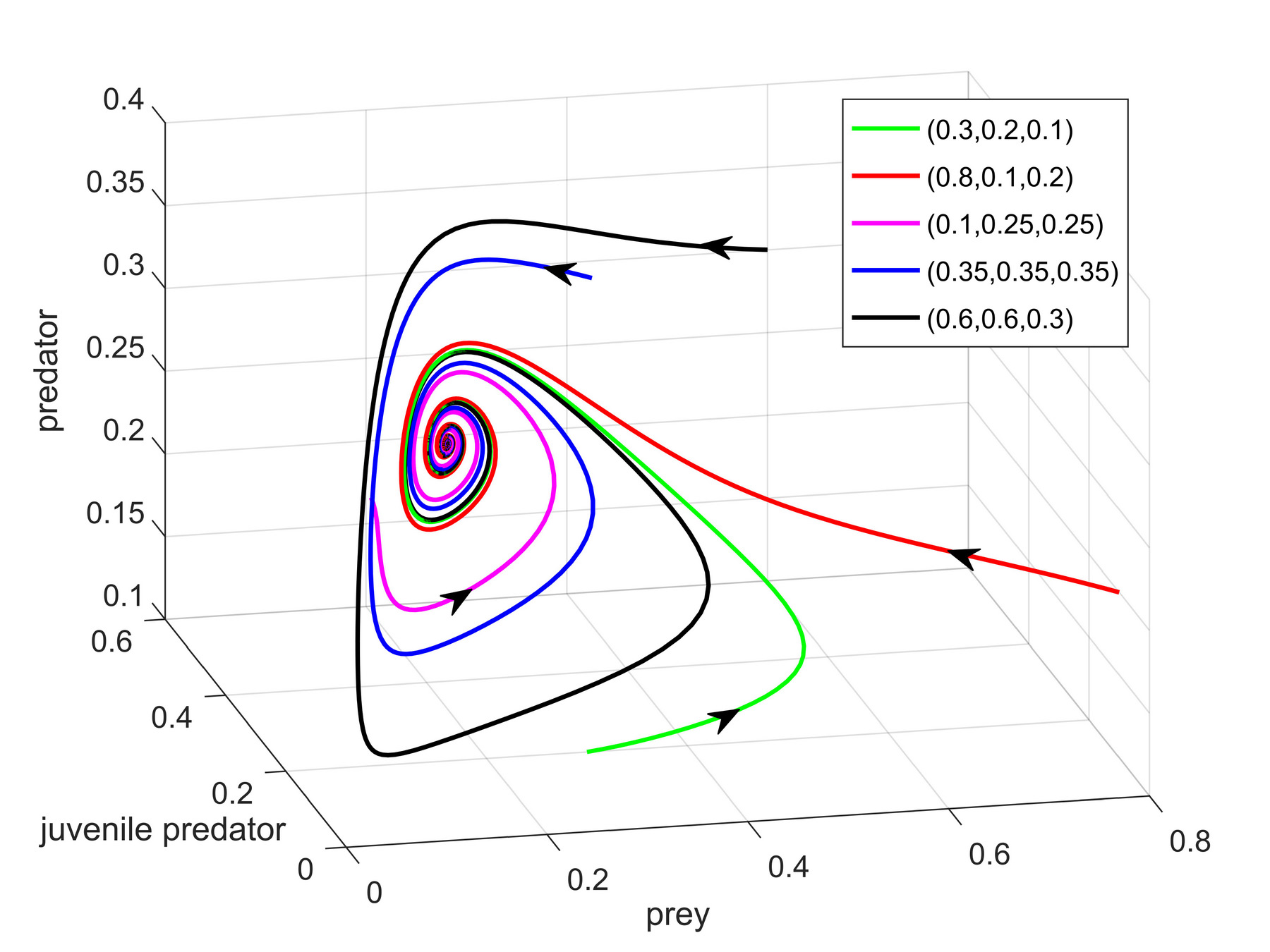}
    \caption{Local Stability of Co-extant Equilibrium $E_3(x^*,y^*,z^*)$}
    \label{lo-coeq}
\end{figure}
\begin{figure}[H]
    \centering
    \includegraphics[width=10cm]{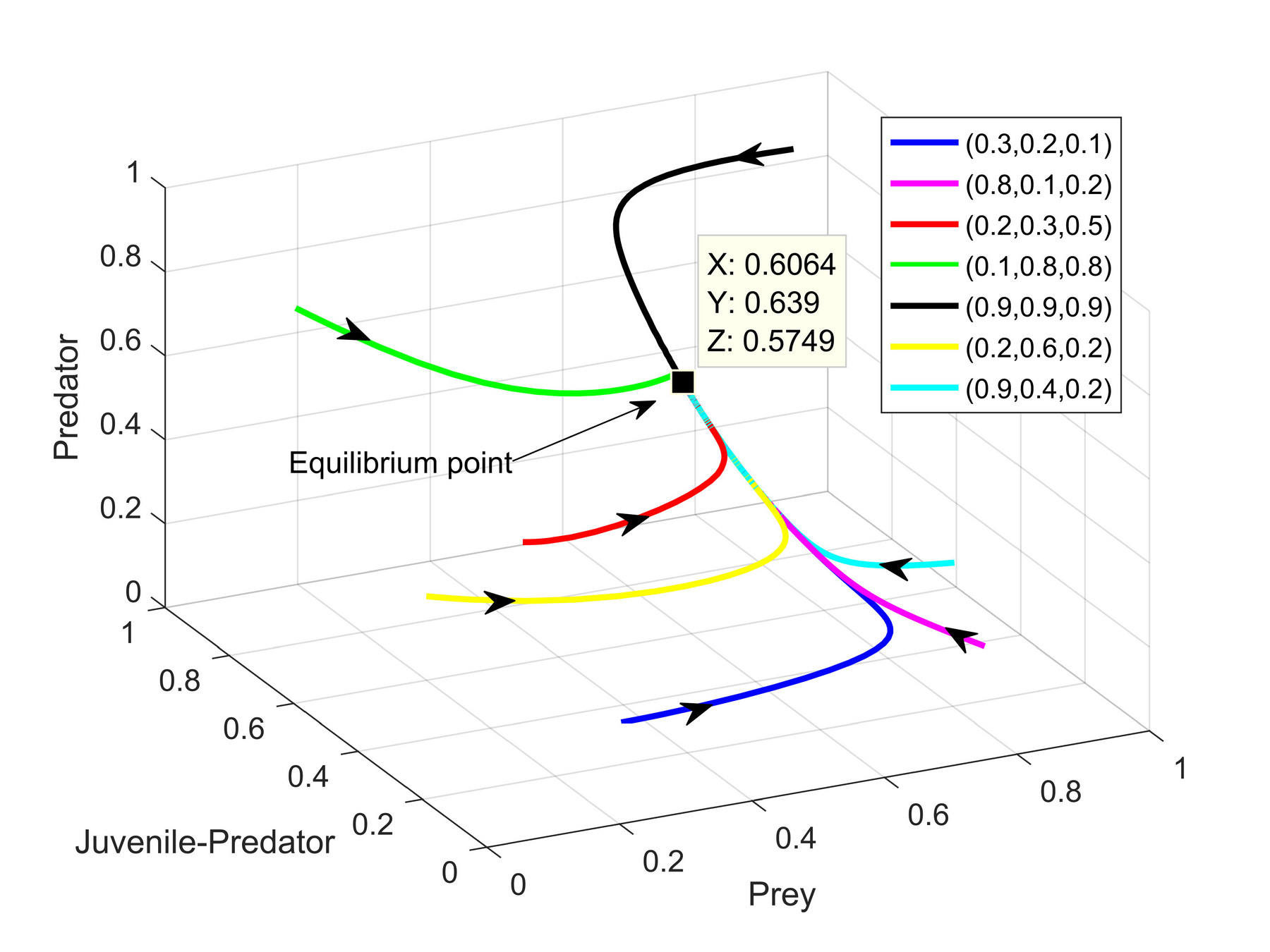}
    \caption{Global Stability of Co-extant Equilibrium}
    \label{global}
\end{figure}
\begin{multicols}{2}
\subsection{Numerical verification of the equilibrium points}
We consider $b=1.5,n=0.9$ and rest of the parameters from Table-1. Different initial points as shown in figure\ref{lo-axial} containing co-extant population lead to predator free equilibrium through different trajectories, which confirms the presence of global stability of the axial equilibrium. It stands to reason that presence of global stability implies local stability. The parameter values satisfy the local stability condition of the predator free equilibrium validating the numerical simulation. Furthermore, the condition of global stability of axial equilibrium point(theorem \ref{glo-axial}) being a sufficient one, other trajectories, not satisfying the condition also leads to the point$(1,0,0)$.
\par But as the value of the parameters \textit{b} and \textit{n} are decreased and are considered as is in the table, $E_1$ loses its stability and  the co-extant equilibrium point $E_3(0.189539,0.297824,0.268042)$ is found to be existent attributable to $0 < \frac{(c d_2 + d_1 d_2 + b d_2 n)}{c}<a_3$(theorem \ref{loca-sta1} and theorem \ref{exis} respectively). The local stability, indicated by the Routh-Hurwitz criteria developed in theorem \ref{loca-sta2} is satisfied as we have, $\xi_1=0.646, \xi_3=0.004064, \xi_4=0.008969$, i.e $\xi_1,\xi_3,\xi_4>0$. Figure \ref{lo-coeq} attests the numerical reasoning. However for this set of values, neither the conditions of global stability nor of perseverance is obeyed numerically.
\par For the verification of the condition of global-stability of co-extant equilibrium point(theorem \ref{glo-coex}), we have taken another set of parameter values, $a_1=0.75$, $b=0.7$, $m=1.95$, $n=0.28$ and the rest, as given in the table  For these  values, we acquire $\xi_1=1.0927, \xi_3=0.014187, \xi_4=0.32469$ along with $\vartheta= -0.0027871<0$ which is the condition of global stability. Graphical illustration( figure \ref{global}) validates the existence of a neighbourhood satisfying global stabilty.
Condition of perseverance (theorem \ref{persev}) too is satisfied at these parameters,so theoretically, the initial co-extant population will coexist perpetually. Stability of the coexisting equilibrium supports this fact as the trajectories taking different initial points stay in the coexisting equilibrium in the long run.
\end{multicols}
\begin{figure}[H]
    \centering
    \begin{subfigure}[b]{0.5\textwidth}
         \centering
    \includegraphics[width=\textwidth]{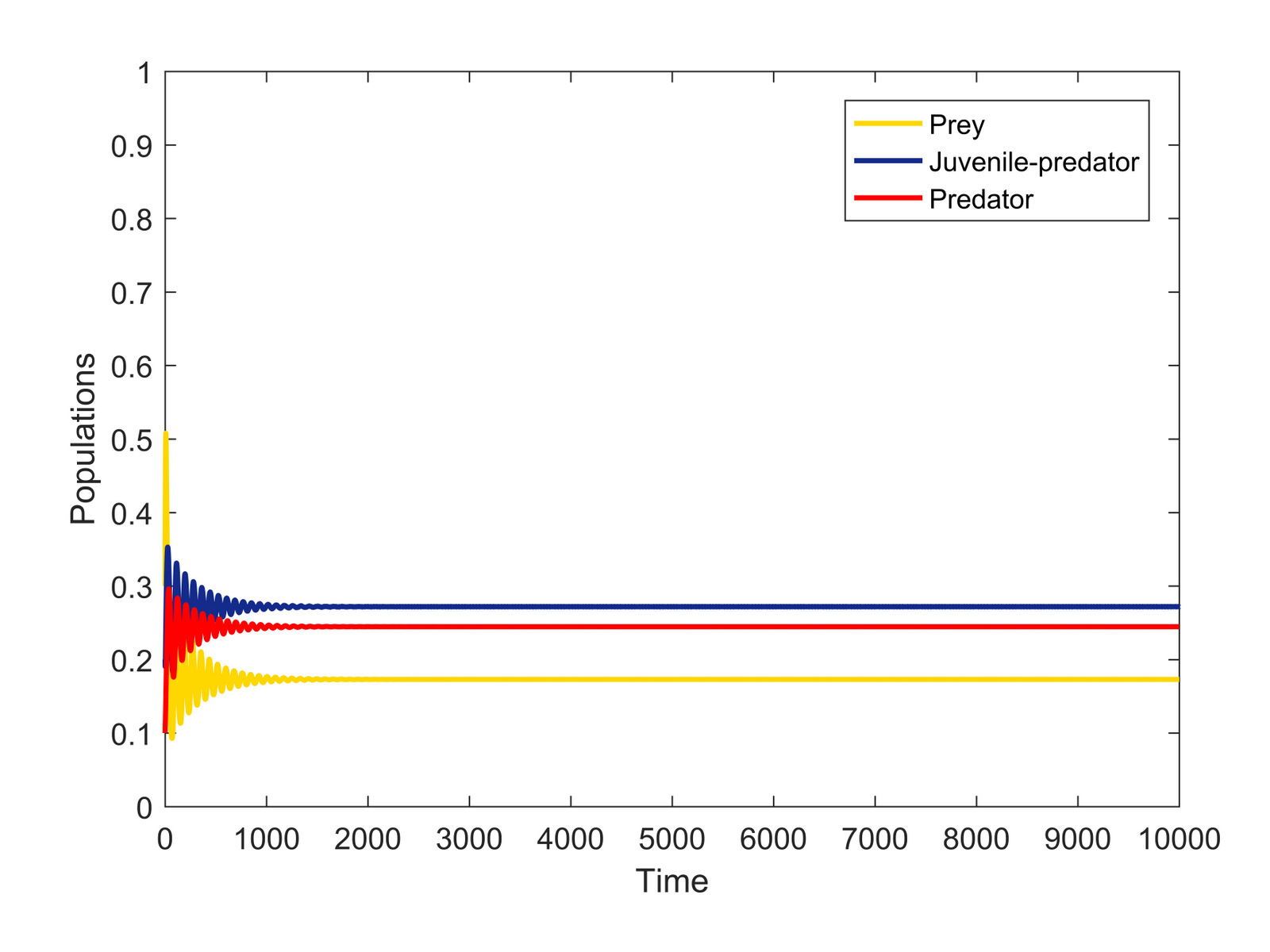}
    \caption{}
    \label{a1-1a}
    \end{subfigure}\hfill
     \begin{subfigure}[b]{0.5\textwidth}
         \centering
    \includegraphics[width=\textwidth]{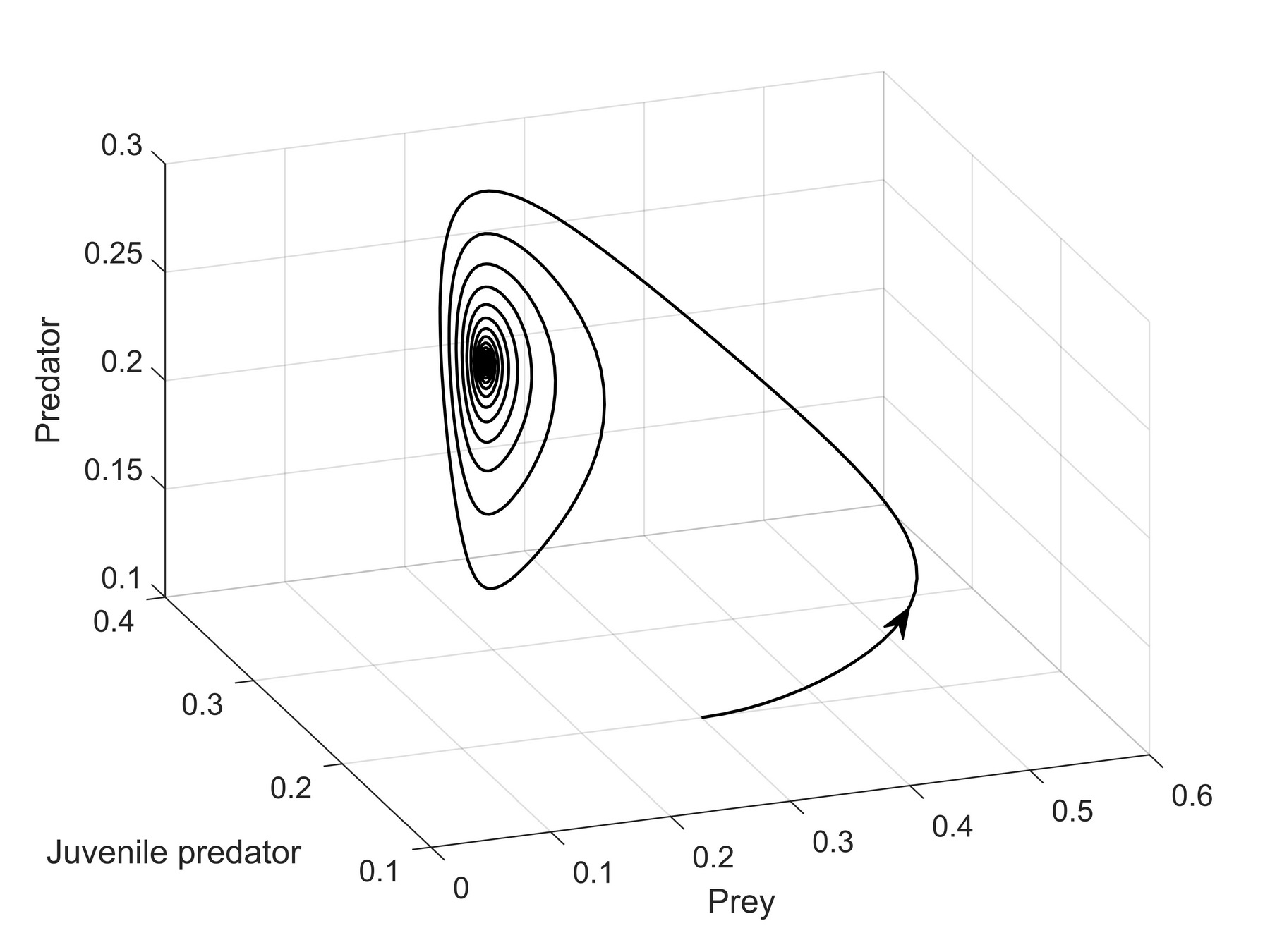}
    \caption{}
    \label{a1-1b}
    \end{subfigure}
    \caption{ At $a_1 = 1.32 < a_{1h}$ (\ref{a1-1a}) shows time series of prey, juvenile-predator and predator populations and (\ref{a1-1b}) depicts phase portrait}
    \label{a1-1}
\end{figure}
\begin{figure}[H]
    \centering
    \begin{subfigure}[b]{0.5\textwidth}
         \centering
    \includegraphics[width=\textwidth]{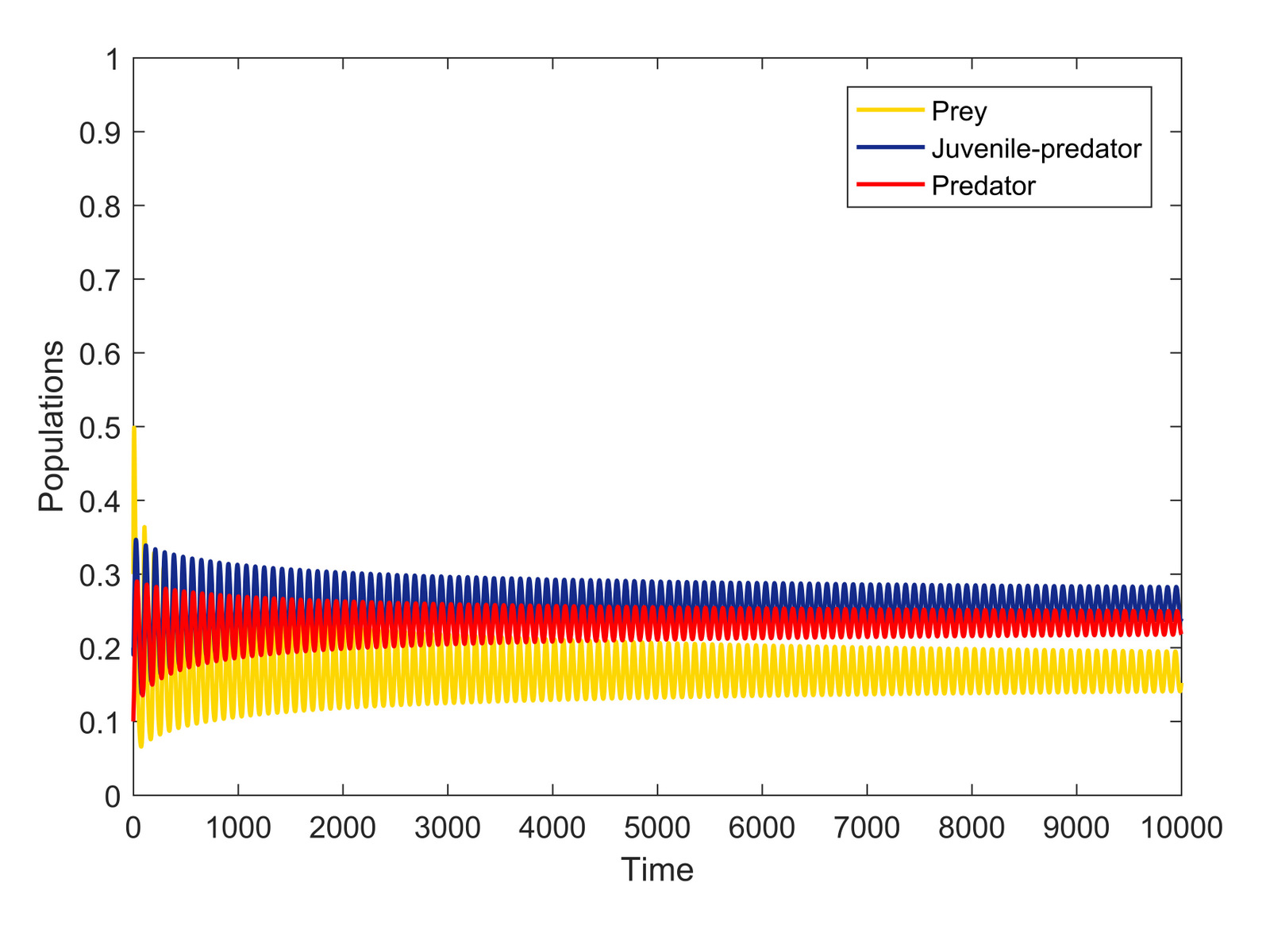}
    \caption{}
    \label{a1-2a}
    \end{subfigure}\hfill
     \begin{subfigure}[b]{0.5\textwidth}
         \centering
    \includegraphics[width=\textwidth]{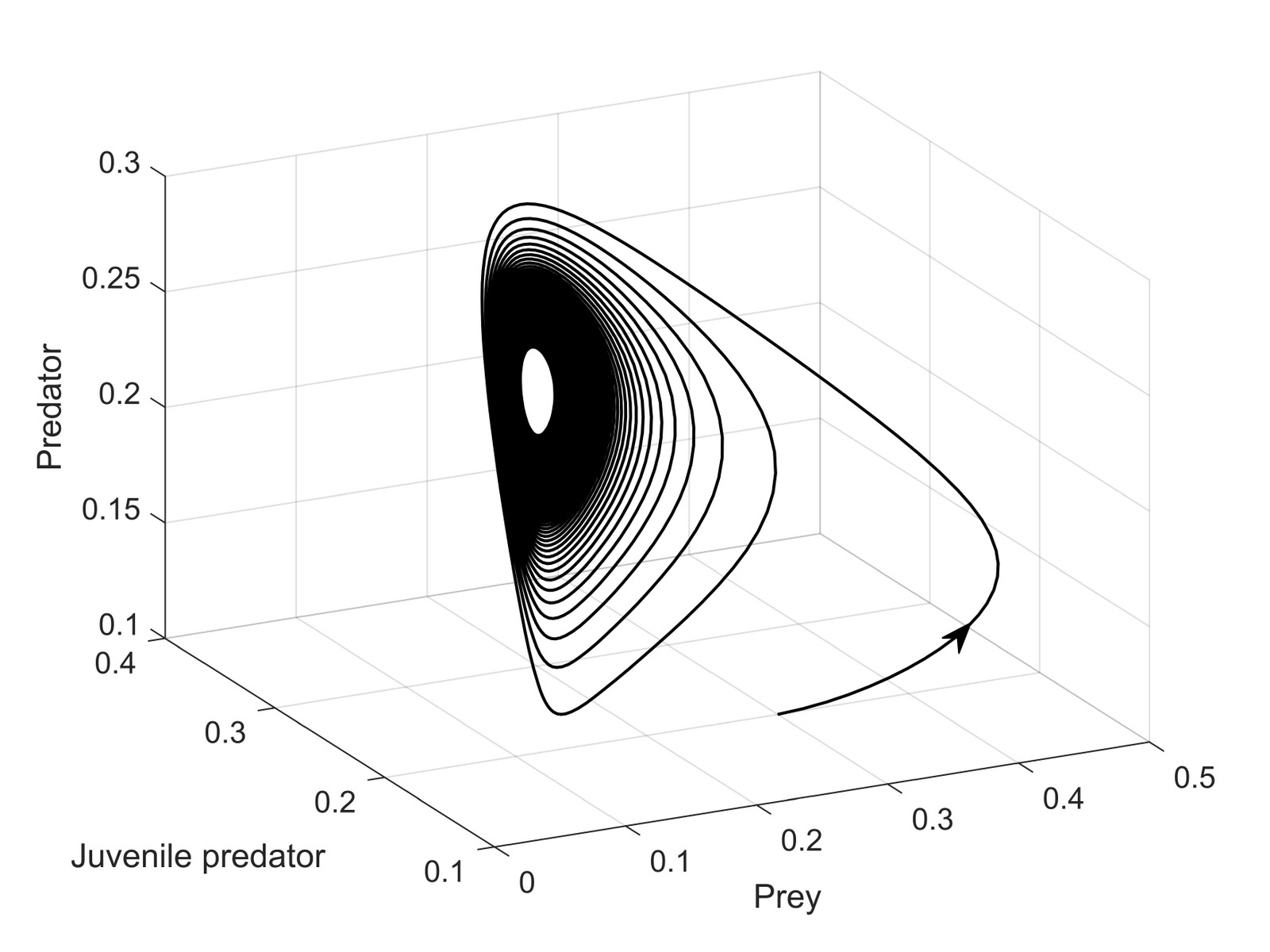}
    \caption{}
    \label{a1-2b}
    \end{subfigure}
    \caption{The solutions of the bio-system \ref{eq2} exhibiting emergence of periodic oscillations around the equilibrium
point at $a_1 = a_{1h} = 1.350115$}
    \label{a1-2}
\end{figure}
\begin{figure}[H]
    \centering
    \begin{subfigure}[b]{0.5\textwidth}
         \centering
    \includegraphics[width=\textwidth]{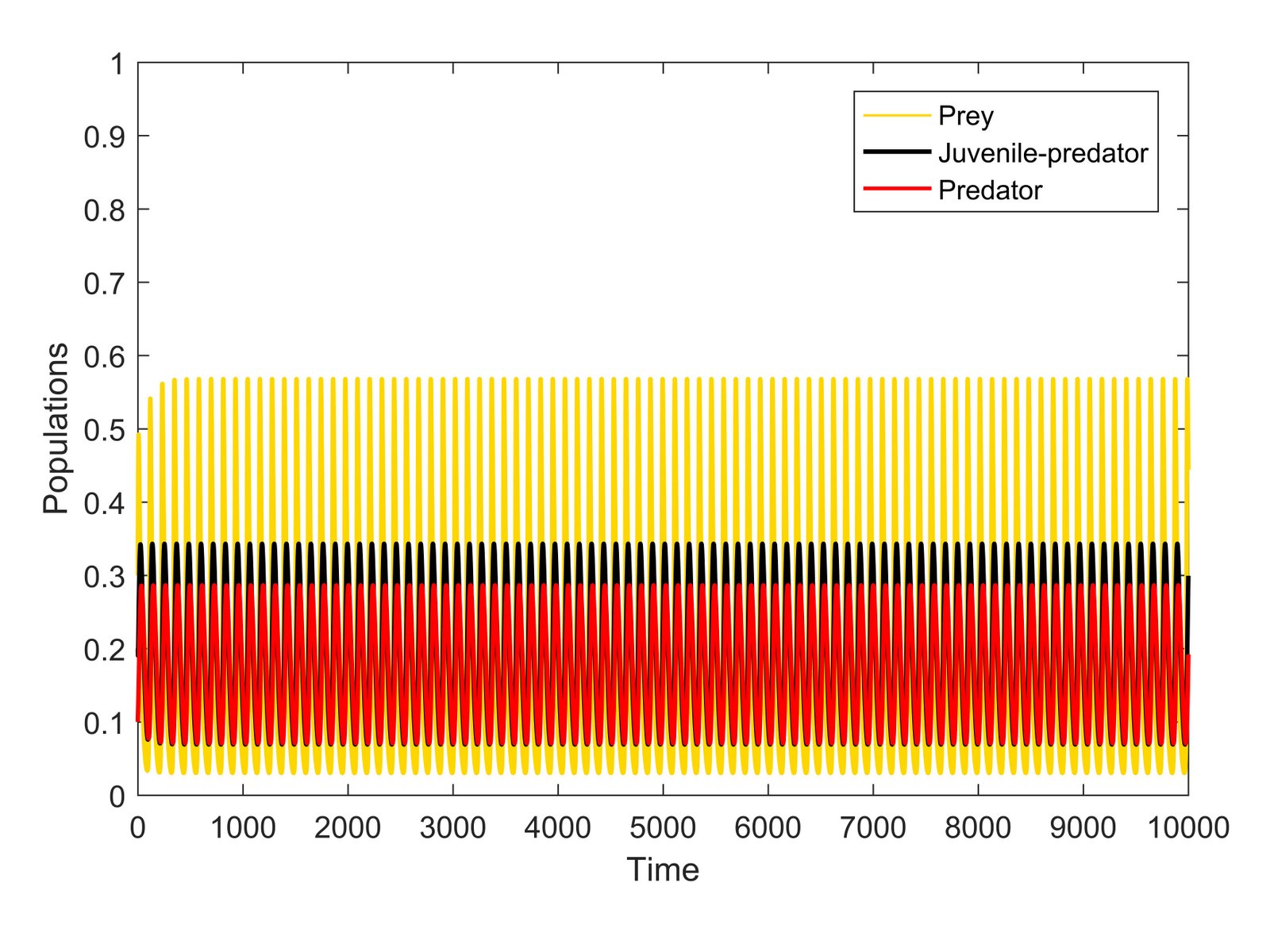}
    \caption{}
    \label{a1-3a}
    \end{subfigure}\hfill
     \begin{subfigure}[b]{0.5\textwidth}
         \centering
    \includegraphics[width=\textwidth]{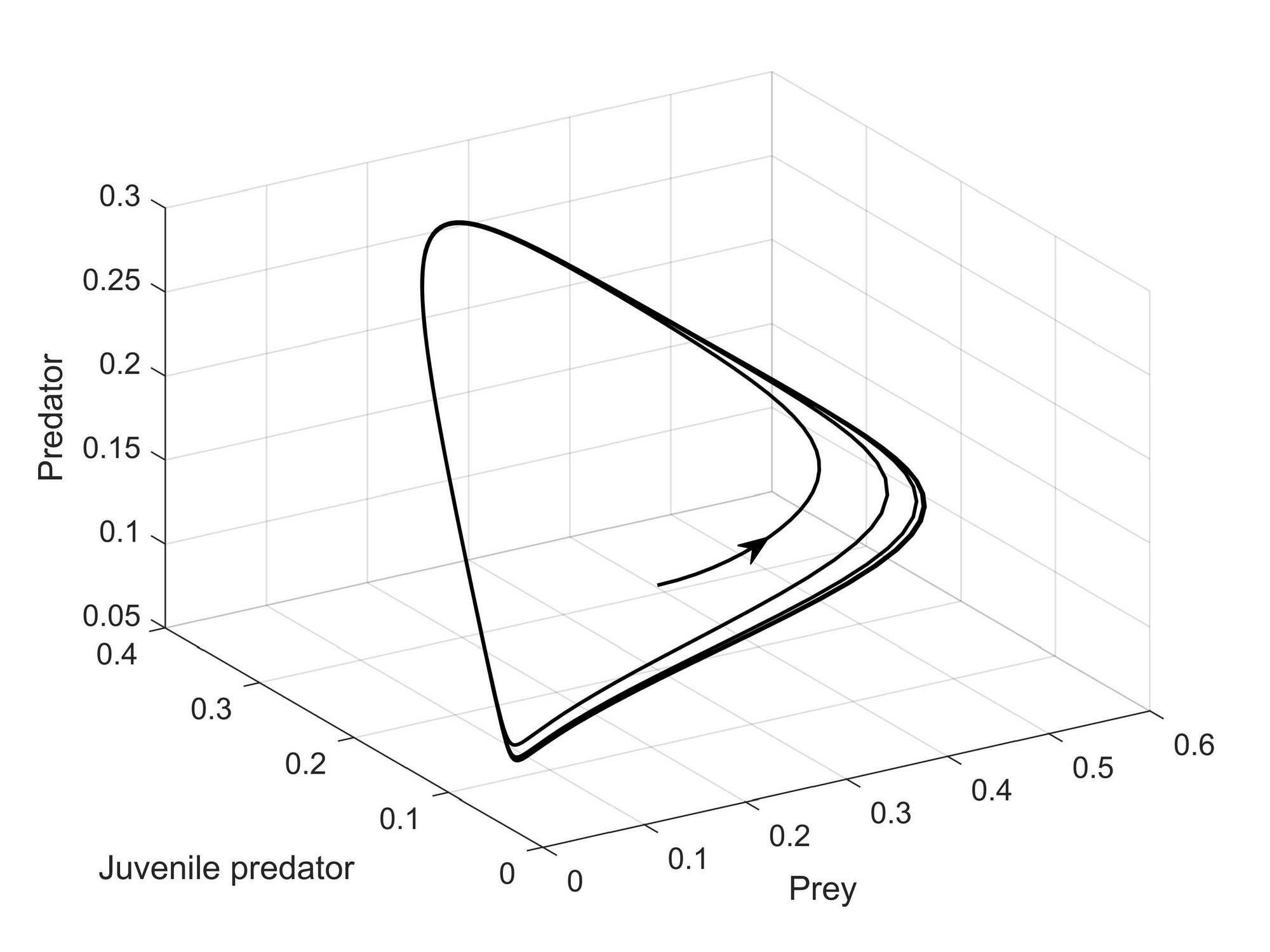}
    \caption{}
    \label{a1-3b}
    \end{subfigure}
    \caption{At $a_1 = 1.37 > a_{1h}$ (\ref{a1-3a}) depicts time series of the three populations and (\ref{a1-3b}), the phase portrait of equilibrium point $E_3(x^*,y^*,z^*)$ where it is unstable but the limit cycle is stable}
    \label{a1-3}
\end{figure}
\begin{figure}[H]
    \centering
    \begin{subfigure}[b]{0.33\textwidth}
         \centering
    \includegraphics[width=\textwidth]{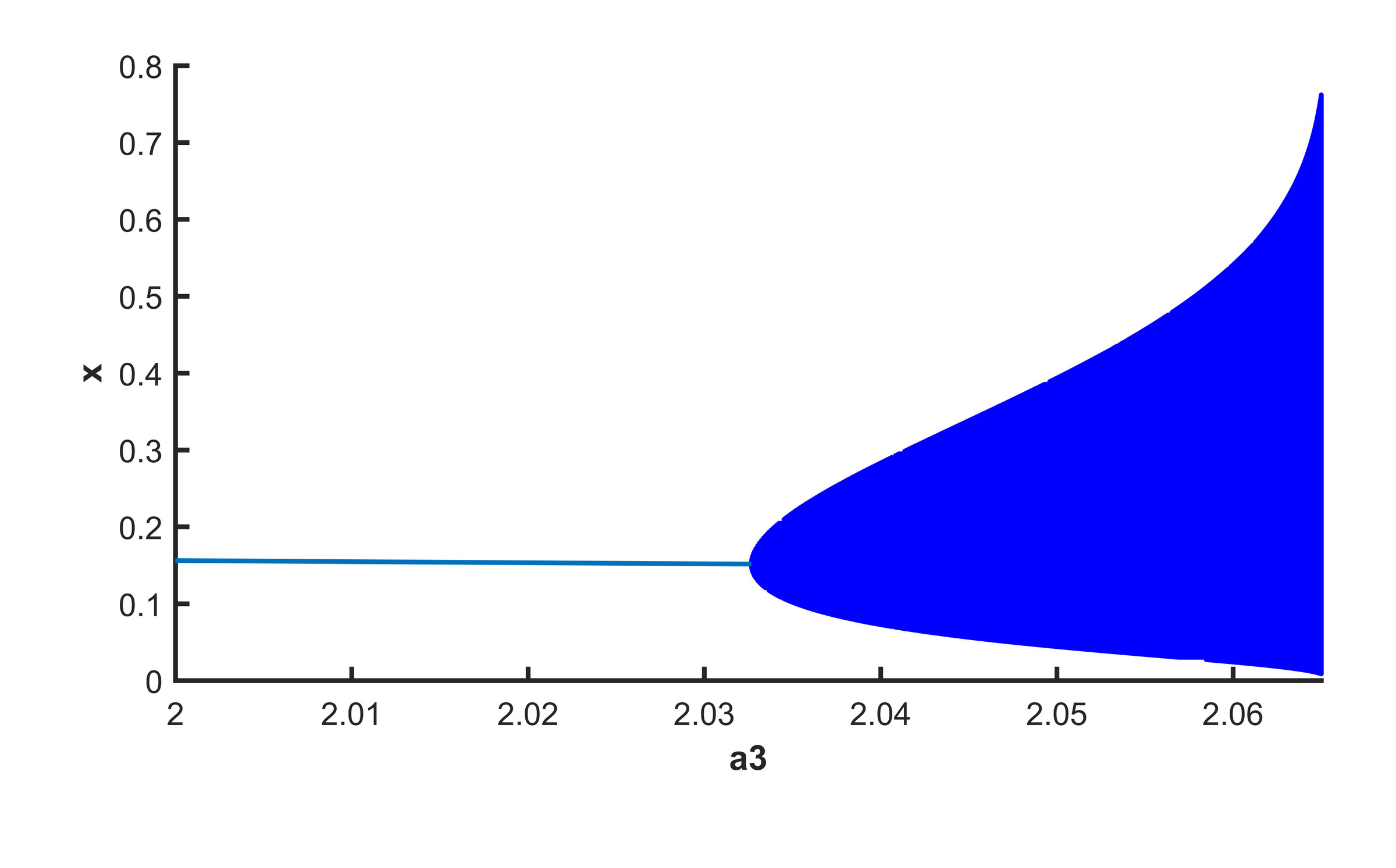}
    \caption{}
    \end{subfigure}\hfill
     \begin{subfigure}[b]{0.33\textwidth}
         \centering
    \includegraphics[width=\textwidth]{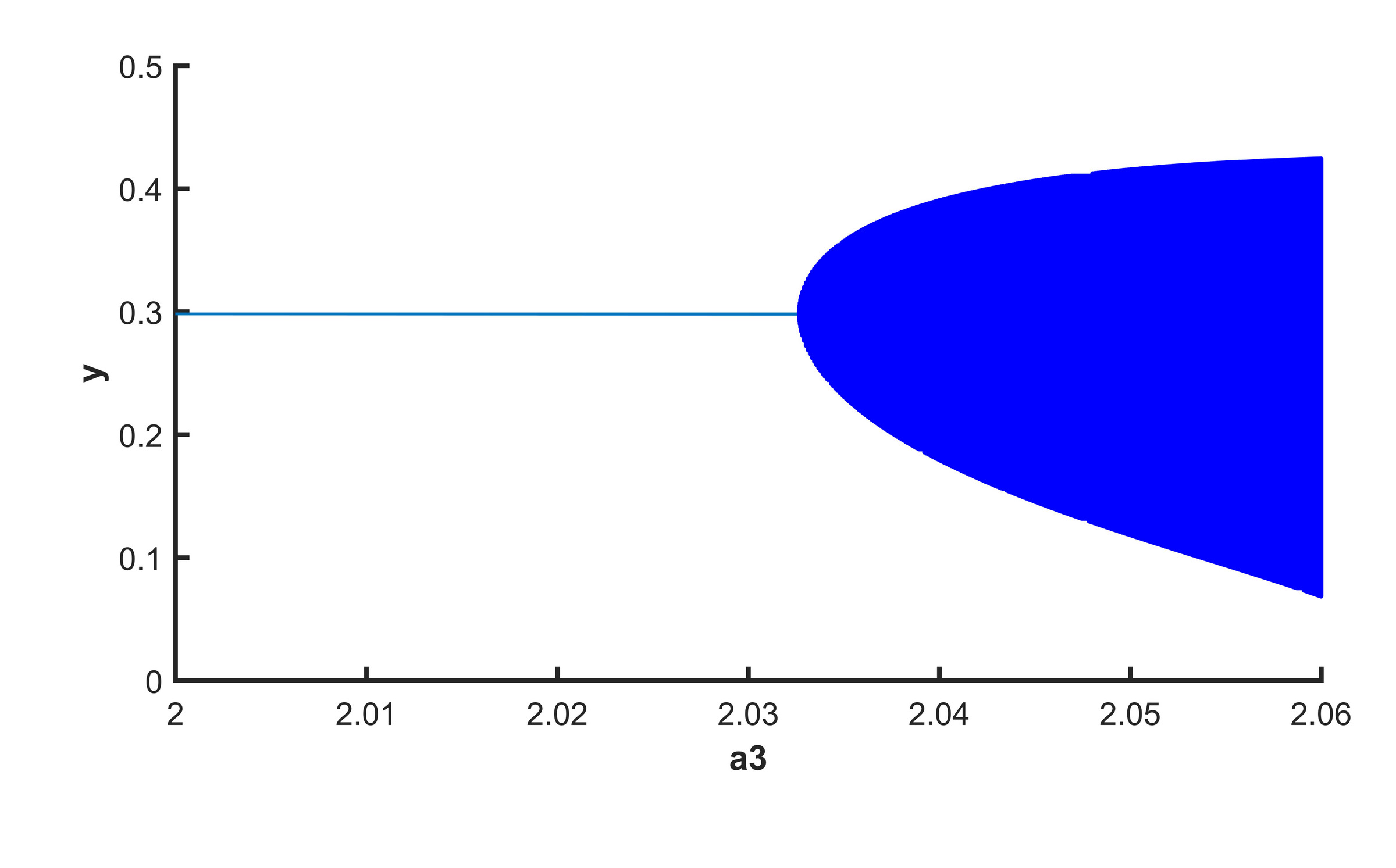}
    \caption{}
    \end{subfigure}\hfill
     \begin{subfigure}[b]{0.33\textwidth}
         \centering
    \includegraphics[width=\textwidth]{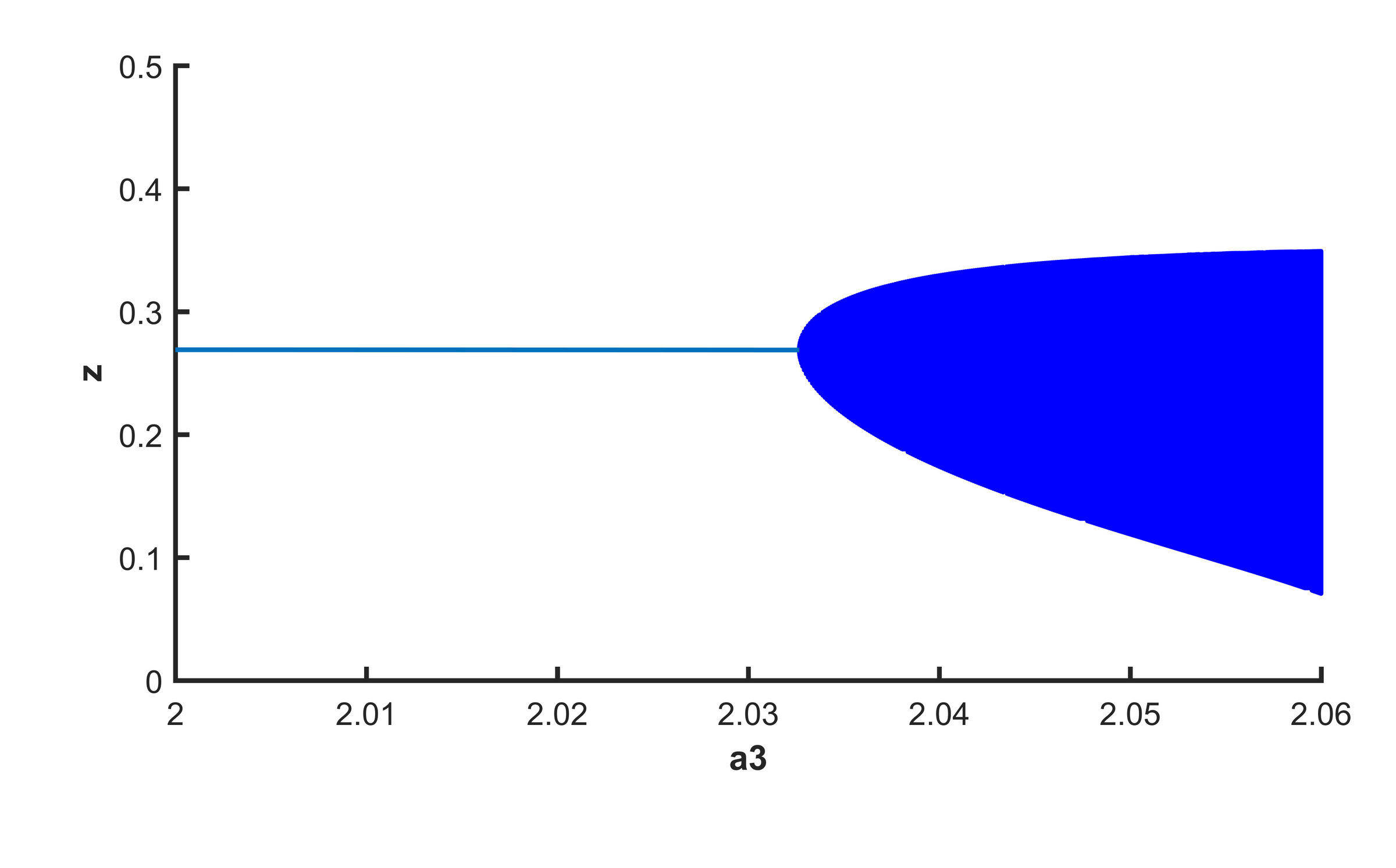}
    \caption{}
    \end{subfigure}
    \caption{Forward Bifurcation with respect to conversion rate of
juveniles ($a_3$)}
    \label{a3}
\end{figure}
\begin{figure}[H]
    \centering
    \begin{subfigure}[b]{0.33\textwidth}
         \centering
    \includegraphics[width=\textwidth]{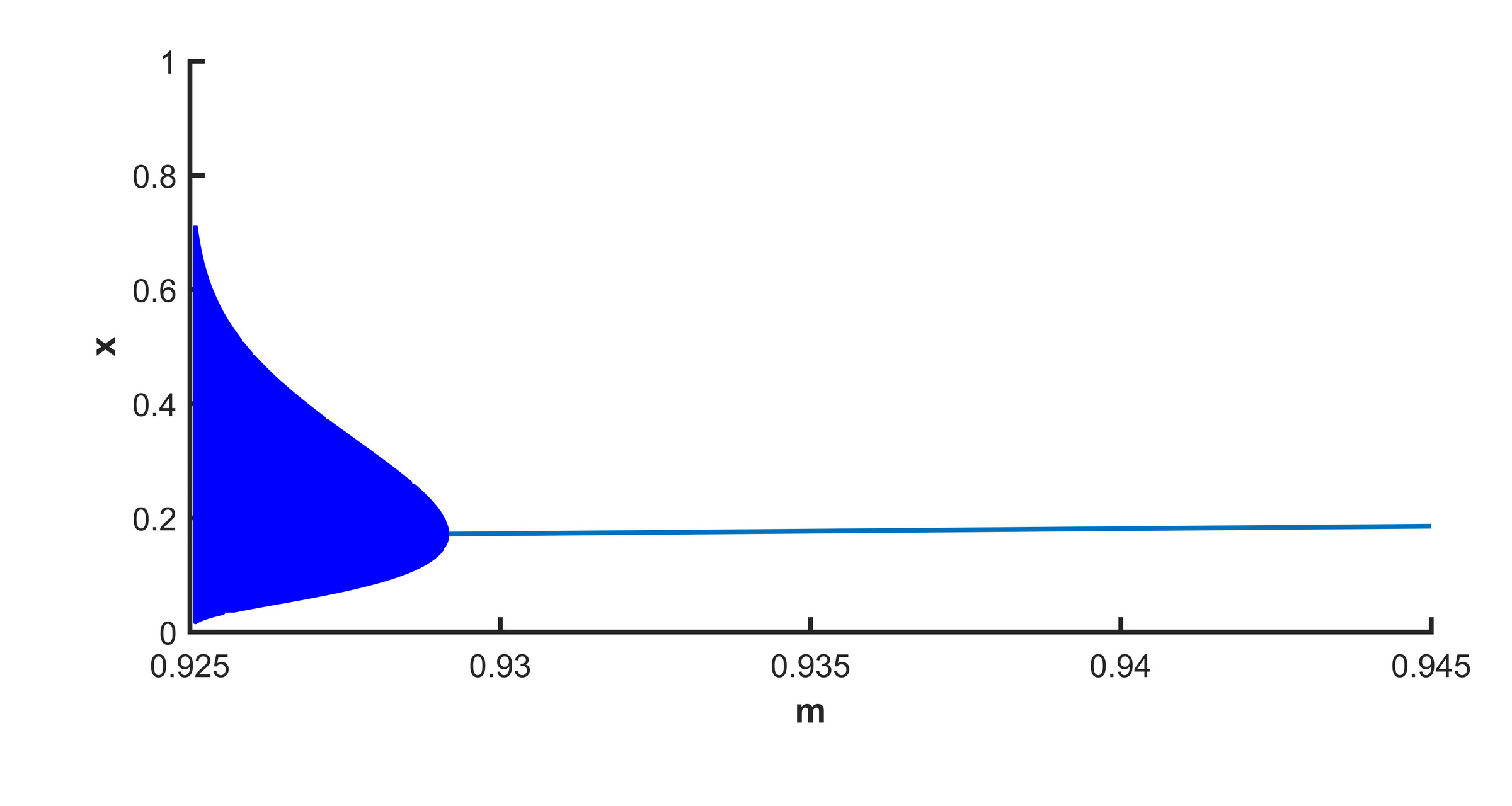}
    \caption{}
    \end{subfigure}\hfill
     \begin{subfigure}[b]{0.33\textwidth}
         \centering
    \includegraphics[width=\textwidth]{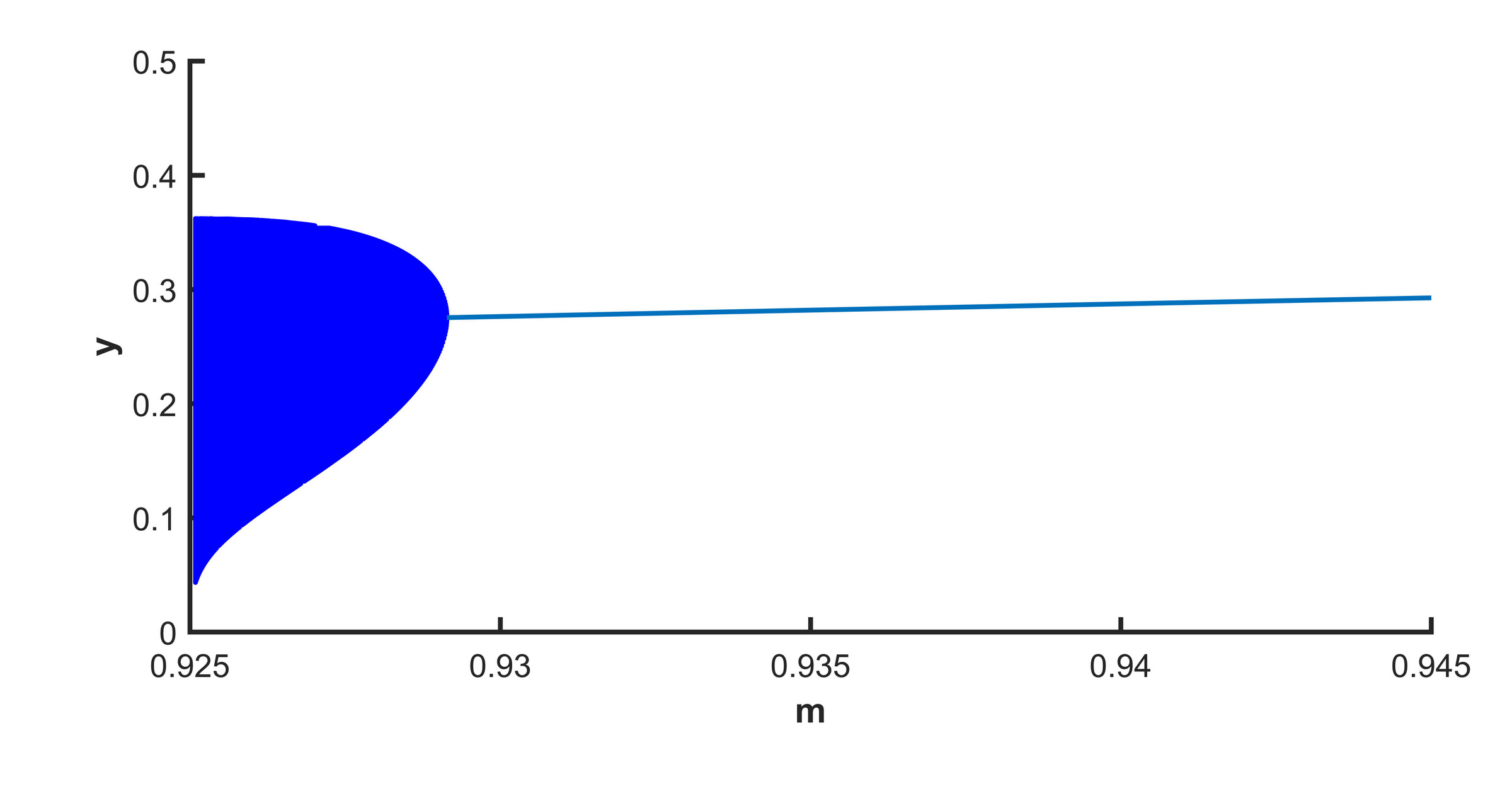}
    \caption{}
    \end{subfigure}\hfill
     \begin{subfigure}[b]{0.33\textwidth}
         \centering
    \includegraphics[width=\textwidth]{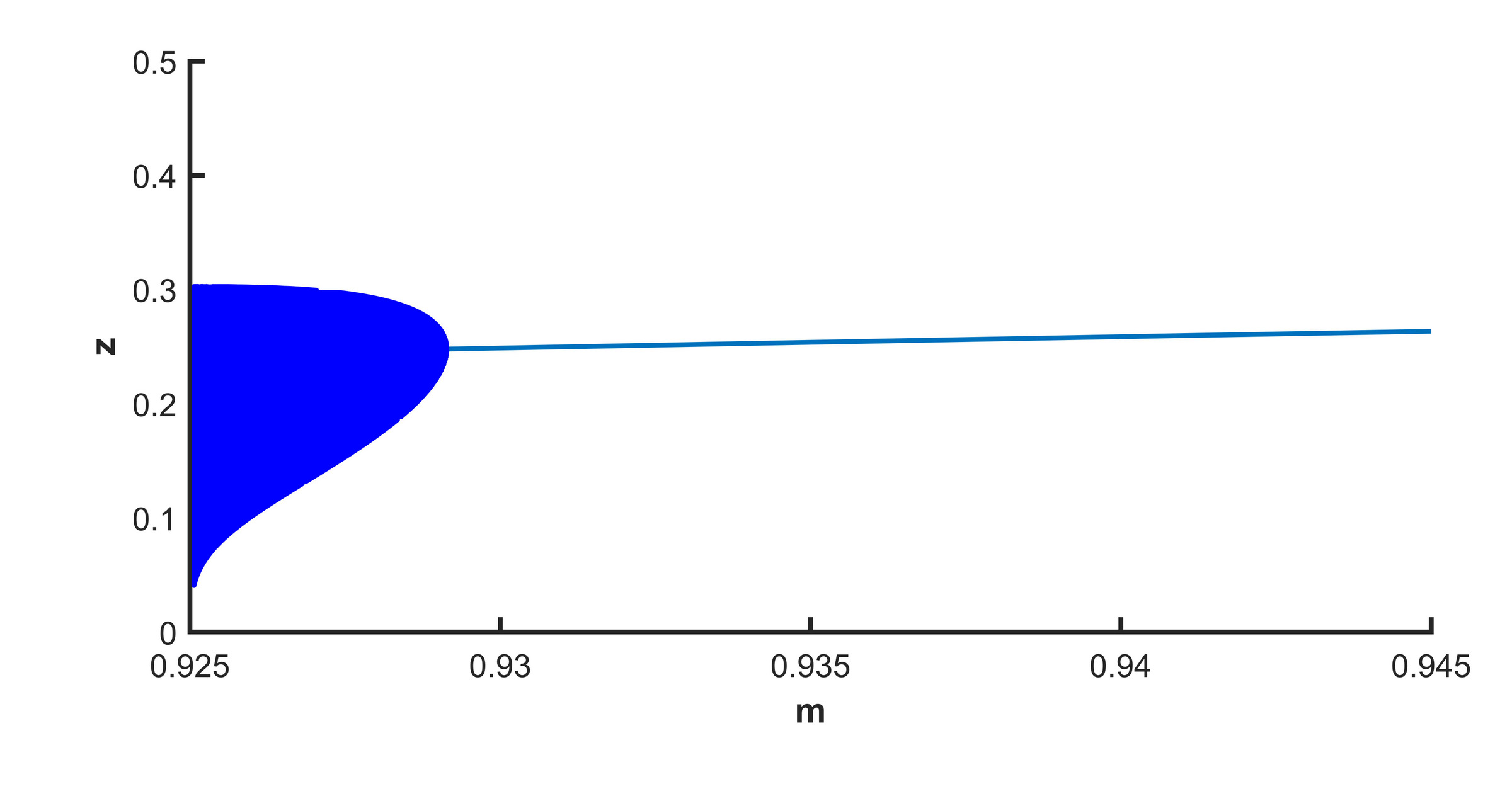}
    \caption{}
    \end{subfigure}
    \caption{Backward Bifurcation with respect to search and conquer rate (\textit{m})}
    \label{m}
\end{figure}
\begin{multicols}{2}
\subsection{Bifurcations}
Verification of the Hopf bifurcation conditions and its direction is performed prior to the exploration of the occurrence of periodic oscillations and its nature among the three populations with respect to the different bifurcation parameters.\\
By reason of evaluation of the parameters for Hopf bifurcation and its direction, numerically, we first consider the parameter $a_1$,the predation rate by juveniles.
The bio-system \ref{eq2} undergoes Hopf bifurcation at $a_1=a_{1h}=1.350114$ as it satisfies the NASC condition for Hopf-bifurcation i.e. $\xi_1=0.618917,\xi_3=0.00355832,\xi_4=0$ and $\frac{d\xi_4}{da_1}=-0.0856855(\neq 0)$ at $a_1=a_{1h}$\\
As specified in the theorem \ref{ho-thm}, we are able to find the nature and direction of bifurcating periodic solution. From those theories with the above mentioned set of parametric values we get,
$$g_{11}=-0.199376 - 0.0174382i,\hspace{1pt} g_{02}=-0.493577 - 0.043236i$$
$$g_{20}=-0.120705 + 0.180489i, \hspace{1pt} g_{21}-0.929725 + 0.771532i$$
And hence we get,
$C_1(0)=-0.0996479 - 0.874509i, \hspace{3pt}
\mu_2=0.904326\\ \beta_2=-0.199296 .$\\
Jere, $\mu_2$ being greater than 0 attests the bifurcation to be a supercritical one and as $\beta_2<0$, thus the periodic cycle is stable.\\
Figures \ref{a1-1},\ref{a1-2}, and \ref{a1-3}, are agreeing with the numerical calculations, shows the phenomenon of bifurcations. At $a_1=a_{1h}=1.350114$, there is a birth of limit cycle(figure \ref{a1-2}) as the co-existing initial point follows an oscillation. The amplitude of this oscillation keeps on diminishing as the value of $a_1$ diminishes and the system evinces stable focus point, figure \ref{a1-1} shows this at$a_1=1.32<a_{1h}$. There is a visible increase in the amplitude of oscillation at $a_1=1.37>a_{1h}$ (figure \ref{a1-3}), the co-extant equilibrium point becomes unstable as parameter crosses the threshold value $a_1=a_{1h}$ and the system exhibits limit cycle oscillations.
\vspace{5pt}\par Next, the parameter $a_3$ is examined, this parameter is pertinent to growth factor of juvenile predator. It harbours the capacity to cause both transcritical and hopf-bifurcation(see section \ref{hopf}). Considering given set of parametric values from table-1, at $a_3=0.8861111$ there is reciprocity of stability between axial equilibrium point and co-extant equilibrium point  due to transcritical bifurcation. At $a_3=a_{3h}=2.0326485$, there is a supercritical Hopf bifurcation,the first Lyapunov coefficient being $-3.601898e^{-02}$, exchanging the stability from the co-existing equilibrium to a  stable limit cycle being born around the said equilibrium point. As the parametric value of $a_3$ is continually increased after the critical value $a_{3h}$, the amplitude of fluctuation in all the populations too would increase, figure \ref{a3} visualises this very scenerio. \\
Similarly, for the searching rate of both the predator, taken as average,(\textit{m}) figure \ref{m} visualises the backward bifurcation with respect to \textit{m}. The point of bifurcation being $m=0.9291448$. The first lyapunov coefficient is found to be  $-1.110749e^{-01}<0$, hence a supercritical hopf-bifurcation.
As \textit{m} decreases the fluctuation in prey, juvenile predator and matured predator populations increases. Now, if the parametric value is further decreased, the limit cycle that arises too would cease to exist after a certain value of the parameter. The co-extant equilibrium point being unstable, the trajectory goes to the vanishing equilibrium point.
\end{multicols}
\begin{figure}[H]
    \centering
    \begin{subfigure}[b]{0.3\textwidth}
         \centering
    \includegraphics[width=\textwidth]{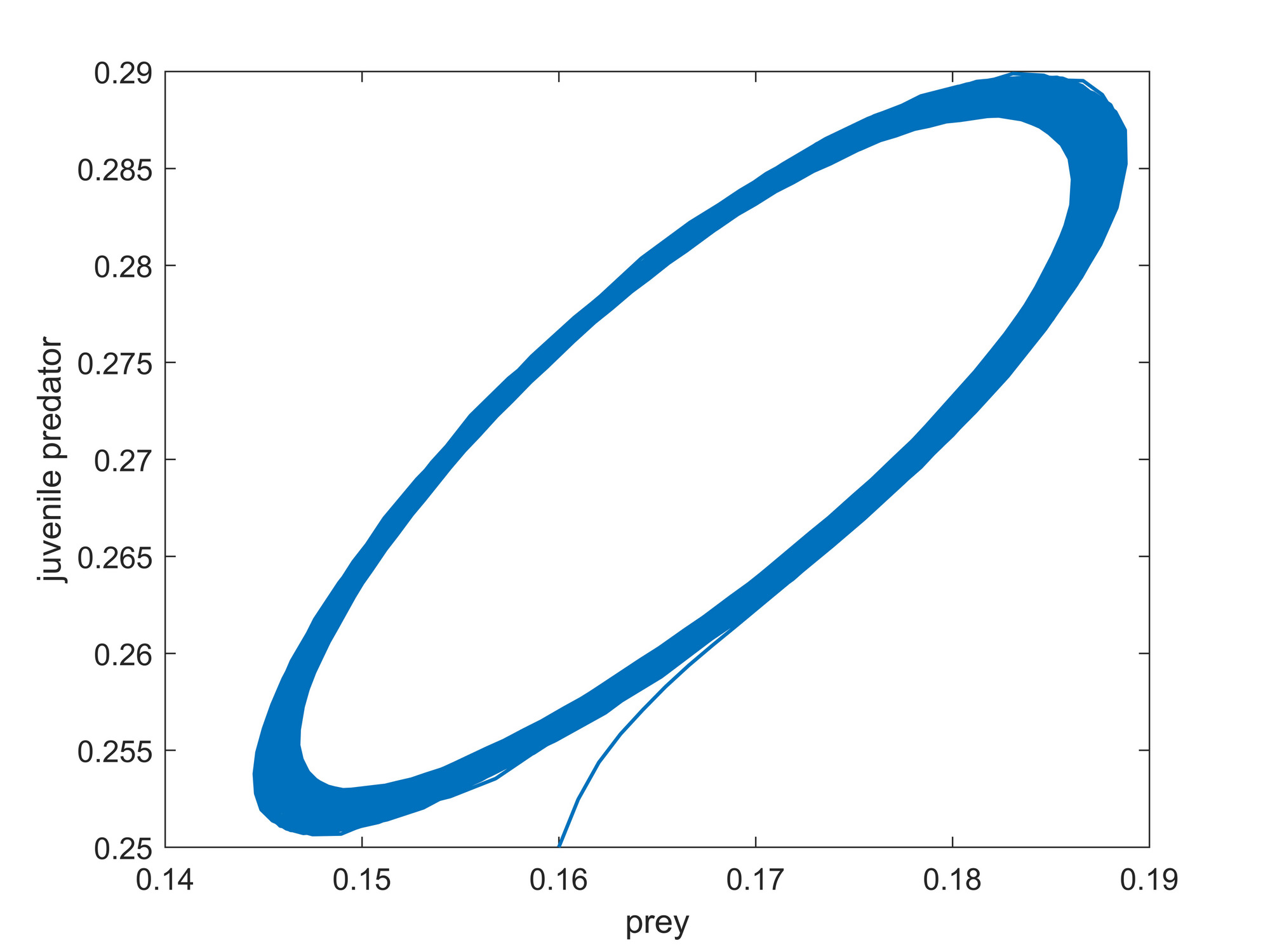}
    \includegraphics[width=\textwidth]{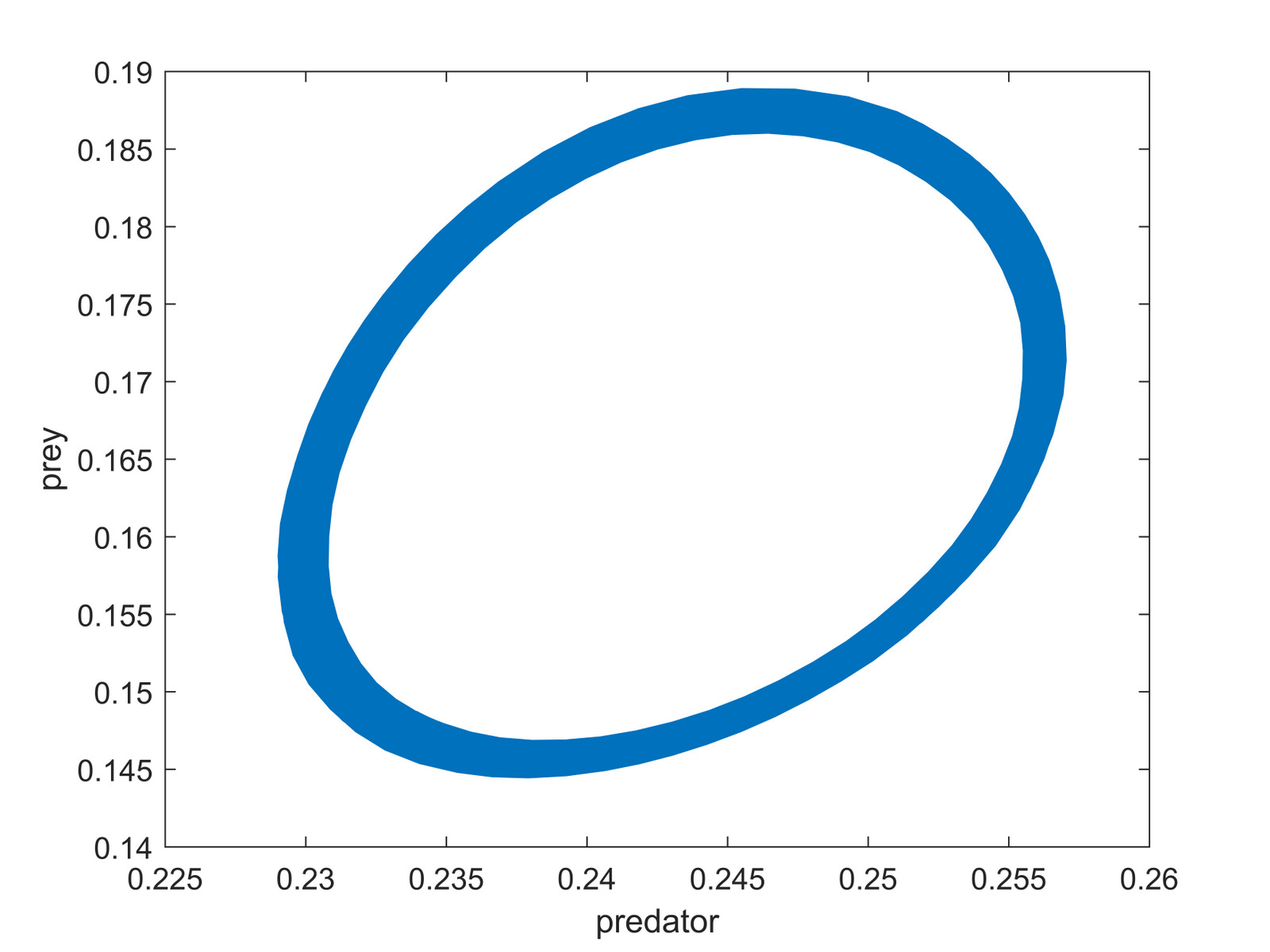}
    \includegraphics[width=\textwidth]{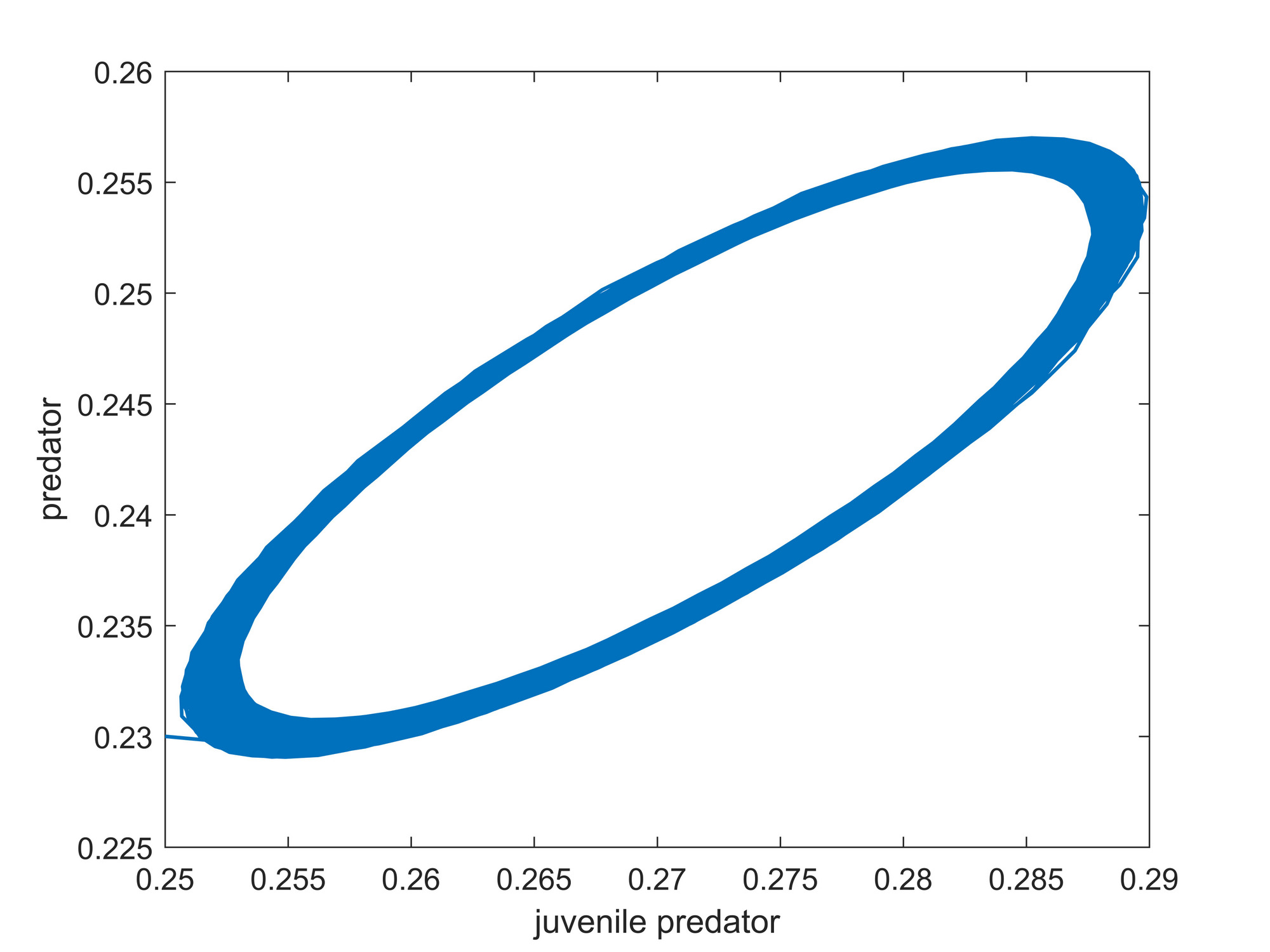}
    \caption{at $n=0.42$}
    \label{n-dim-a}
    \end{subfigure}
    \begin{subfigure}[b]{0.3\textwidth}
         \centering
    \includegraphics[width=\textwidth]{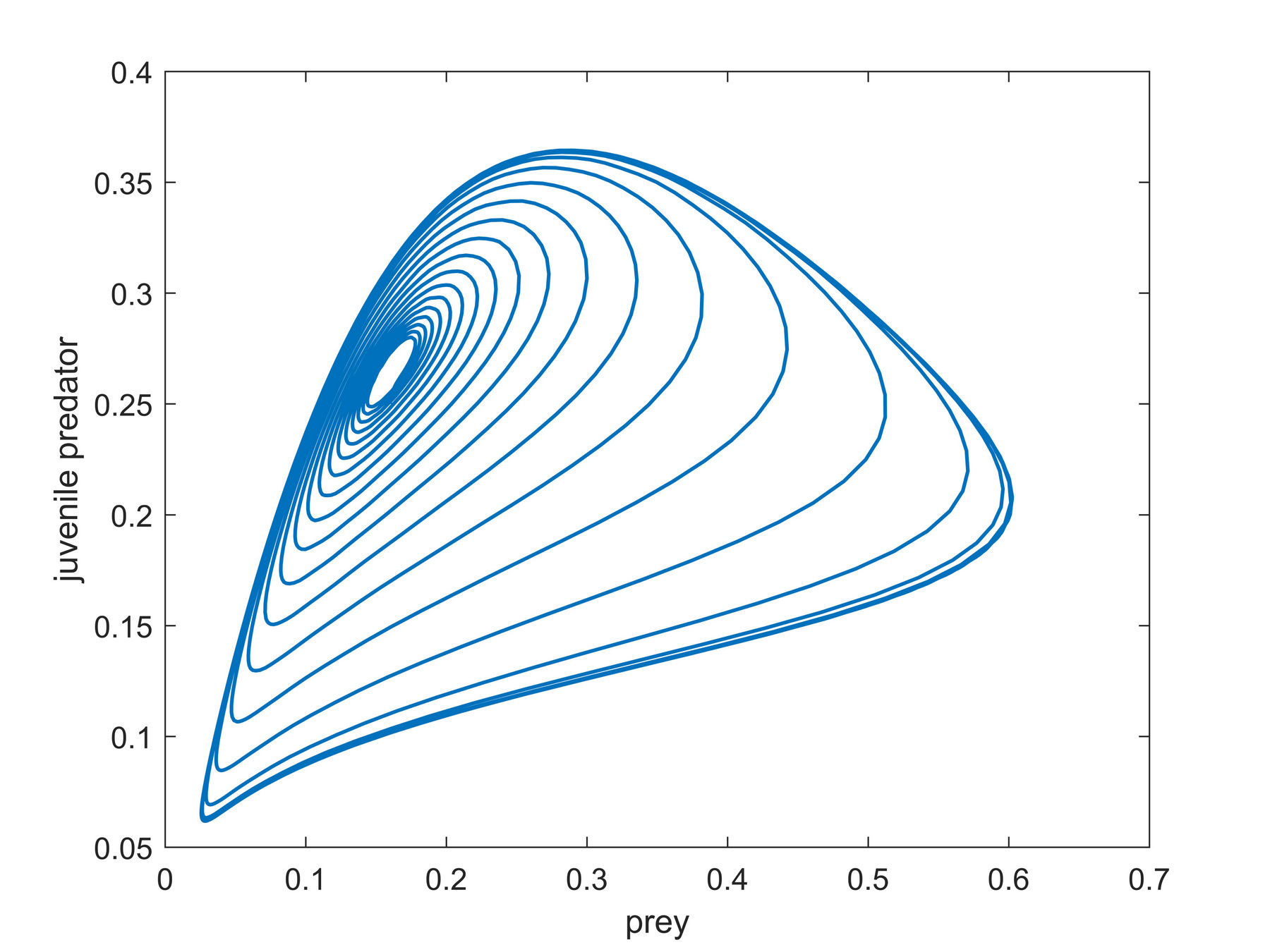}
    \includegraphics[width=\textwidth]{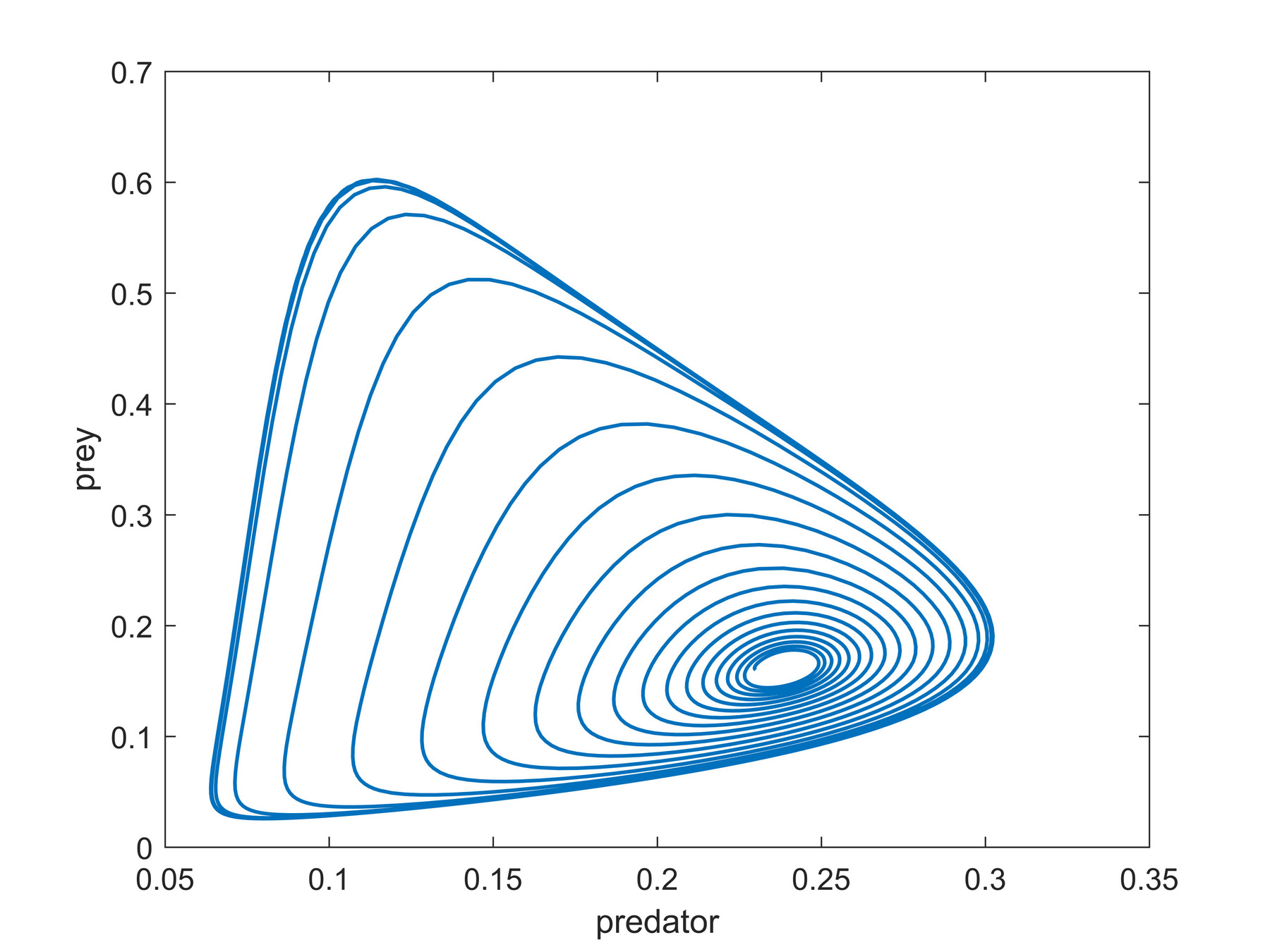}
    \includegraphics[width=\textwidth]{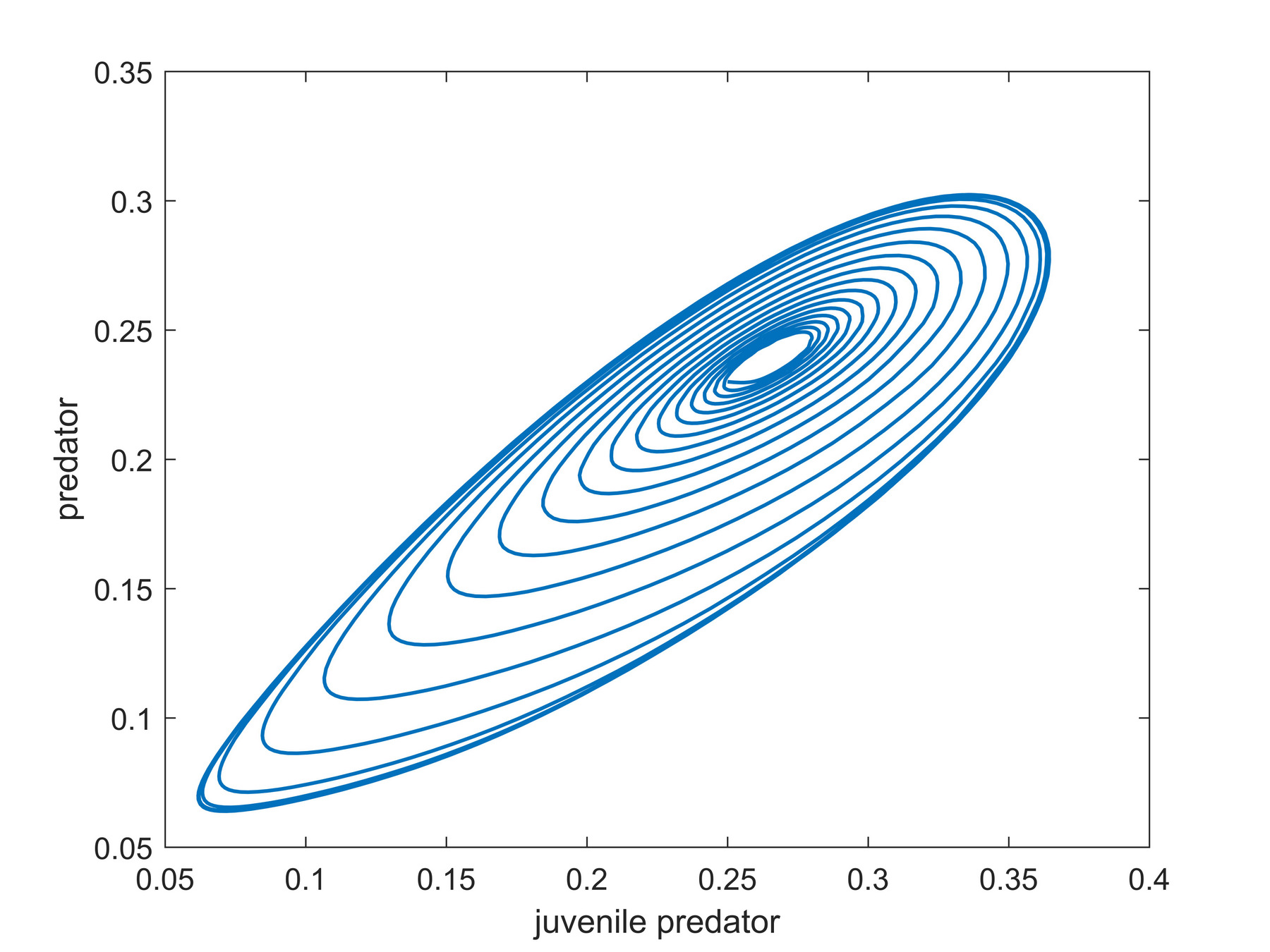}
    \caption{at $n=0.46633256$}
    \label{n-dim-b}
    \end{subfigure}
    \begin{subfigure}[b]{0.3\textwidth}
         \centering
    \includegraphics[width=\textwidth]{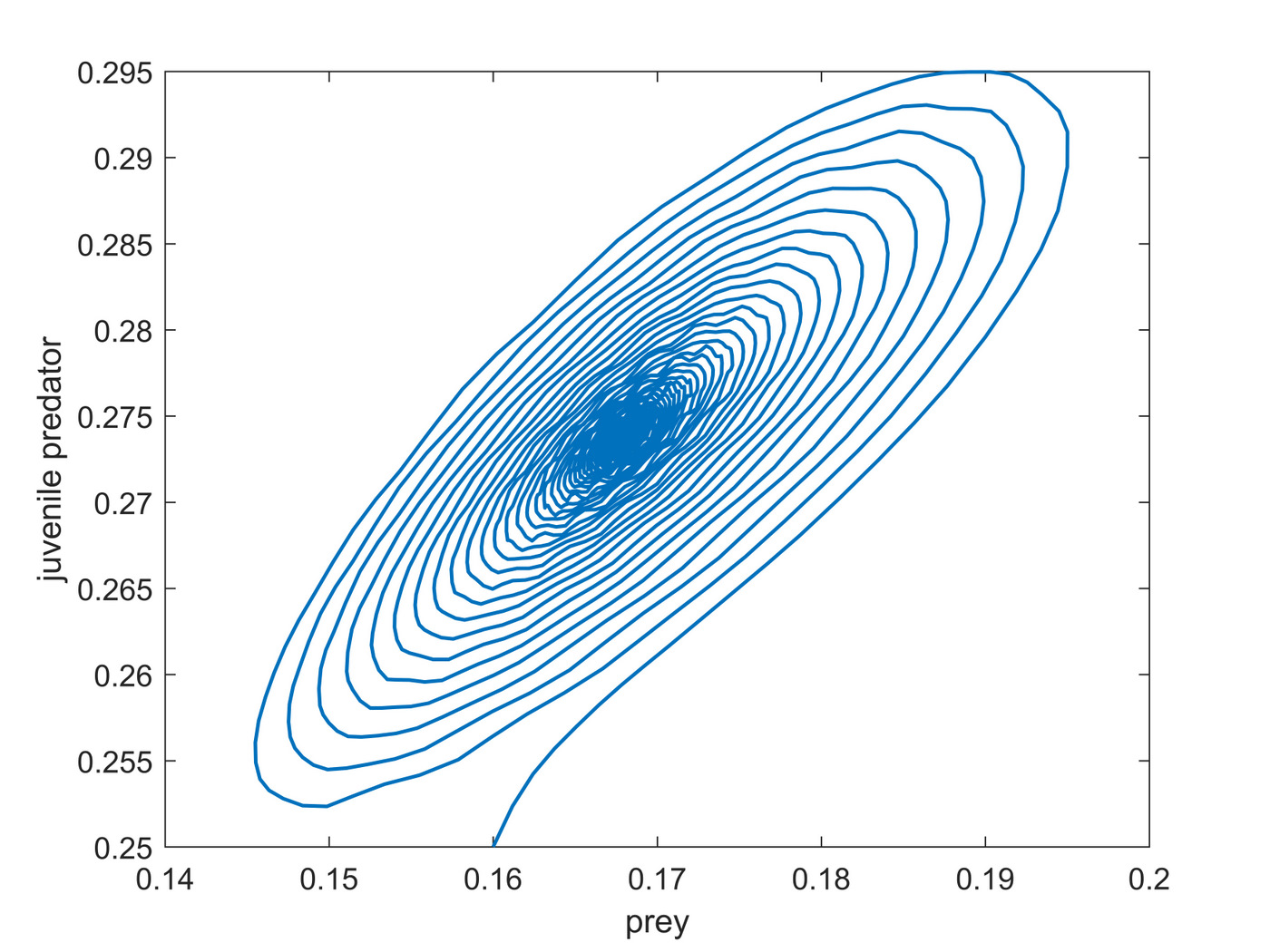}
    \includegraphics[width=\textwidth]{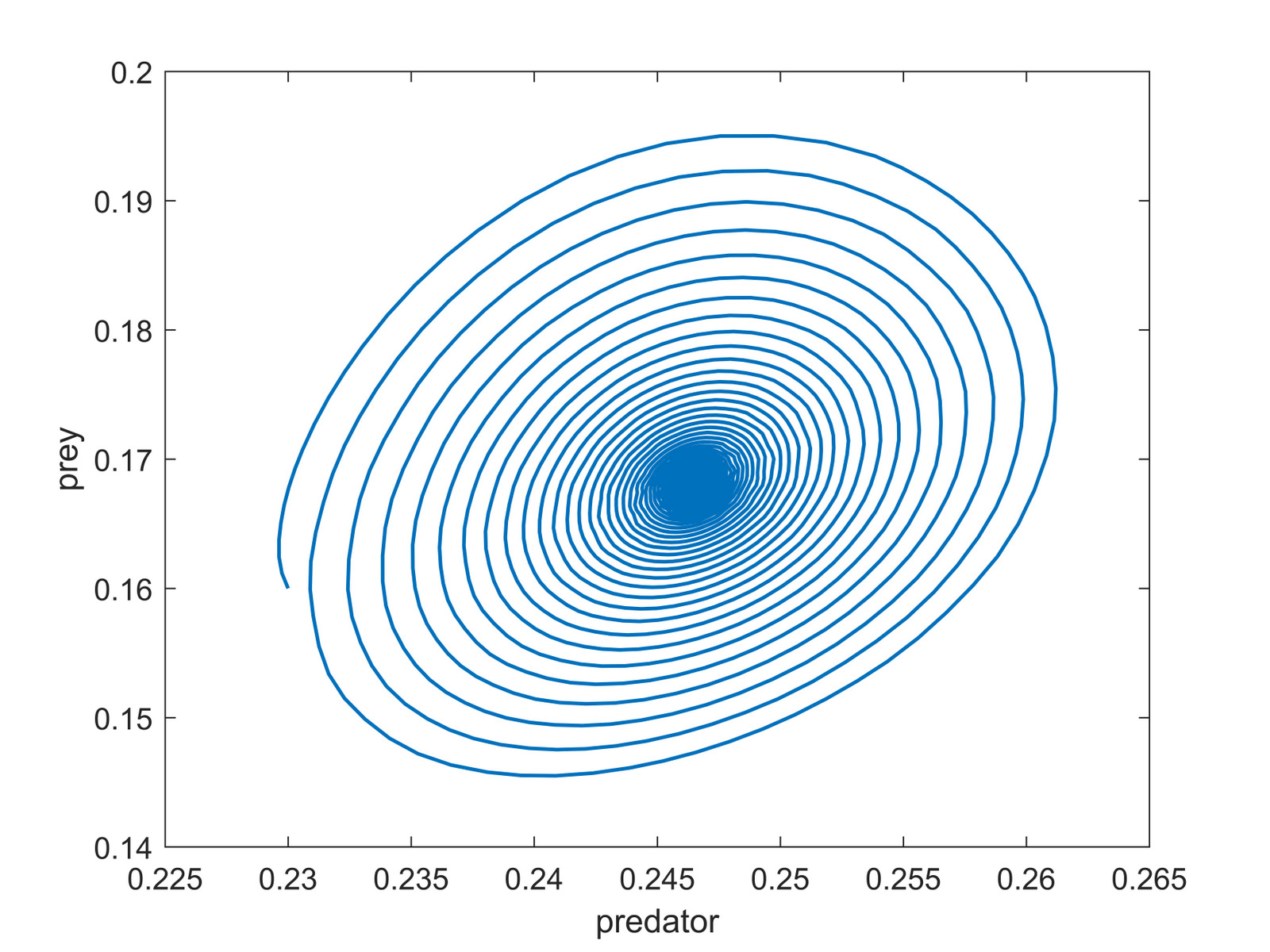}
    \includegraphics[width=\textwidth]{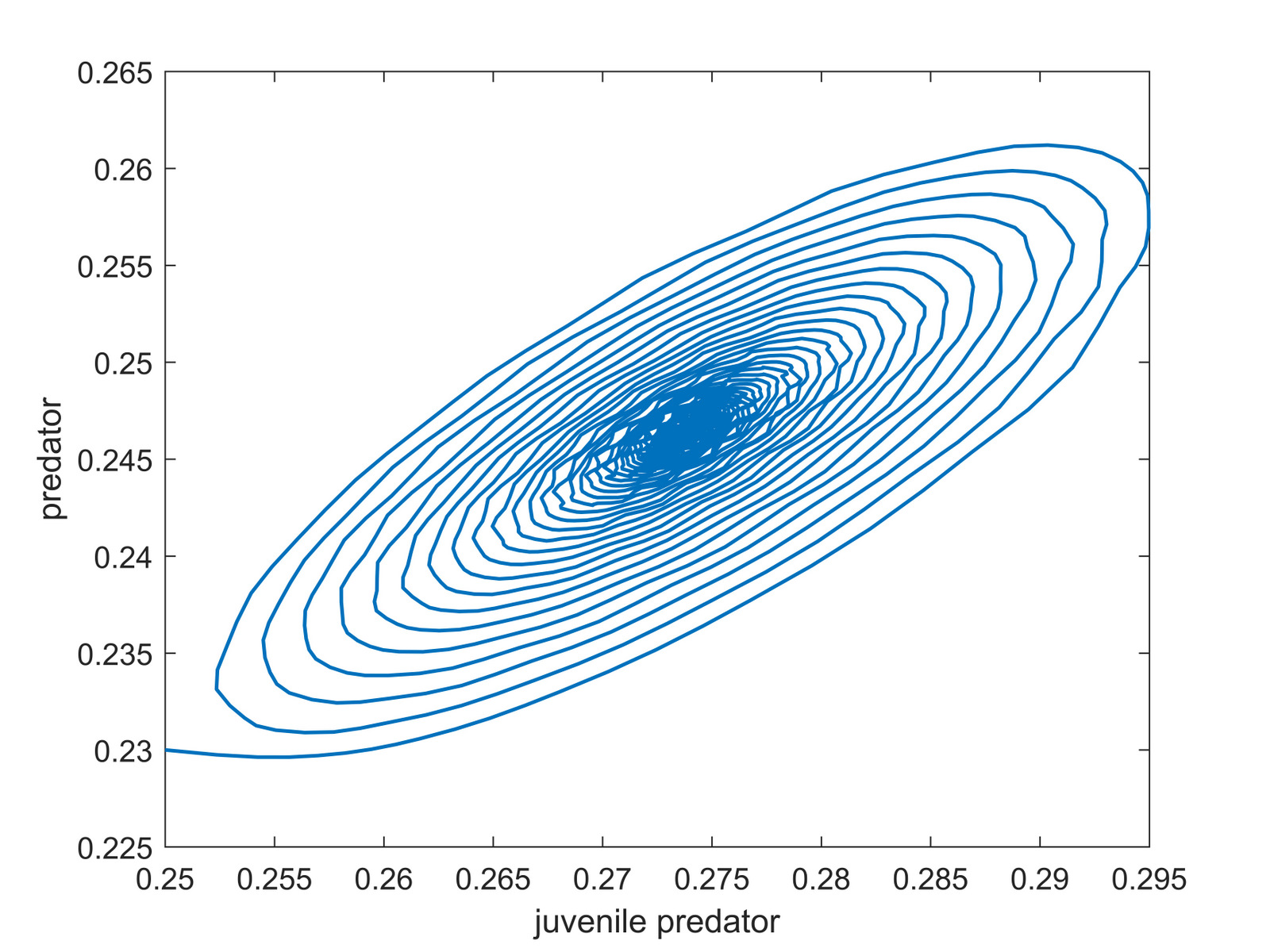}
    \caption{at $n=0.49$}
    \label{n-dim-c}
    \end{subfigure}
    \caption{Portraying prey population, juvenile and matured predator population dynamics in a bi-dimensional space,where $n =0.46633256$ is Hopf-bifurcation point}
    \label{n-2dim}
\end{figure}
\begin{figure}[H]
   \begin{subfigure}[b]{0.5\textwidth}
         \centering
    \includegraphics[width=\textwidth]{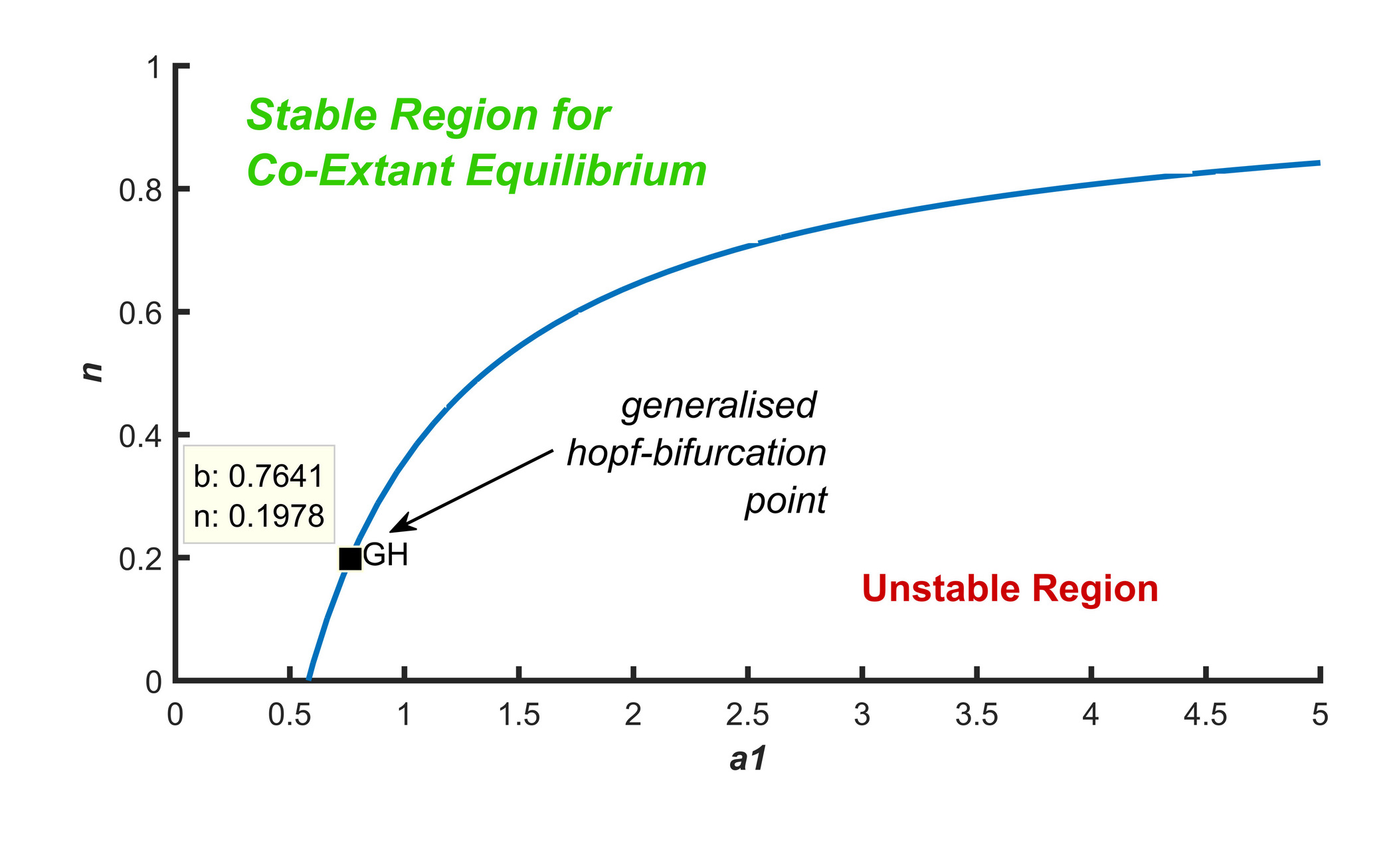}
   \caption{Predation rate by juveniles and habitat complexity}
    \label{ho-bi-1}
    \end{subfigure}\hfill
    \begin{subfigure}[b]{0.5\textwidth}
         \centering
    \includegraphics[width=\textwidth]{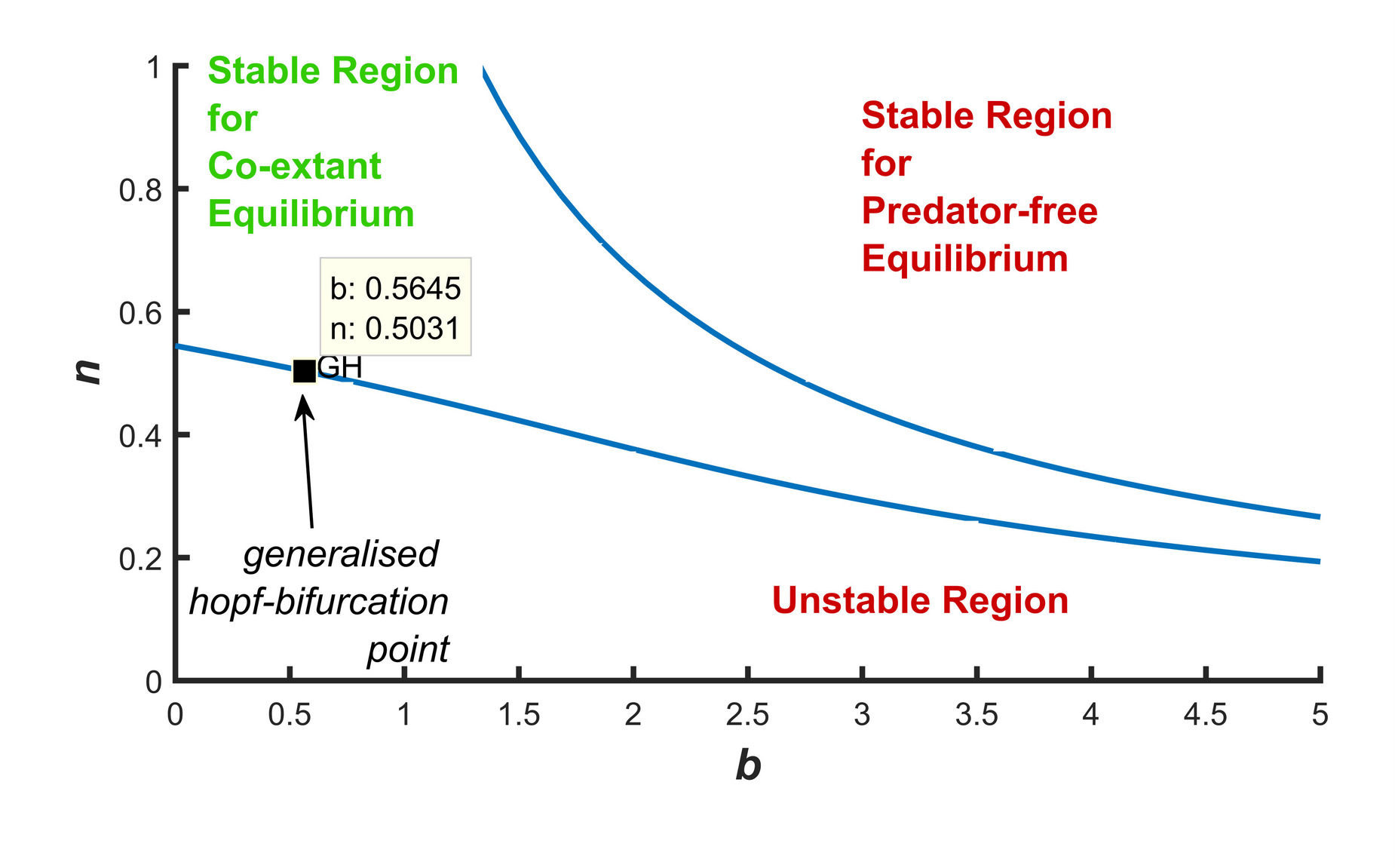}
    \caption{Inefficiency rate and habitat complexity}
    \label{ho-bi-2}
    \end{subfigure}
    \caption{Bifurcation curves in bi-parametric region}
    \label{ho-bi-curve}
\end{figure}
\begin{figure}[H]
   \begin{subfigure}[b]{0.33\textwidth}
         \centering
    \includegraphics[width=\textwidth]{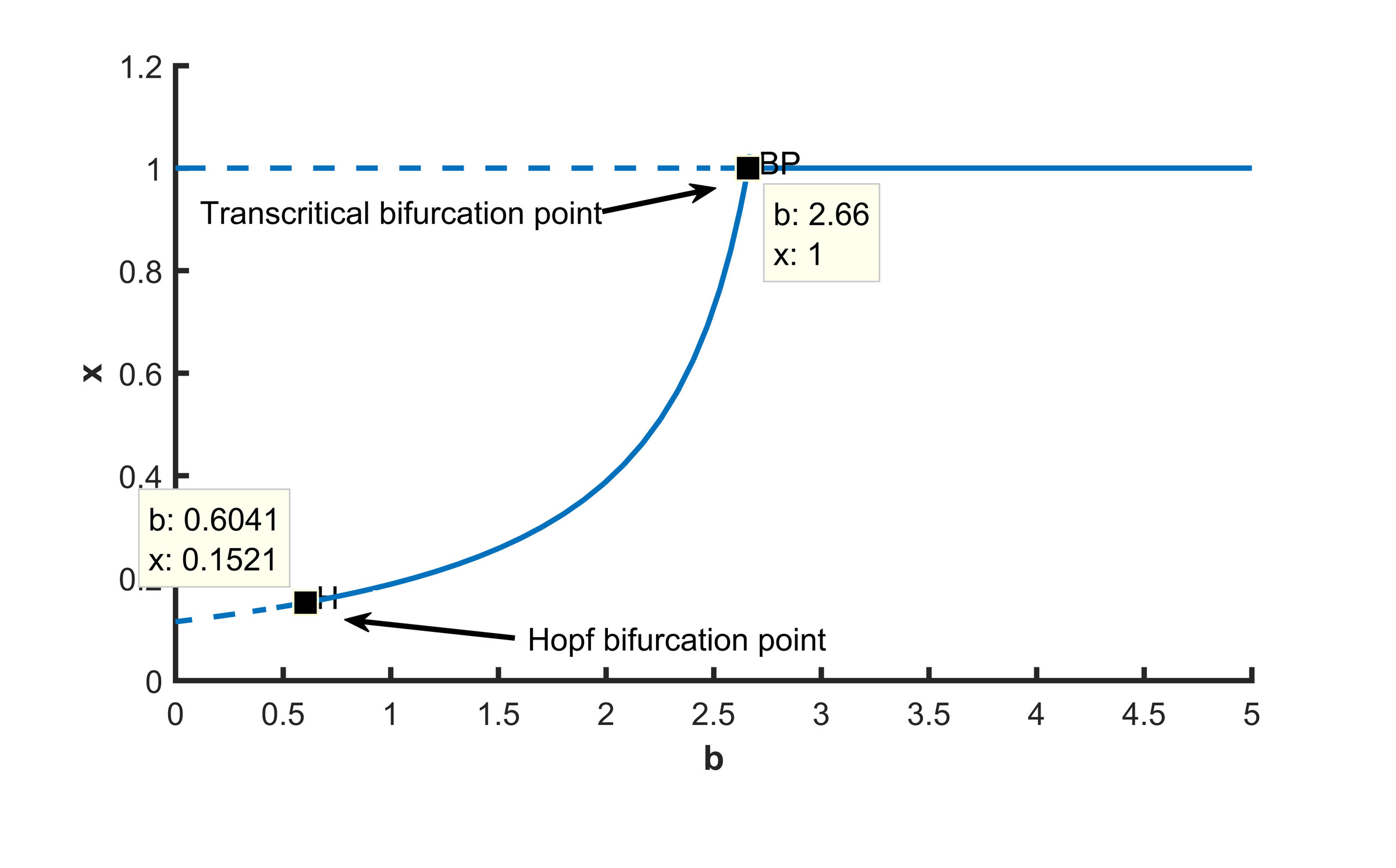}
    \caption{}
    \end{subfigure}\hfill
    \begin{subfigure}[b]{0.33\textwidth}
         \centering
    \includegraphics[width=\textwidth]{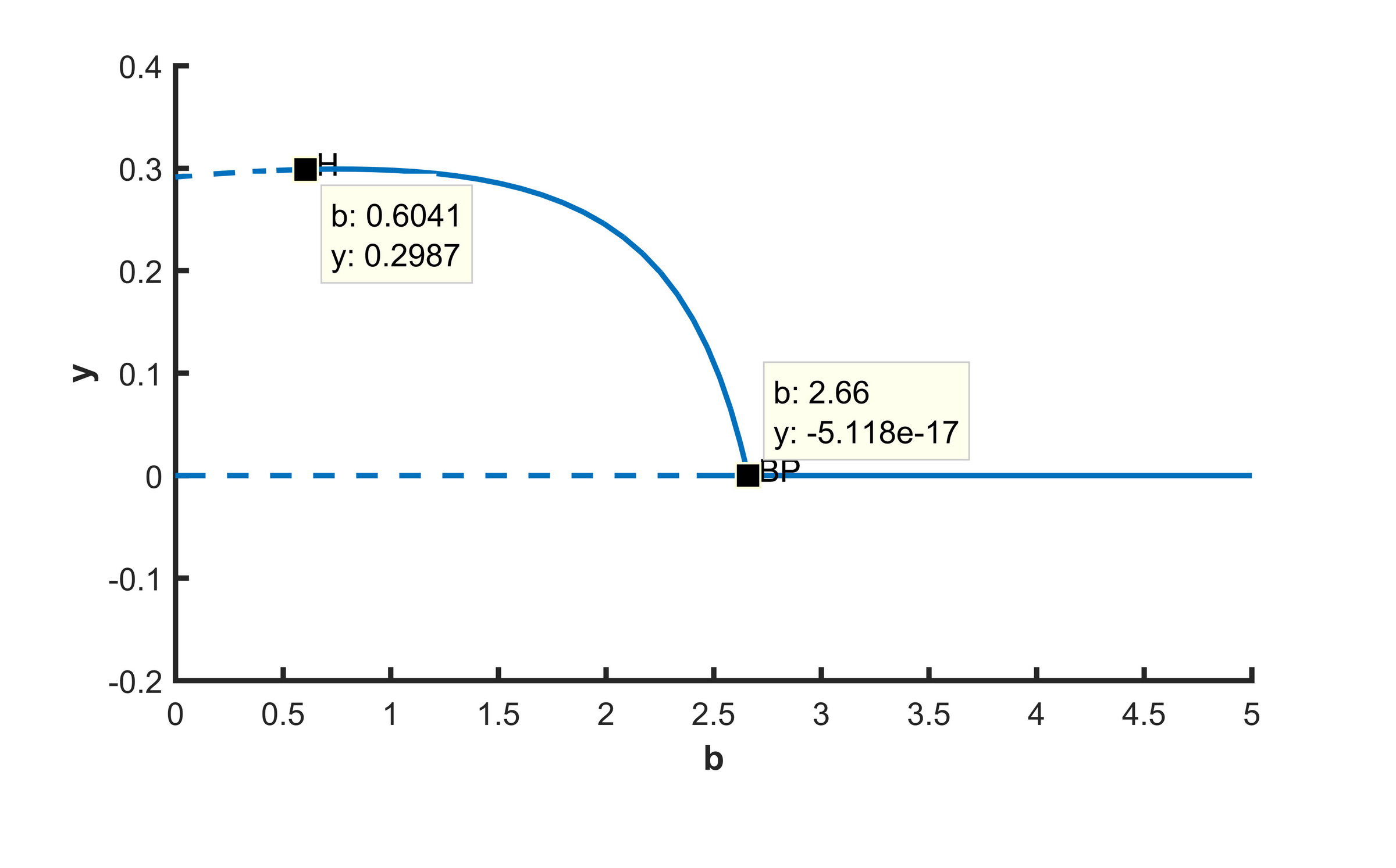}
    \caption{}
    \end{subfigure}\hfill
    \begin{subfigure}[b]{0.33\textwidth}
         \centering
    \includegraphics[width=\textwidth]{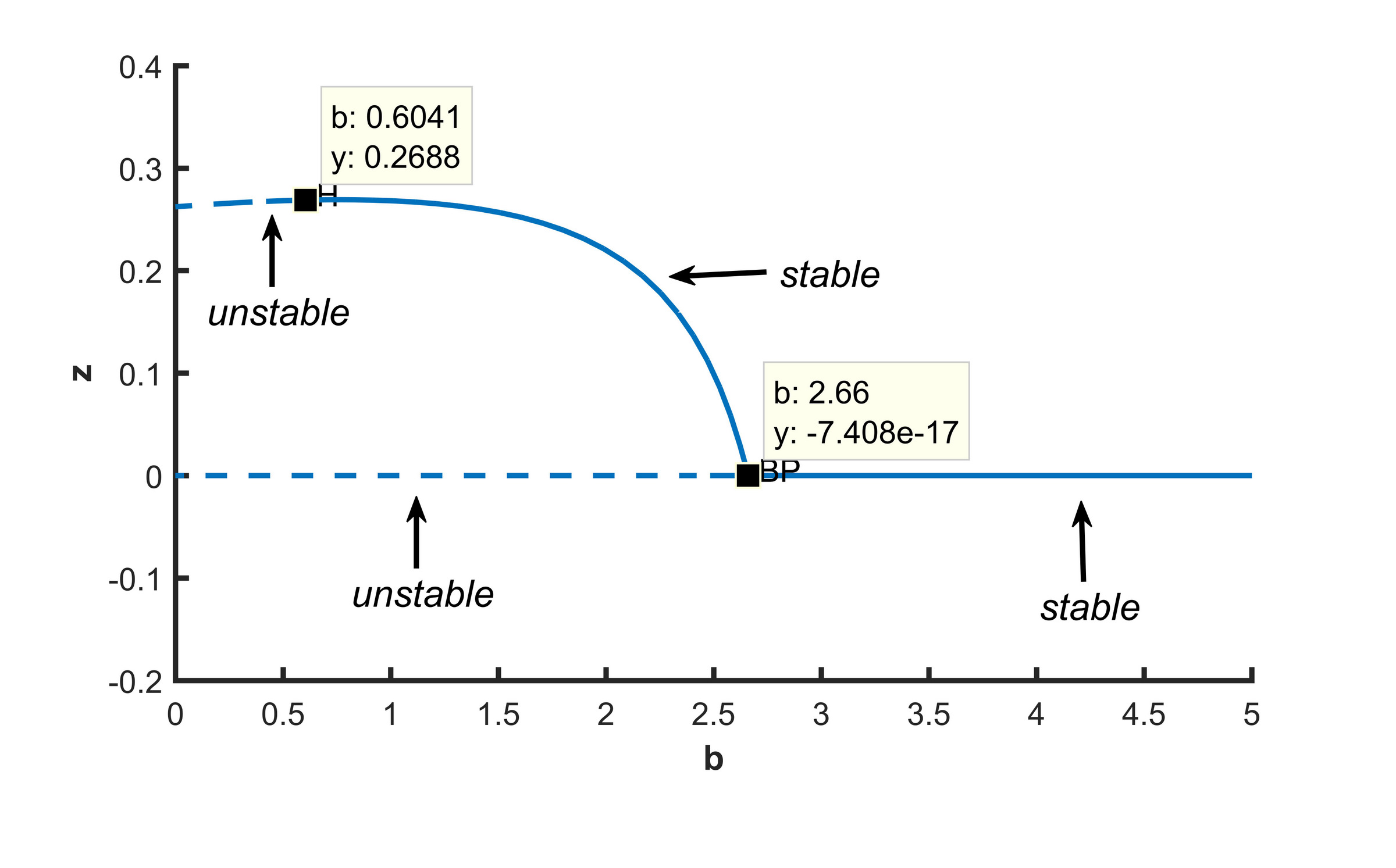}
    \caption{}
    \end{subfigure}
    \caption{Equilibrium curves with respect to inefficiency rate}
    \label{b-equ-curve}
\end{figure}
\begin{figure}[H]
   \begin{subfigure}[b]{0.5\textwidth}
         \centering
    \includegraphics[width=\textwidth]{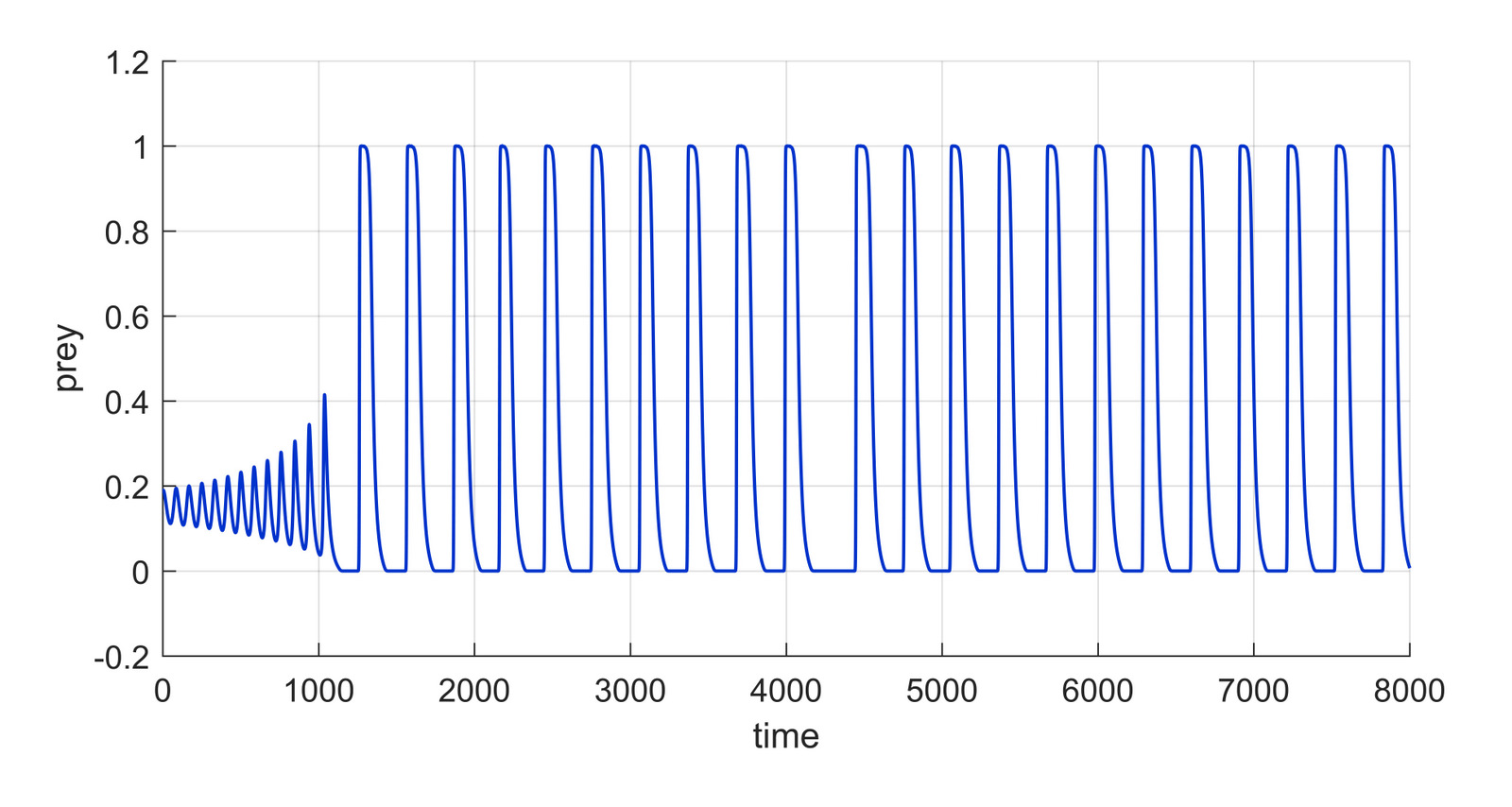}
    \caption{with respect to prey population}
    \label{b-1}
    \end{subfigure}\hfill
    \begin{subfigure}[b]{0.5\textwidth}
         \centering
    \includegraphics[width=\textwidth]{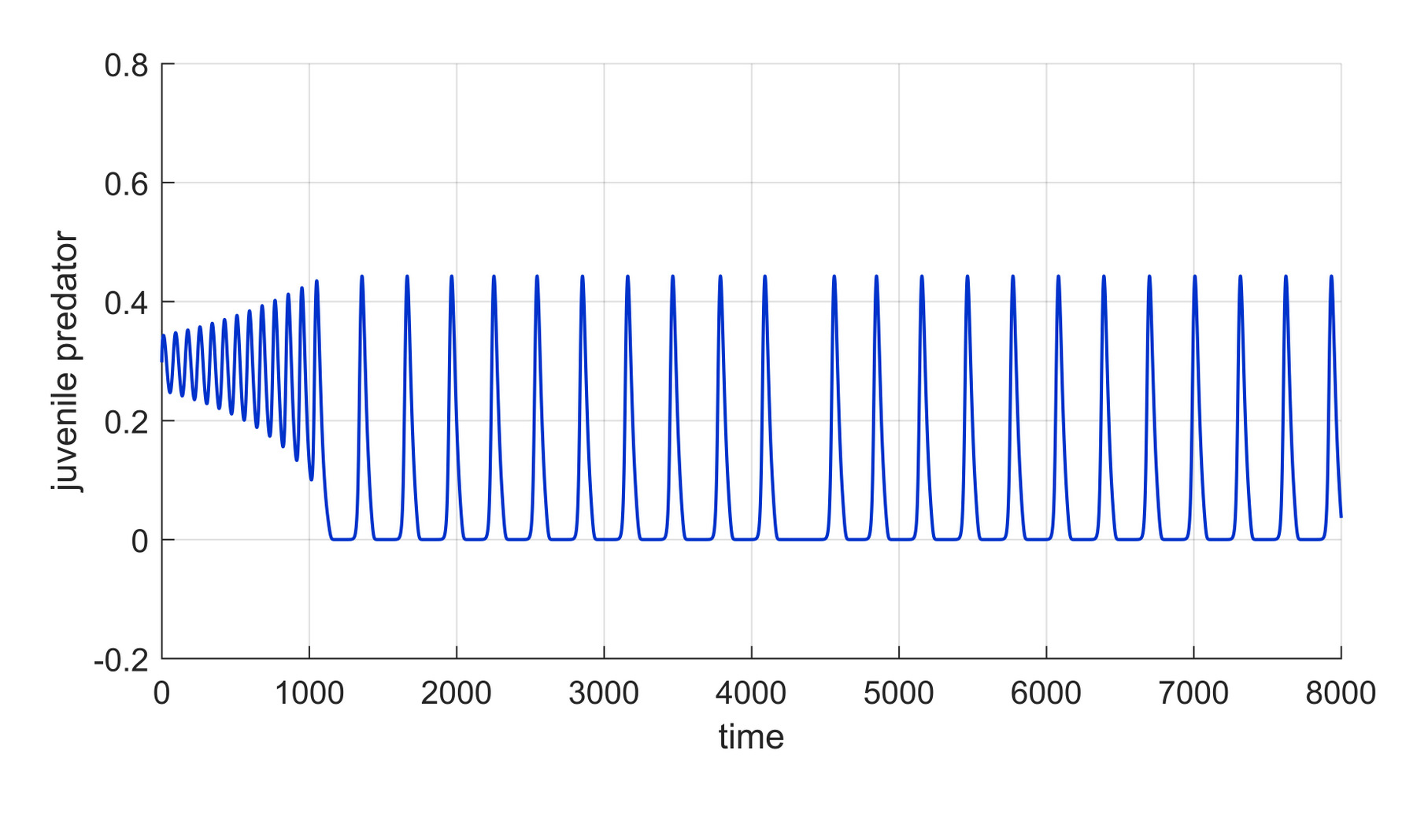}
    \caption{with respect to juvenile predator population}
    \label{b-2}
    \end{subfigure}\hfill
    \begin{subfigure}[b]{0.5\textwidth}
         \centering
    \includegraphics[width=\textwidth]{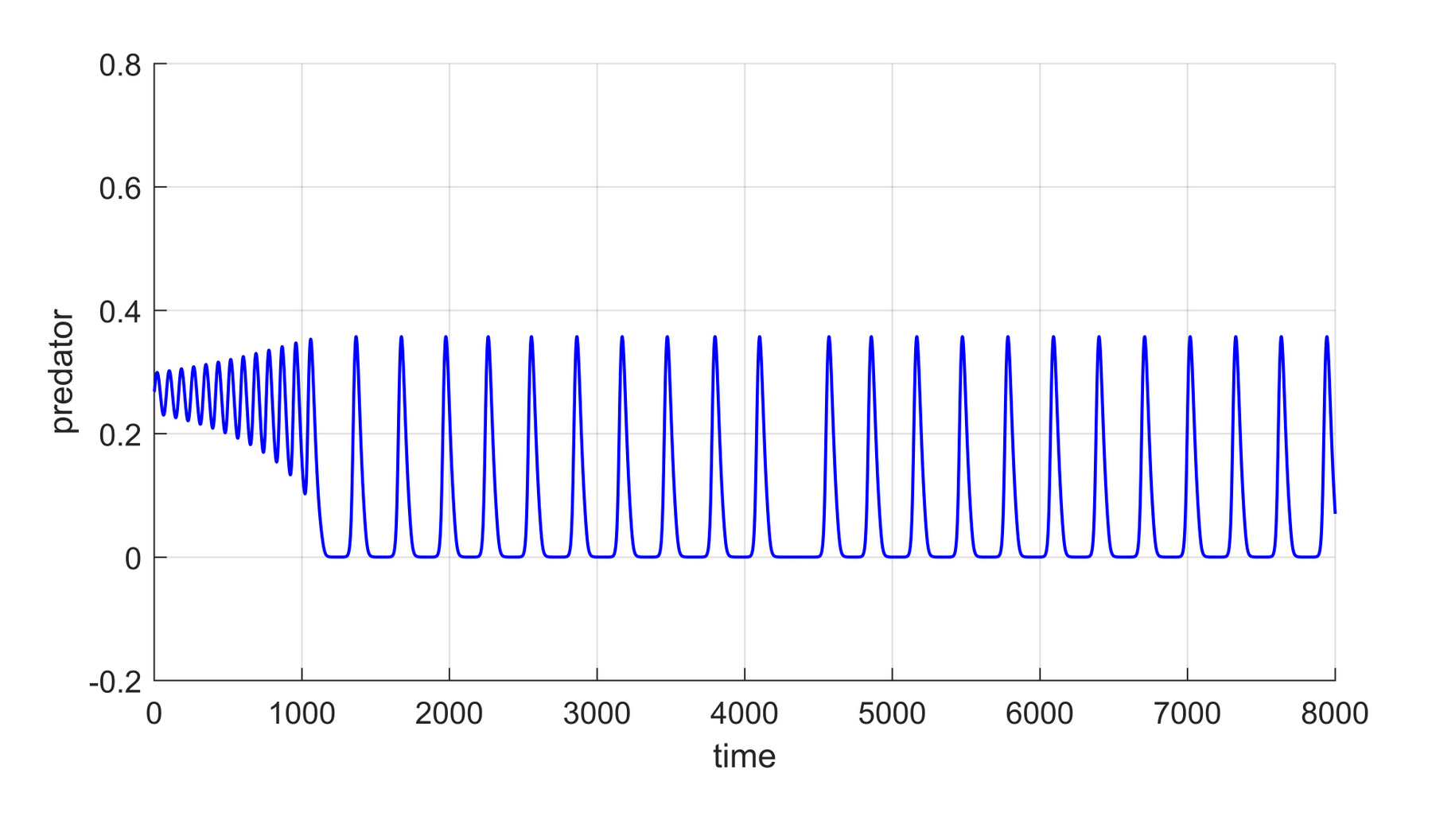}
    \caption{with respect to predator population}
    \label{b-3}
    \end{subfigure}\hfill
    \begin{subfigure}[b]{0.5\textwidth}
         \centering
    \includegraphics[width=\textwidth]{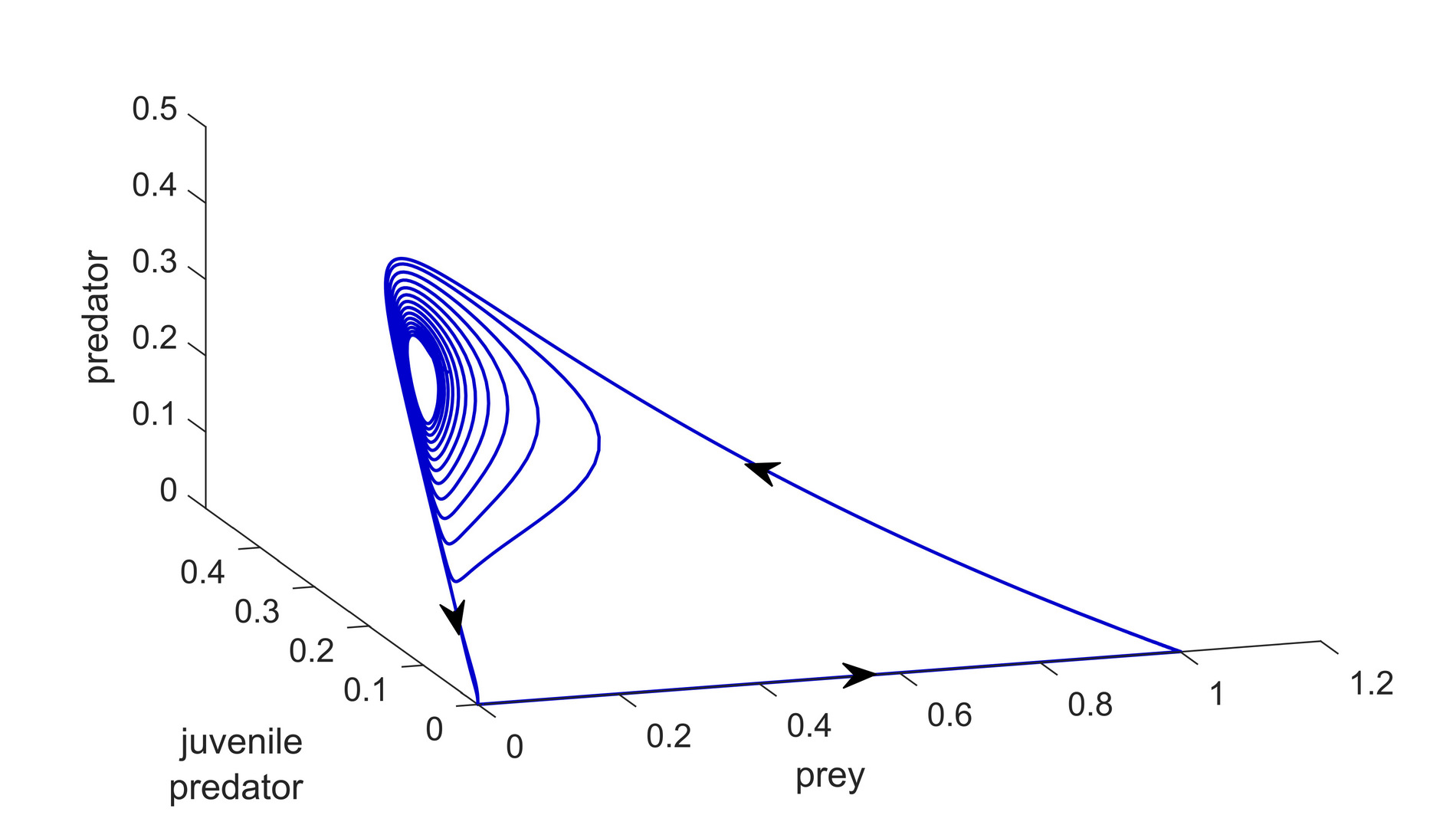}
    \caption{phase portrait}
    \label{b-4}
    \end{subfigure}
    \caption{Phase portrait and time series of all three populations depicting stable limit cycle involving extinction of all (0,0,0), extinction of only predators (1,0,0) and also the co-existence of prey, juvenile predator and matured predators}
    \label{b-curve}
\end{figure}
\begin{multicols}{2}
\subsection{Influence of Ineffiency and Habitat complexity}
To comprehend the significance of habitat complexity(\textit{n}) in determining the behaviour of the bio-system, figure \ref{n-time} is drawn at different set of values of \textit{n}. Other parameters are taken as is in the table except for $b=1.5$. For $n=0$, the system goes towards the vanishing equilibrium(figure\ref{a}), while at $n=1$ the system harbours a stable predator free equilibrium(figure \ref{d}). At $n=n_{h1}=0.42286752$ a supercritical Hopf-bifurcation is found. For $n\in(n_{h1}n_{h2})$ the co-extant equilibrium is stable(figures \ref{b} and \ref{c}), where $n_{h2}=0.886667$. The system stability is reassigned from co-extant equilibrium point to predator free equilibrium point at $n_{h2}$ as a result of transcritical bifurcation.
\par The Hopf bifurcation plot with respect to \textit{n} is portrayed in figure \ref{n-2dim} (with $b=1.015$), the bifurcating parametric value is at $n=0.46633256$. The initial co-existing population follows a trajectory that spirals into the co-extant equilibrium point at $n=0.49$ (figure \ref{n-dim-c}) and spirals around the said equilibrium point at $n=0.46633256$ (figure \ref{n-dim-b}). A stable limit cycle with unstable equilibrium point at its focus can be seen at $n=0.42$ (figure \ref{n-dim-a}). Biologically, with the habitat complexity too high, the survival of juvenile predators becomes really difficult leading to extinction of the whole predator species, while with less habitat complexity, the juvenile predator can easily survive and kill the prey which would initially increase the biomass density of the predator species but would lead to collapse of the whole bio-system in the long run.\\
\par Portrayal of dynamical scenerio with respect to the bifurcation in bi-parametric region is given in figure \ref{ho-bi-curve}. Hopf-bifurcation curve along the parametric axes is drawn thereby dividing the whole region into stable and unstable parts. Figure \ref{ho-bi-1} shows the relation between predation by juveniles($a_1$) and habitat complexity(\textit{n}). The Generalised Hopf bifurcation is at $(a_1,n)=(0.764105,0.197769)$, the first Lyapunov coefficient being zero while the second Lyapunov coefficient is $-3.700813e-01$, the Generalised Hopf-bifurcation divides the Hopf curve into two parts, supercritical and subcritical parts. Similarly, the Hopf bifurcation curve with respect to the two parameters inefficiency rate(b) and habitat complexity depicts the stable and unstable region (figure \ref{ho-bi-2}), $(b,n)=(0.564633,0.503071)$ is the Generalised Hopf bifurcation point. The stable region is further divided into two regions of stability, each for axial and co-extant equilibrium point by transcritical bifurcation curve.\\
\par The equilibrium curve with respect to ineffiency rate is visualised in figure \ref{b-equ-curve}. The Hopf bifurcation point of b is at 0.6040816,$b>0.6040816$, the bio-system possess a satble co-extant equilibrium point otherwise unstable, ultimately going to vanishing equilibrium point; which implies the juveniles becoming highly efficient is detrimental to the bio-system. Also, in the contrary,them becoming highly inefficient would cause the whole predator species too be annihilated as the bio-system has a transcritical bifurcation at $b=2.66$\\
\par For $b=0.55$, the initial co-existing population is seen to follow a particular trajectory where a cycle spiraling around the co-extant equilibrium point $E_3(0.14803,0.29835,0.268515)$ is created between all the three different kinds of equilibrium achievable in our bio-system; from periodic co-existing point $E_3'(x',y',z')$ to vanishing equilibrium $E_1(0,0,0)$ and then to predator free equilibrium $E_2(1,0,0)$ then again to $E_3'(x',y',z')$. At this set of values neither the axial equilibrium(1,0,0) nor the co-extant equilibrium (0.14803,0.29835,0.268515) satisfy their stability conditions. Figure(\ref{b-curve}) shows the all the three populations fluctuating between the three equilibrium scenerios.
\section{Discussion and Conclusion}
In this paper, a bionetwork comprising of two species, the prey and the predator with stage structure in the predator species has been set forth with ratio dependent functional response. A new perspective of predation by juvenile predators and the circumstances that effectuates the fatalities of only juvenile predators due to their inefficiency in handling preys along with the prey's habitat, has been explored through both theoretical and numerical methodology.The prey's antipredator behaviour is inferred to be its habitat, without which the prey is innocuous. During predatism, juvenile predators may be inefficient to deal this habitat complexity, ensuing their death. The habitat complexity may also be taken as a territorial intermediate predator, who are at the same stage in the food chain as the juveniles of the top predator, if not higher; in such a scenerio barring the territory of intermediate predator, the juveniles are at liberty to hunt without any repercussions. The intermediate predators are envisaged only as the functioning of the habitat complexity and not as a species in the bio-system.
\par The biosystem \ref{eq2} is constricted in a particular region in the first octant, hence none of the populations can increase rampantly, which is in conformity of nature's state of affairs. The system of equations manifests three equilibrium points; all of whose local stability as well as global stability of axial and co-extant equilibrium points has been investigated and then visualized graphically.
\par It is observed that the predator free equilibrium point $E_2(1,0,0)$ doesn’t possess a Hopf-bifurcation point. As absenteeism of predators gives carte blanche to the prey whereby they attain their carrying capacity. However transcritical bifurcation does inhabit this equilibrium point. It is notable that stability of axial equilibrium, occurence of transcritical bifurcation,and existence of co-extant equilibrium point are all related to the same expression $d_2(bn+c+d_1)-a_3c$ being greater than, equal to, or lesser than zero respectively.
\par Results concomitant to perseverance of the biosystem has been looked into; ample amount of the work is done on co-extant equilibrium point due to its ecological importance. Different bifurcation graphs around the co-extant equilibrium $E_3$ for various parameters are shown.\\
Predation by juveniles creates a supplementary slaughter of the prey's population. An increase in $a_1$ (predation rate by juvenile) after a critical point is reciprocated by the equilibrium point $E_3$ by becoming unstable through hopf-bifurcation, which is observed to be supercritical. Juveniles predating ensures their nutriment while also making them highly efficient for their later stage of life.
The more efficient juvenile predators are, the more their survivability would be into the matured stage. Henceforth, decreased inefficiency rate(\textit{b}) after a critical value would indicate increase in the biomass density of both juveniles and the matured predators initially, correspondingly the biomass of prey would rapidly decrease, which would automatically after a certain period of time result into extinction of all, on the flip side, a great amount of inefficiency would correspond to predator getting completely annihilated and the prey reaching its carrying capacity.
Habitat complexity(\textit{n}) is appertained to continuity of both the prey and predator species. Decreased complexity of the habitat conforms much to the previous scenerio of more predators surviving in their juvenile stage, along with increased deaths of the prey in the hands of the juveniles, whereby leading to complete extermination the prey. If either the juvenile predators becomes as efficient as the matured predator (i.e. $b=0$) or in the absence of habitat complexity (i.e. $n=0$), the constrains(their inefficiency in dealing with habitat complexity) effectuating the death of juveniles are eliminated,and hence the term implementing the negative repercussions of juveniles hunting vanishes, resulting in both co-extant equilibrium as well axial equilibrium point to be unstable. That being the case, inefficiency rate as well as habitat complexity is essential for the survivablity of the species. An increase in rate of habitat complexity is allied with the increase in the biomass density of the prey species, and on the flip side decreases the  biomass of juvenile and matured predators (figure\ref{n-time}). \\  
It requires special mention that for some distinctive values of \textit{b}, the bio-system becomes fickle, the vanishing equilibrium, predator-free equilibrium along with the periodic co-existing point all comes into play one after another thereby creating a unique stable cycle between them (figure\ref{b-curve}). From biological point of view, this can be interpreted as- for low inefficiency rate, the total biomass density of all three populations deceases and gradually moves towards the collapse of the whole bio-system, but as soon as the predators of both the stages approaches extinction, the prey, who themselves were nearing extinction quickly revives and reaches the carrying capacity of the environment, the almost extinct predators now having the availability of plenty of food too revives rapidly. Once all three populations reaches its full potential while co-existing, again starts decreasing and going towards extinction of all, thereby creating this particular cycle where all three kinds of equilibrium can achieved one after another.
\par Existence of predators largely relies on the conversion rate of juvenile predator $a_3$, which refers to the amount of nutrients required by the matured predators for reproduction. Diminishing conversion rate insinuates the equilibrium point going towards the predator free equilibrium point due to the presence of transcritical bifurcation point. Also, excessive conversion rate culminates into extinction for all the three species, as with the increase of juvenile predators, more prey would be killed as a consequence but the prey are not being able to further increase their number, hence the extinction. The search and conquer rate of both stages of predators(\textit{m}) is observed to have the ability to destablize the bio-system for a certain value of m. As m decreases, the biomass density of predator species would initially increase due to being them able to kill prey faster and then decrease due to unavailability of prey \\
\par On that account we can assert that rate of predation by juvenile ($a_1$),their inefficiency (\textit{b}), habitat complexity (\textit{n}), search and conquer rate(\textit{m}) and the conversion rate ($a_3$) contributes greatly in changing the dynamics of  our proposed bio-system. All the parameters should be in fair amount for all the species to co-exist.
\begin{center}
    \Large{\textbf{Appendix}}
\end{center}
\begin{appendix}
\section{\texttt{Proof of Positivity}} \label{proof-posi}
The initial conditions of the biosystem \ref{eq2} i.e. $x_0,y_0,z_0$ are all positive, therefore at any time t, we would acquire 
\begin{align*}
 x(t) &=x_0 exp[ \int_0^t ((1-x(s))-\frac{a_1(1-n)y(s)}{m(y(s)+z(s))+(1-n)x(s)}\\
 &-\frac{a_2 z(s)}{m(y(s)+z(s))+x(s)}) \, ds] \geq0 \\
 y(t) &=y_0 exp[\int_0^t (\frac{a_3 x(s) z(s)}{y(s) (m(y(s)+z(s) )+x(s) )}\\
 &-\frac{bn x(s)}{m(y(s)+z(s) )+x(s)}-c-d_1) \,ds] \geq0 \\
z(t) &= z_0 exp[\int_0^t(\frac{cy(s)}{z(s)}-d_2) \,ds] \geq0 
\end{align*}
Hence, the non-negative solution exists.
\section{\texttt{Proof of Boundedness:}} \label{proof-bound}
Defining p as, $p=a_3x+a_2y+a_2z$\\ \\
Taking the derivative of p with respect to time t, we have the following equation:\\
\begin{equation*}
\frac{dp}{dt}=a_3\frac{dx}{dt}+a_2\frac{dy}{dt}+a_2\frac{dz}{dt}
\end{equation*}
Now, putting the values of  derivatives of x, y, z, we have\\
\begin{align*}
\frac{dp}{dt}&=a_3x(1-x)-\frac{a_1a_3(1-n)xy}{(m(y+z)+(1-n)x)}\\
&-\frac{a_2bnxy}{(m(y+z)+x)}-a_2d_1y-a_2d_2z\\
\frac{dp}{dt}&\leq a_3(1-x)x-d_1a_2y -d_2a_2z
\end{align*}
\begin{align*}
i.e \frac{dp}{dt} +d_3 p& \leq a_3(1-x)x+d_3a_3x-a_2y(d_1-d_3)\\
&-a_2z(d_2-d_3)\hspace{0.5cm} [where\hspace{0.1cm} d_3=min(d_2,d_1)]\\
i.e \frac{dp}{dt}+d_3p &\leq a_3 x(1+d_3 -x)\\
&=a_3(-x^2 +(1+d_3)x)\\
&=-a_3(x-\frac{1+d_3}{2})^2+\frac{a_3(1+d_3)^2}{4}\\
i.e. \frac{dp}{dt}+d_3p &\leq\frac{a_3(1+d_3)^2}{4} =H(say)
\end{align*}
Therefore, we have $ p \leq \frac{H}{d_3}(1-exp(-d_3 t))+p(x_0, y_0, z_0)\exp(-d_3 t)$. 
When $t\rightarrow \infty$ , we have $p\leq \frac{H}{d_3}$. 
Hence all solutions of the system \ref{eq2} are bounded in the region \{(x, y, z) $\in {R^3}_+$ : p=$\frac{H}{d_3}-\tau$, for any $\tau>0\}$ 
\section{\texttt{Stability of vanishing\\ equilibrium point}} \label{proof-vanish}
The Jacobian matrix being undefined at (0,0,0), the usual method for inspecting the stability cannot be done, hence the method developed  by Arino et al\cite{equ} is used.
The biosystem \ref{eq2}\\
$\dot{X}=G(X)$    where 
$X=\left( \begin{array}{c}
    x \\
    y\\
    z
\end{array}\right)$ and \\ $G(X)=\left( \begin{array}{c}
     x(1-x)-\frac{a_1(1-n)xy}{m(y+z)+(1-n)x}-\frac{a_2xz}{m(y+z)+x} \\
     \frac{a_3xz}{m(y+z)+x}-\frac{bnxy}{m(y+z)+x}-cy-d_1y\\
     cy-d_2z
\end{array}\right)$\\
is transformed to \begin{equation} \label{lo1}
    \frac{dX}{dt}=L(X(t))+K(X(t))
\end{equation}
Where L(.) is a homogeneous function of degree 1 which is continuous everywhere except the origin; K is a $C^1$ function with K(V)=o(V) in the vicinity of (0,0,0),\\
$L=\left( \begin{array}{c}
     x-\frac{a_1(1-n)xy}{m(y+z)+(1-n)x}-\frac{a_2xz}{m(y+z)+x} \\
     \frac{a_3xz}{m(y+z)+x}-\frac{bnxy}{m(y+z)+x}-cy-d_1y\\
     cy-d_2z
\end{array}\right),\hspace{3pt} and\\ K=\left( \begin{array}{c}
    -x^2 \\
    0\\
    0
\end{array}\right)$\\
If $X_1(t)$ is the bounded solution of (\ref{lo1}) such that $\liminf{||X_1(t)||}=0$ and $X_1(t_n,+,.)$ is the subsequence converging to zero, then we define a sequence $\{x_n\}$ by\\
$x_n=\frac{X_1(t_n+s)}{||X_1(t_n+s)||}$ with $||x_n||=1$ 
where $||.||$ is the euclidean norm.\\
A sub-sequence of $\{x_n\}$ is found to be convergent to a function w(t) that satisfies \\
\begin{equation}\label{lo-2}
    \frac{dw}{dt}=L(w(t))-(w(t),L(w(t)))w(t)
\end{equation}
with $||w(t)||=1$, with the aid of Ascoli-Arzela theorem.\\
The vector $v(t)=(v_1,v_2,v_3)$ assuages the steady state of L by 
$L(v)=(v,L(v))v$, which is rewritten as\\
$L(v)=\eta v$\\
Now, we are at the stage where we can determine the conditions for reaching the origin with the help of the following:\\
$v_1 (a_1 (-1 + n) v_2 (v_1 + m (v_2 + v_3))-a2 v_3 (v_1 - n v_1 + m (v_2 + v_3)) + (v_1 + m (v_2 +v_3)) (v_1-nv_1 +m(v_2+v_3))-(v_1+m(v_2+v_3))(v_1-nv_1 +m(v_2+v_3))\eta)=0$\\
$a_3 v_1v_3-b v_1v_2-v_2 (c+d_1+\eta ) (m (v_2+v_3)+v_1)=0$\\
$c v_2- v_3(d_2 + \eta)=0$\\ \\
For which, we arrive at the following cases:\\
\textbf{Case1}  Along x-axis \hspace{10pt}i.e. $v_1 \neq0, v_2=v_3=0$\\
We have $(\eta-1) (n-1) v_1^3=0$\\
$\eta=1$ hence (0,0,0) can't be reached along x-axis\\
\textbf{Case2} Along y-axis \hspace{10pt}i.e. $v_1=v_3=0, v_2 \neq0$\\
We have $-m v_2^2 (c+d_1+\eta )=0$\\
For $\eta=-(c+d_1)$ origin is reachable along y-axis.\\
\textbf{Case3}  Along z-axis\hspace{10pt}i.e. $v_1= v_2=0, v_3 \neq0$\\
We have $(\eta+d_2)v_3=0$\\
For $\eta=-d_2$ origin is reachable along z-axis.\\ 
\textbf{Case4} Along xy-plane \hspace{10pt}i.e. $v_3=0, v_1\neq0, v_2\neq0$\\
There is no possibility of reaching along this plane.\\
\textbf{Case5} Along zx-plane\hspace{10pt}i.e. $v_2=0, v_1\neq0, v_3\neq0$\\
The eigen-value along this plane is $\eta=1$, hence unstable\\\
\textbf{Case6} Along yz-plane \hspace{10pt}i.e. $v_1=0, v_3\neq0, v_2\neq0$\\
In the direction of (0,-z,z), the eigen-value is $-c-d_2$\\
In the direction of (0,y',z), the eigen-value is $-c-d_1$, hence stable for both directions.\\
where $y'=\frac{-c z-d_1 z+d_2 z}{c}$
\section{\texttt{Proof of local stability of \\$E_1(1,0,0)$}} \label{proof-stab-axi}
\begin{align}\label{lo-jaco}
    \left(
    \begin{array}{ccc}
 -1 & -a_1 & -a_2 \\
 0 & -c-d_1-b n & a_3 \\
 0 & c & -d_2 \\
\end{array}
\right)
\end{align}
is the Jacobian matrix of the biosystem \ref{eq2} at axial equilibrium point $E_2(1,0,0)$, whose eigen values are:\\
$-1,\\
\frac{1}{2} (-\sqrt{(b n+c+d_1+d_2)^2-4 (-a_3 c+b d_2 n+c d_2+d_1 d_2)}\\
-b n-c-d_1-d_2),\\
\frac{1}{2} (\sqrt{(b n+c+d_1+d_2)^2-4 (-a_3 c+b d_2 n+c d_2+d_1 d_2)}\\
-b n-c-d_1-d_2)$\\
The equilibrium point will be stable if all the eigen values are less than 0. The first two eigen values are negative. For the third one to be as well, we need\\
$$ a_3 < \frac{(c d_2 + d_1 d_2 + b d_2 n)}{c}$$
Hence, our proof is done.
\section{\texttt{Proof of global stability of\\ (1,0,0)}} \label{proof-glo-axi}
To investigate the globally asymptotically stability of $ (x_1, y_1, z_1)$= (1, 0, 0), the following  positive definite Lyapunov function is considered.\\ \\
V(x, y, z) =$L_1 (x- x_1 - x_1 ln\frac{x}{x_1}) +L_2y + L_3z$ \\
where $L_1, L_2,  L_3$ are positive constants, that are to be chosen appropriately. \\
Now taking the time derivative of V(x, y, z) along the solution of the biosystem \ref{eq2} , we have
\begin{align*}
\frac{dV(x, y, z)}{dt} &= L_1\frac{(x- x_1)}{x} \frac{dx}{dt} +L_2 \frac{dy}{dt} +L_3\frac{dz}{dt},\\
&= L_1\frac{(x- x_1)}{x} ( x (1-x) - \frac{ a_1 (1-n) xy}{m(y+z) + (1-n)x}\\
&-\frac{a_2 xz}{m(y+z)+x})+L_2(\frac{a_3 xz}{m(y+z)+x}\\
&-\frac{bnxy}{m(y+z)+x} –cy – d_1y)+L_3( cy – d_2z)
\end{align*}
Suppose $L_2=L_3$, and putting $x_1=1$ we have, \\
\begin{align*}
\frac{ dV(x, y, z)}{dt}&= L_1 (x-1) ( (1-x) - \frac{ a_1 (1-n) y}{m(y+z) + (1-n)x}\\ &-\frac{a_2z}{m(y+z)+x})+L_2(\frac{a_3 xz}{m(y+z)+x} \\
&-\frac{bn xy}{m(y+z)+x} –cy – d_1y) +L_2( cy – d_2z)\\
&= -L_1(x-1)^2 +\frac{L_1(1-x) a_1 (1-n) y}{m(y+z) + (1-n)x}\\
&+\frac{L_1a_2(1-x)z}{m(y+z)+x})+ L_2[\frac{a_3 xz}{m(y+z)+x}\\
&-\frac{bn xy}{m(y+z)+x} – d_1y– d_2z]\\
&\leq \frac{L_1a_1(1-n)(1-x)y}{m(y+z)+(1-n)x}-L_2d_1y-L_2d_2z+\\
&\frac{L_1a_2(1-x)z}{m(y+z)+x}+\frac{L_2a_3xz}{m(y+z)+x}-\frac{L_2bnxy}{m(y+z)+x}
\end{align*}
Now, putting $L_1a_2=L_2a_3$, we have,
\begin{align*}
    \frac{dV}{dt}&\leq L_2[\frac{a_1a_3(1-n)(1-x)y}{a_2(m(y+z)+x(1-n))}-d_1y-d_2z\\
    &\frac{a_3z}{m(y+z)+x}-\frac{bnxy}{m(y+z)+x}]\\
    \implies\frac{dV}{dt}&<0 \hspace{6pt} if\\
    &\frac{a_1a_3(1-n)(1-x)y}{a_2my}<d_1y+d_2z\\
    i.e.\hspace{4pt} &a_1a_3(1-n)(1-x)<a_2m(d_1y+d_2z)\\
    \text{and}\hspace{8pt} & \frac{a_3z-bnxy}{m(y+z)+x}<0\\
    i.e.\hspace{4pt} &a_3z<bnxy
\end{align*}
Hence, we obtain the proof.
\section{\texttt{Proof of local stability of\\ co-extant equilibrium}} \label{proof-sta-co}
The Jacobian matrix of our bio-system is given by
\begin{equation}
J(x, y, z)= \left(
\begin{array}{ccc}
 E_{11} & E_{12} & E_{13} \\
E_{21} & E_{22} & E_{23} \\
 0 & c & -d_2
\end{array}
\right) \label{jacobian}
\end{equation}
Hence, 
\begin{align*}
E_{11}&=\frac{a_1 (1-n)^2 x y}{(m (y+z)+(1-n) x)^2}-\frac{a_1 (1-n) y}{m (y+z)+(1-n) x}\\
&-\frac{a_2 z}{m (y+z)+x}+\frac{a_2 x z}{(m (y+z)+x)^2}-2 x+1,\\
E_{12}&=-\frac{a_1 (1-n) x}{m (y+z)+(1-n) x}+\frac{a_1 m (1-n) x y}{(m (y+z)+(1-n) x)^2}\\
&+\frac{a_2 m x z}{(m (y+z)+x)^2},\\
E_{13}&=\frac{a_1 m (1-n) x y}{(m (y+z)+(1-n) x)^2}-\frac{a_2 x}{m (y+z)+x}\\
&+\frac{a_2 m x z}{(m (y+z)+x)^2},\\
E_{21}&=\frac{a_3 z}{m (y+z)+x}-\frac{a_3 x z}{(m(y+z)+x)^2}\\
&-\frac{b n y}{m (y+z)+x}+\frac{b n x y}{(m (y+z)+x)^2},\\
E_{22}&=-\frac{a_3 m x z}{(m (y+z)+x)^2}-\frac{b n x}{m (y+z)+x}\\
&+\frac{b m n x y}{(m (y+z)+x)^2}-c-d_1,\\
E_{23}&=\frac{a_3 x}{m (y+z)+x}-\frac{a_3 m x z}{(m(y+z)+x)^2}\\
&+\frac{b m n x y}{(m (y+z)+x)^2}.
\end{align*}
Jacobian matrix \ref{jacobian} at the co-extant equilibrium point can be obtained by swapping (x,y,z) with $(x^*,y^*,z^*)$,and whose characteristic equation is given below
\begin{equation} \label{cheq}
    \lambda^3+\chi_1 \lambda ^2+\chi_2 \lambda +\chi_3 =0
\end{equation}
where,
\begin{align*}
\chi_1&=-(E_{11}+E_{22}-d_2),\\
\chi_2 &=E_{11}E_{22}-E_{22}d_2-d_2E_{11}-(E_{12}E_{21}+E_{23}c),\\
\chi_3 &=E_{11}(E_{22}d_2+E_{23}c)-E_{12}E_{21}d_2-E_{13}E_{21}c,\\
\text{and }\chi_4 &=\chi_1\chi_2 -\chi_3.
\end{align*}
By using Routh-Hurwitz criteria, we can see that the point $E_2(x^*, y^*, z^*)$ is locally asymptotically stable if $\chi_1>0, \chi_3>0, \text{and }\chi_4>0$ 
\section{\texttt{Proof of global stability of\\ co-extant equilibrium point}} \label{proof-glo-co}
The mathematical model can be written as \\
$$\frac {dX} {dt} = G(X)$$
 where,
\begin{align} \label{equa}
G(X)&=\left(
\begin{array}{c}
 x(1-x)-\frac{a_1(1-n)xy}{m(y+z)+(1-n)x}-\frac{a_2xz}{m(y+z)+x} \\
 \frac{a_3xz}{m(y+z)+x}-\frac{bnxy}{m(y+z)+x}-cy-d_1y\\
 cy-d_2z \\
\end{array}
\right),\\
\text{and }&\hspace{5pt} X=\left(
\begin{array}{c}
 x \\
 y \\
 z \\
\end{array}
\right)
\end{align}
For the Jacobian J as given in (\ref{jacobian}), its second additive compound matrix $J^{|2|}$, can be written as
$$ J^{|2|}=\left(
\begin{array}{ccc}
 E_{11}+E_{22} & E_{23} & -E_{13} \\
 c & E_{11}-d_2 & E_{12} \\
 0 & E_{21} & E_{22}-d_2 \\
\end{array}
\right)$$
Now, let us consider a function M (X) in $C^1$(D) in such a way that
$M =\left(\begin{array}{ccc}
    z/x &0&0  \\
     0&z/x&0\\
     0&0&z/x
\end{array}\right)$.\\
Again, we define,\\
$$ M_G = \frac {dM} {dX} =\left(\begin{array}{ccc}
   z/x-\dot{x}z/x^2   & 0&0 \\
     0 &z/x-\dot{x}z/x^2 &0\\
     0&0&z/x-\dot{x}z/x^2
 \end{array}\right)$$
$$ \therefore M_G M^{-1} =diag \{\frac {\dot{z}} {z} - \frac {\dot{x}} {x},\frac {\dot{z}} {z} - \frac {\dot{x}} {x}, \frac {\dot{z}} {z} - \frac {\dot{x}} {x}\} $$
$$ M J^{|2|} M^{-1} = J^{|2|}$$
We have,
\begin{align*}
N &= M_G M^{-1} + M J^{|2|} M^{-1}\\
&=(\frac{\dot{z}}{z}-\frac{\dot{x}}{x})I+J^{|2|}\\
&=\left(
\begin{array}{cc}
    N_{11} & N_{12} \\
   N_{21} & N_{22}
\end{array}
\right)
\end{align*}
where,
\begin{align*}
I&=\text{identity matrix order 3},\\
N_{11}&=\frac{dz/dt}{z}-\frac{dx/dt}{x}+E_{11}+E_{22}, \\ 
N_{12}&=\left(\begin{array}{cc} 
 E_{23} & -E_{13} \\ 
\end{array} \right),\\
N_{21}&=\left(
\begin{array}{c} 
 c \\ 
 0 \\ 
\end{array} 
\right),\\
N_{22}&=\left(
\begin{array}{cc} 
 E_{11}-d_2+\frac{\dot{z}}{z}-\frac{\dot{x}}{x} & E_{12} \\ 
 E_{21}&\frac{\dot{z}}{z}-\frac{\dot{x}}{x}+ E_{22}-d_2 \\ 
\end{array} 
\right).
\end{align*}
Considering a vector \textit{(i, j, k)} in $R^3$ whose vector norm can be writren as $|(i, j, k)|$=max\{ $|i|, |j|+|k|$\} and Lozinskii measure with respect to the norm is denoted by $\Gamma$.\\
So, $\Gamma(N) \leq sup\{\xi_1, \xi_2\}$, where $\xi_1$= $ \Gamma _1(N_{11})+|N_{12}|$, and $\xi_2 = \Gamma _1(N_{22})+|N_{21}| $.\\
Here $|N_{12}|, |N_{21}|$ are the matrix norms with respect to the $L^1$ vector norm, and $\Gamma_1$ is the Lozinskii measure with respect to that norm. Then, the required values can be obtained as\\ 
\begin{align*}
\Gamma _1(N_{11})&=\frac{dz/dt}{z}-\frac{dx/dt}{x}+E_{11}+E_{22},\\
|N_{12}|&=Q_1=max\{E_{23}, |-E_{13}|\}\\
|N_{21}|&=max\{c, 0\}=c\text{ and}\\
\Gamma _1(N_{22})&=\frac{\dot{z}}{z}-\frac{\dot{x}}{x} +max\{E_{11}+E_{21}-d_2,E_{22}+E_{12}-d_2\}\\
&=\frac{dz/dt}{z}-\frac{dx/dt}{x} +Q_2\\
& \text{where}\hspace{2pt} Q_2=max\{E_{11}+E_{21}-d_2, E_{22}+E_{12}-d_2\}.
\end{align*}
Again, from the first equation of the biosystem \ref{eq2}
\begin{align*}
    \frac{dx/dt}{x}=1-x-\frac{a_1(1-n)y}{m(y+z)+x(1-n)}-\frac{a_2z}{m(y+z)+x}
\end{align*}
Now,
\begin{align*}
 \xi_ 1&= \Gamma _1(N_{11})+|N_{12}| \\
=&\frac{dz/dt}{z}-\frac{dx/dt}{x}+E_{11}+E_{22}+Q_1,\\
\text{and }\xi_ 2&= \Gamma _1(N_{22})+|N_{21}| \\
&=\frac{dz/dt}{z}-\frac{dx/dt}{x}+Q_2+c\\
\end{align*}
Next,
$$\Gamma(N)\leq max\{\xi_ 1, \xi_ 2\}$$
i.e.
\begin{align*}
\Gamma(N)&\leq \frac{dz/dt}{z}- \frac{dx/dt}{x}+max\{E_{11}+E_{22}+Q_1, Q_2 +c\}\\
i.e\hspace{3pt} \Gamma(N)&\leq \frac{dz/dt}{z}-\frac{dx/dt}{x}+Q\\
&\text{where}\hspace{3pt} Q=max\{E_{11}+E_{22}+Q_1, Q_2 +c\}
\end{align*}
\begin{align*}
\text{Thus, } \Gamma(N)\leq \frac{\dot{z}}{z}+\vartheta \text{ where} \vartheta= Q-\frac{\dot{x}}{x}
\end{align*}
We consider $\alpha$ which is a positive real number and $T\geq 0$ such that $\alpha=inf\{x(t),y(t),z(t)\}$ when $t>T$\\
Also, we would have
\begin{align*}
Q_1&=max\{|E_{23}|, |-E_{13}|\}\\
\therefore Q_1&= max\{|\frac{a_3}{2m+1}-\frac{a_3 m}{(2m+1)^2}+\frac{b m n}{(2m +1)^2}|,\\
&|-\frac{a_1 m (1-n)}{(m2+(1-n))^2}+\frac{a_2}{m2+1}-\frac{a_2 m}{(m2+1)^2}|\}\\
\text{at } \hspace{3pt}\alpha&=inf\{x(t),y(t),z(t)\}
\end{align*}
Similarly,
\begin{align*}
Q_2 &=max\{E_{11}+E_{21}-d_2, E_{22}+E_{12}-d_2\}\\
&=max\{\frac{a_1 (1-n)^2 }{(m2+(1-n))^2}-\frac{a_1 (1-n)}{m2+(1-n)}-\frac{a_2}{m2+1}\\
&+\frac{a_2}{(m2+1)^2}-2\alpha+1+\frac{a_3}{2m+1}-\frac{a_3}{(2m+1)^2}-\frac{b n}{2m +1}\\
&+\frac{b n}{(2m+1)^2}-d_2,-\frac{a_3 m}{(2m+1)^2}-\frac{b n}{2m+1}+\frac{b m n}{(2m+1)^2}\\
&-c-d_1-\frac{a_1 (1-n)}{m2+(1-n)}+\frac{a_1 m (1-n)}{(m2+(1-n))^2}+\frac{a_2 m}{(m2+1)^2}\\
& -d_2\}\hspace{3pt} \text{at} \hspace{3pt}\alpha=inf\{x(t),y(t),z(t)\}
\end{align*}
Also, at $\alpha$
\begin{align*}
    Q&=max\{E_{11}+E_{22}+Q_1, Q_2 +c\}\\
    &=max\{\frac{a_1 (1-n)^2}{(m 2+(1-n))^2}-\frac{a_1 (1-n)}{m2+(1-n)}-\frac{a_2 }{m2+1}\\
    &+\frac{a_2}{(m2+1)^2}-2 \alpha+1-\frac{a_3 m}{(2m+1)^2}-\frac{b n}{2m+1}\\
    &+\frac{b m n}{(2m+1)^2}-c-d_1+\beta_1, \beta_2+c\},\\
    \text{and }\dot{x}/x&=1-\alpha-\frac{a_1(1-n)}{2m +1-n}-\frac{a_2}{2m+1}
\end{align*}
Hence, we would  get $\vartheta$ at $\alpha$\\
Therefore, we have\\
\begin{align*}
\Gamma(N)&\leq \frac{\dot{z}}{z}+\vartheta\\
\int_0^t  \Gamma(N)  \,ds &\leq \int_0^t \frac{dz/dt}{z} \,dt+\vartheta \int_0^t 1 \,dt \\
i.e. \int_0^t  \Gamma(N)  \, ds &\leq  log\frac{z(t)}{z(0)} +\vartheta t\\
i.e.\frac{1}{t} \int_0^t  \Gamma(N)  \,& ds\leq \frac{1}{t}log\frac{z(t)}{z(0)}+\vartheta\\
\limsup_{t\to\infty}sup\frac{1}{t}  \int_0^t  \Gamma(N)  \, ds&<\vartheta <0
\end{align*}
 Thus the system of equations is globally asymptotically stable around its co-extant equilibrium point, when $\vartheta<0$
\section{\texttt{Proof of perseverance}} \label{proof-perse}
At the outset, we shall find the condition for which $x(t)>0$ for all $t>0$, from 
the first equation of our bio-system \ref{eq2} we have the following: 
\begin{align*}
    \dot{x}&=x(1-x)-\frac{a_1(1-n)x y}{m(y+z)+x(1-n)}-\frac{a_2xz}{m(y+z)+x}\\
    &\geq x(1-x)-\frac{a_1(1-n)x y}{my}-\frac{a_2xz}{mz}\\
    &=x(1-x-a_1(1-n)/m-a_2/m).\\
\end{align*}
So, $\liminf_{t\to \infty} x(t)\geq 1-\frac{a_1(1-n)+a_2}{m}=L(say)$
Thus $L>0 \hspace{5pt} i.e.,\hspace{3pt} m>a_1(1-n)+a_2$
If the above condition is satisfied then, for the bio-system the vanishing equilibrium point cannot be achieved as the prey would always exists. The next plausible manifestation would be either the predator free equilibrium point or the co-extant equilibrium point. Now from the previous stability theorem of the axial equilibrium point (\ref{loca-sta1}), if we take $a_3c>d_2(bn+c+d_1)$, then the predator free equilibrium point too would be unstable. Hence, the only possible scenerio left would be the co-existence of all three populations. Moreover without the stability of the co-extant equilibrium point, there would exist some  bounded trajectory in the positive octant showing  co-existence of the populations. 
 Therefore, we can say, the biosystem \ref{eq2} perseveres if
 $$ m>a_1(1-n)+a_2 \text{  and } a_3c>d_2(bn+c+d_1).$$
\section{\texttt{Criteria of Hopf-bifurcation}}
$n_h$ is assumed to be the point of Hopf-bifurcation for the parameter \textit{n}.
A necessary and sufficient condition for hopf bifurcation at $n_h$ is 
\begin{enumerate}
    \item[(i)] $\chi_1,\chi_2>0,\chi_3=\chi_1\chi_2$. $\chi_i,i=1,2,3$ are from (\ref{cheq}).
    \item[(ii)] $ \frac{d}{dn} (Re(\lambda_i (n)))_{n = n_h} \neq 0 \hspace{5pt} for\hspace{5pt} i=1,2,3 $
\end{enumerate}

The characteristic equation given in \ref{cheq}:\\
\begin{align}
    \notag &\lambda^3+\lambda^2\chi_1+\lambda \chi_2+\chi_3=0\\
    \notag &\text{for} \hspace{5pt} \chi_3=\chi_1 \chi_2 \text{ at } n=n_h\\
    \notag\text{So, } &\lambda^3+\lambda^2\chi_1+\lambda \chi_2+\chi_1 \chi_2=0\\
    \notag i.e.&\lambda^2(\lambda+\chi_1)+\chi_2(\lambda+\chi_1)=0\\
    \label{1} i.e.&(\lambda^2+\chi_2)(\lambda+\chi_1)=0\\
    \notag\text{Thus, } &\lambda=-\chi_1, \hspace{4pt}\pm \sqrt{\chi_2}i\text{ if } \chi_1,\chi_2\text{ are positive}
\end{align}
For  $n \in (n_h- \epsilon, n_h+ \epsilon)$, the general form of the roots are \\
$\lambda_1= \Psi_1 (n) +i \Psi_2(n)$, \\
$\lambda_2= \Psi_1 (n) -i \Psi_2(n)$, and\\
$\lambda_3 = - \chi_1 (n)$\\ 
Now, for transversality condition, we need to establish
$$ \frac{d}{dn} (Re(\lambda_i (n)))_{n = n_h} \neq 0 \hspace{5pt} for\hspace{5pt} i=1,2,3 $$
Substituting  $\lambda_1= \Psi_1 (n) +i \Psi_2(n)$ in (\ref{1}), we have
\begin{equation} \label{2}
(\Psi_1 ^2 - \Psi_2 ^2 +2 i \Psi_1 \Psi_2 +\chi_2) (\Psi_1 +i \Psi_2 +\chi_1)=0
\end{equation}
Differentiating (\ref{2}) and separating the real and imaginary part we get the following\\
$$P_1(n) \Psi_1' (n) – P_2 (n) \Psi’{_2} +R_1(n)=0, $$
$$\text{and }P_2(n) \Psi’{_1} (n)+ P_1 (n) \Psi’{_2} +R_2(n)=0. $$
here
$P_1=  3 \Psi_1 ^2 -3 \Psi_2^2  + \chi_2 +2 \chi_1 \Psi_1$,\\
$P_2= 6 \Psi_1 \Psi_2 + 2\Psi_2 \chi_1$,\\
$R_1= \chi_1’ \Psi_1^2 - \chi_1’ \Psi_2^2+ \chi_2’ \Psi_1 +\chi_3' $,\\
and $R_2= \chi_2’\Psi_2 +2 \chi_1’\Psi_1\Psi_2$.\\
Using $\Psi_1(n_h)=0$ and $\Psi_2 (n_h)= i \sqrt{\chi_2 (n_h)}$ we get \\
$$P_1(n_h)=-3 \chi_2 +\chi_2 = -2 \chi_2,\hspace{3pt}P_2(n_h)=2 \chi_1 \sqrt{\chi_2},$$
$$R_1(n_h)= \chi_3'- \chi_1' \chi_2,\hspace{3pt}\text{and }R_2(n_h)= \sqrt{\chi_2} \chi_2’ .$$
Hence, 
\begin{align*}
    \frac{d}{dn}[Re(\lambda_{1, 2}(n))]_{n= n_h}&=\Psi_1\\
    &= -\frac{P_1 R_1 +P_2 R_2}{P_1^2+P_2^2}\\
    &= - \frac{-2 \chi_2 (\chi_3' -\chi_1' \chi_2) +2 \chi_1 \chi_2 \chi_2'}{4 \chi_2^2+ 4 \chi_1^2 \chi_2}\\
    &= - \frac{\chi_1' \chi_2 -\chi_3'+\chi_1 \chi_2'}{2 \chi_2 +2 \chi_1^2}\\
    & \neq 0
\end{align*}
$$\text{if }\chi_1' \chi_2 -\chi_3'+\chi_1 \chi_2' \neq 0$$
$$ i.e \hspace{5pt} \frac{d}{dn} (\chi_1 \chi_2 -\chi_3) \neq 0 \hspace{5pt} at \hspace{5pt} n=n_h. $$
Therefore, the transversality conditions hold.
Hence, the biosystem \ref{eq2}  undergoes Hopf-bifurcation around the positive interior equilibrium point $E_2(x^*, y^*, z^*)$. when the habitat complexity(\textit{n}) crosses the critical value $n=n_h$. 
\section{\texttt{Direction and stability of\\ bifurcating solutions}}
Here we endevour to determine the direction and stability criterion of the bifurcating periodic solution arising from Hopf-bifurcation with the help of procedure given by Hassard et al \cite{hassa}, where the set of differential equations of the system \ref{eq2} is reduced into its normal form.\\
We introduce new variables ($x', y',z'$) as\\
$x= x' +x^*, y=y'+y^*, z=z'+z^*$ 
For simplicity taking: $x'\to x,y'\to y,z'\to z$.  
The system of equation \ref{eq2} is transformed to
\begin{align*}
    \frac{dx}{dt}&=(x+x^*)(1-x+x^*)-\frac{a_2(x+x^*)(z+z^*)}{m(y+z+y^*+z^*)+x+x^*}\\
    &-\frac{a_1(1-n)(x+x^*)(z+z^*)}{m(y+z+y^*+z^*)+(x+x^*)(1-n)},\\
    \frac{dy}{dt}&=\frac{a_3(x+x^*)(z+z^*)}{m(y+z+y^*+z^*)+x+x^*}-c(y+y^*)\\
    &-\frac{bn(x+x^*)(z+z^*)}{m(y+z+y^*+z^*)+x+x^*}-d_1(y+y^*),\\
    \frac{dz}{dt}&=c(y+y^*)-d_2(z+z^*).
\end{align*}
Partioning the above equations into linear and non-linear part, we get
\begin{equation} \label{pa-ho}
    \dot{x}=ln_1+nln_1;\hspace{2pt}\dot{y}=ln_2+nln_2;\hspace{2pt}\dot{z}=ln_3+nln_3
\end{equation}
Here, $ln_i,(i=1,2,3)$ and $nln_i(i=1,2,3)$ are the linear and non-linear part respectively.
(\ref{pa-ho}) is written as
\begin{equation}\label{b1}
    \dot{X}=RX+S
\end{equation}
where A is the Jacobian matrix of the linear part and B is the non linear part
$$ X= \left(
\begin{array}{c}
 x\\
 y \\
 z \\
\end{array}
\right), R= \left(
\begin{array}{ccc}
 r_{11} & r_{12} & r_{13} \\
 r_{21} & r_{22} & r_{23} \\
 r_{31} & r_{32} & r_{33} \\
\end{array}
\right), S= \left(
\begin{array}{c}
 S_1 \\
 S_2 \\
 S_3 \\
\end{array}
\right)$$
where,
\begin{align*}
    r_{11}&=\left(\frac{a_1 (n-1) y^*m (y^*+z^*)}{(m (y^*+z^*)-n x^*+x^*)^2}-\frac{a_2 z^*m (y^*+z^*)}{(m(y^*+z^*)+x^*)^2}\right)\\
    &-2 x^*+1,\\
    r_{12}&=\left(\frac{a_2 m z^*x^*}{(m (y^*+z^*)+x^*)^2}-\frac{a_1x^*((n-1)^2 x^*-(1-n)m z^*)}{(m (y^*+z^*)-n x^*+x^*)^2}\right),\\
    r_{13}&=\left(-\frac{a_1 m (n-1)y^*x^*}{(m(y^*+z^*)-n x^*+x^*)^2}-\frac{a_2 (m y^*+x^*)x^*}{(m(y^*+z^*)+x^*)^2}\right),
\end{align*}
\begin{align*}
    r_{21}&=\frac{-m z^* (b ny^*-a_3 y^*)+a_3 mz^{*2}-b m n y^{*2}}{(m (y^*+z^*)+x^*)^2},\\
    r_{22}&=[-mz^* (a_3 x^*+b n x^*)-b nx^{*2}-c (m (y^*+z^*)+x^*)^2\\
    &-d_1 (m (y^*+z^*)+x^*)^2]/[(m (y^*+z^*)+x^*)^2],\\
    r_{23}&=\frac{a_3 m x^*y^*+a_3x^{*2}+b m n x^*y^*}{(m (y^*+z^*)+x^*)^2},\\
    r_{31}&=0,\hspace{5pt} r_{32}=c,\hspace{5pt} r_{33}=-d_2.
\end{align*}
Now we consider two conjugate imaginary eigen values\\
$\lambda_{1, 2}= \pm \beta$ and other eigen value $\lambda_3=v_1$ [where $\beta = i \sqrt{\eta_2}$] of the characteristic equation (\ref{cheq})\\
We procure a transformation matrix T such that \\
$$ T^{-1} R T = \left(
\begin{array}{ccc}
 0 & -\beta  & 0 \\
 \beta  & 0 & 0 \\
 0 & 0 & v \\
\end{array}
\right)$$
Now, T is a non-singular matrix which is given as\\
$$ T=\left(
\begin{array}{ccc}
 1 & 0 & 1 \\
 t_{21} & t_{22} & t_{23} \\
 t_{31} & t_{32} & t_{33} \\
\end{array}
\right)$$
where
\begin{align*}
    t_{21}&=[r_{23}(r_{21} r_{32} r_{33}+r_{22} r_{31} r_{33}-r_{31} \beta ^2)-r_{21} r_{22} (r_{33}^2\\
    &+\beta ^2)-r_{23}^2 r_{31} r_{32}]/[\left(r_{22}^2+\beta ^2\right) \left(r_{33}^2+\beta ^2\right)\\
    &+2 r_{23} r_{32} \left(\beta ^2-r_{22} r_{33}\right)+r_{23}^2 r_{32}^2],\\
    t_{22}&=\frac{\beta  \left(r_{21} \left(r_{23} r_{32}+r_{33}^2+\beta ^2\right)-r_{22} r_{23} r_{31}-r_{23} r_{31} r_{33}\right)}{\left(r_{22}^2+\beta ^2\right) \left(r_{33}^2+\beta ^2\right)+2 r_{23} r_{32} \left(\beta ^2-r_{22} r_{33}\right)+r_{23}^2 r_{32}^2},\\
    t_{23}&=\frac{r_{21} (v-r_{33})+r_{23} r_{31}}{(r_{22}-v) (r_{33}-v)-r_{23} r_{32}},\\
    t_{31}&=-[-r_{22} r_{32} (r_{21} r_{33}+r_{23} r_{31})+r_{21}r_{32} \left(r_{23} r_{32}+\beta ^2\right)\\
    &+r_{22}^2 r_{31} r_{33}+r_{31} r_{33} \beta ^2]/[\left(r_{22}^2+\beta ^2\right) \left(r_{33}^2+\beta ^2\right)\\
    &+2 r_{23} r_{32} \left(\beta ^2-r_{22} r_{33}\right)+r_{23}^2 r_{32}^2],\\
    t_{32}&=\frac{\beta  \left(-r_{21} r_{22} r_{32}-r_{21} r_{32} r_{33}+r_{22}^2 r_{31}+r_{23} r_{31} r_{32}+r_{31} \beta ^2\right)}{\left(r_{22}^2+\beta ^2\right) \left(r_{33}^2+\beta ^2\right)+2 r_{23} r_{32} \left(\beta ^2-r_{22} r_{33}\right)+r_{23}^2 r_{32}^2},\\
    t_{33}&=\frac{r_{21} r_{32}-r_{22} r_{31}+r_{31} v}{(r_{22}-v) (r_{33}-v)-r_{23} r_{32}}
\end{align*}
Again, we make another change of variable to obtain the normal form of (\ref{b1}).\\
$$\text{We consider } X= TY$$
Hence, $$ \dot{Y} = T^{-1} R T Y + T^{-1} S$$
$$ \dot{Y} = (T^{-1} R T) Y +U $$ 
where $U= \left(
\begin{array}{c}
 U_1 \\
 U_2 \\
 U_3 \\
\end{array}
\right)$
$U_i(y_1, y_2, y_3)$ \textit{i}=1, 2, 3) can be obtained by transforming $B_i$'s using the variables\\
$x = y_1+ y_3, \hspace{5pt} y= t_{21} y_1 +t_{22} y_2+ t_{23} y_3, \hspace{5pt} z= t_{31} y_1+ t_{32} y_2 +t_{33} y_3.$\\ \\
Now we derive the expressions $g_{11}, g_{02}, g_{20}, G_{21}, h^1_{11}, h^1_{20},\\w_{11}, w_{20}, G^1_{110}, G^1_{101}, g_{21}$
at $(y_1,y_2,y_3)=(0,0,0)$
\begin{align*}
    g_{11}&=\frac{1}{4} \left(\frac{\partial ^2U_1}{\partial y_1\, \partial y_1}+\frac{\partial ^2U_1}{\partial y_2\, \partial y_2}+i \left(\frac{\partial ^2 U_2}{\partial y_1\, \partial y_1}+\frac{\partial ^2 U_2}{\partial y_2\, \partial y_2}\right)\right)\\
    g_{02}&=\frac{1}{4} [i \left(2 \frac{\partial ^2U_1}{\partial y_1\, \partial y_2}+\frac{\partial ^2U_2}{\partial y_1\, \partial y_1}-\frac{\partial ^2U_2}{\partial y_2\, \partial y_2}\right)\\
    &+\frac{\partial ^2U_1}{\partial y_1\, \partial y_1}-\frac{\partial ^2U_1}{\partial y_2\, \partial y_2}-2 \frac{\partial ^2U_2}{\partial y_1\, \partial y_2}]\\
    g_{20}&=\frac{1}{4} [i \left(-2 \frac{\partial ^2U_1}{\partial y_1\, \partial y_2}+\frac{\partial ^2U_2}{\partial y_1\, \partial y_1}-\frac{\partial ^2U_2}{\partial y_2\, \partial y_2}\right)\\
    &+\frac{\partial ^2U_1}{\partial y_1\, \partial y_1}-\frac{\partial ^2U_1}{\partial y_2\, \partial y_2}+2 \frac{\partial ^2U_2}{\partial y_1\, \partial y_2}]\\
    G_{21}&=\frac{1}{8}[i (-\frac{\partial ^3U_1}{\partial y_1\, \partial y_1\, \partial y_2}-\frac{\partial ^3U_1}{\partial y_2\, \partial y_2\, \partial y_2}+\frac{\partial ^3U_2}{\partial y_1\, \partial y_2\, \partial y_2}\\
    &+\frac{\partial ^3U_2}{\partial y_1\, \partial y_1\, \partial y_1})+\frac{\partial ^3U_1}{\partial y_1\, \partial y_2\, \partial y_2}+\frac{\partial ^3U_1}{\partial y_1\, \partial y_1\, \partial y_1}\\
    &+\frac{\partial ^3U_2}{\partial y_1\, \partial y_1\, \partial y_2}+\frac{\partial ^3U_2}{\partial y_2\, \partial y_2\, \partial y_2}]
\end{align*}
\begin{align*}
    G^1_{110}&=\frac{1}{2} \left(i \left(\frac{\partial ^2U_2}{\partial y_1\, \partial y_3}-\frac{\partial ^2U_1}{\partial y_2\, \partial y_3}\right)+\frac{\partial ^2U_1}{\partial y_1\, \partial y_3}+\frac{\partial ^2U_2}{\partial y_2\, \partial y_3}\right)\\
    G^1_{101}&=\frac{1}{2} \left(i \left(\frac{\partial ^2U_1}{\partial y_2\, \partial y_3}+\frac{\partial ^2U_2}{\partial y_1\, \partial y_3}\right)+\frac{\partial ^2U_1}{\partial y_1\, \partial y_3}-\frac{\partial ^2U_2}{\partial y_2\, \partial y_3}\right)\\
    h_{11}^1&=\frac{1}{4} \left(\frac{\partial ^2U_3}{\partial y_1\, \partial y_1}+\frac{\partial ^2U_3}{\partial y_2\, \partial y_2}\right)\\
    h_{20}^1&=\frac{1}{4} \left(-2 i \frac{\partial ^2U_3}{\partial y_1\, \partial y_2}+\frac{\partial ^2U_3}{\partial y_1\, \partial y_1}-\frac{\partial ^2U_3}{\partial y_2\, \partial y_2}\right)\\
    &v_1 w_{11} = - h_{11}^1\\
    &(v_1 -2 i \beta) w_{20} =- h^1_{20}\\ 
    and, g_{21}& = G_{21} +2 G_{110}^1 w_{11} +G_{101}^1 w_{20}
\end{align*}
Now, calculating the quantities, we get the following\\ 
$$C_1 (0)= \frac{i}{2 \beta} (g_{11} g_{20} -2 |g_{11}|^2 - \frac{1}{3} |g_{02}|^2) +\frac{1}{2} g_{21}$$
$$\mu_2 = - Re C_1(0)/ \alpha'(0)$$
$$T_2 = -(Im(C_1(0))+ \mu_2 \beta'(0))/(\beta) $$
$$\beta_2 = 2 Re C_1(0)$$
where $\alpha'(0) = Re \lambda'_1(n_h), \vspace{0.5 cm} \beta'(0)= Im \lambda'_1(n_h)$ \hspace{5pt} $n_h$ is the bifurcation point and $\lambda_1(n_h)=i \beta$.
\section{\texttt{Proof of transcritical\\ bifurcation}}
The Jacobian matrix of the bio-system at the equilibrium point $E_1(1,0,0)$ is as given in [\ref{lo-jaco}]. One of its eigen-values will be zero if $a_3 c=d_2(c+d_1+b n)$  and other two negative. Taking \textit{b} as the bifurcating parameter i.e $b=b_t=\frac{a_3c-cd_2-d_1d_2}{nd_2}$.\\
For $b=b_t$, the variational matrix at $E_1$ would be\\
\begin{align*}
J_1=\left(
\begin{array}{ccc}
 -1&a_1&-a_2\\
 0&-\frac{a_3c}{d_2}&a_3 \\
 0&c&-d_2\\
\end{array}\right)
\end{align*}
Then, $W_1=(v_1,v_2,v_3)^t=(-\left(\frac{a_2c+a_1d_2}{d_2}\right)v_2, v_2, \frac{cv_2}{d_2})^t$ and\\
$W_2=(w_1,w_2,w_3)^t=(0,w_2,\frac{a_3w_2}{d_2})^t$  are the eigen vectors corresponding to the zero eigen-value for the matrices  $J_1$ and $J_1^t$ respectively. 
The biosystem \ref{eq2} is written in the form $\dot{X}=G(X)$ as is in (\ref{equa})
Applying Sotomayor theorem \cite{trans} for transcritical behavior, we should have,\\
$$W_2^t G_b(E_1;b_t)=0,\hspace{5pt}  W_2^t(DG_b(E_1;b_t)V)\neq0$$
and $$W_2^t[D^2G(E_1;b_t)V]\neq0$$
$G_b(E_1;b_t)= \left(\begin{array}{c} 0 \\  -\frac{nxy}{m(y+z)+x} \\  0 \\ \end{array} \right) _{(at \hspace{3pt}(1,0,0), b_t)}=\left(\begin{array}{c} 0 \\  0 \\  0 \\ \end{array} \right)$\\ $DG_b(E_1;b_t)= \left(\begin{array}{ccc}
 0 & 0 & 0 \\
 0 & -n & 0 \\
 0 & 0 & 0 \\
\end{array}
\right)$ $ D^2G(E_1;b_t)V =\left(\begin{array}{c} \gamma_1 \\
\gamma_2 \\
\gamma_3 \end{array}\right)$\\
where\\
$\gamma_1=v_2^2 \left(-\frac{2 (a_1 d_2+a_2 c)^2}{d_2^2}+\frac{c m \left(\frac{a_1 (n-1)^2}{(1-n)^3}+a_2\right)}{d_2}-\frac{2a_1 m}{n-1}+\frac{2 a_2 c^2 m}{d_2^2}\right)$,\\
$\gamma_2=-\frac{m v_2^2 (a_3 c (c-d_2)+d_2 (c+d_1) (c+2 d_2))}{d_2^2}$,\\
$\gamma_3=0$\\
Hence,
\begin{align*}
W_2^t G_b(E_1;b_t)&=0\\
W_2^t(DG_b(E_1;b_t)V)&=-nw_2v_2 \neq0\\
W_2^t[D^2G(E_1;b_t)V]&=-\frac{m v_2^2 (a_3 c (c-d_2)+d_2 (c+d_1) (c+2 d_2))}{d_2^2} w_2\\
&\neq 0
\end{align*}
Hence our result is proven.
\end{appendix}

{\bf Conflict of interest: } The authors declare that there is no conflict of interest.

\end{multicols}
\end{document}